\documentclass[twocolumn, tighten, times]{aastex701}
\usepackage[utf8]{inputenc}
\usepackage{color}
\usepackage{longtable}
\usepackage{xspace}
\usepackage{amsmath}
\usepackage{amssymb}
\usepackage{mathptmx,txfonts,tikz,bm} 
\usepackage[multidot]{grffile}
\usepackage{xcolor, fontawesome}
\usepackage{graphicx}
\usepackage{comment}
\usepackage{color}
\usepackage{booktabs}
\usepackage{tabularx}
\usepackage{threeparttable}
\usepackage{multirow}
\usepackage{makecell}

\DeclareRobustCommand{\VAN}[3]{#2}
\let\VANthebibliography\thebibliography
\def\thebibliography{\DeclareRobustCommand{\VAN}[3]{##3}\VANthebibliography}

\usepackage{academicons}
\usepackage{svg}
\newcommand{\mmode}[1]{\ifmmode{#1}\else{$#1$}\fi}

\newcommand{\Teff}[0]{\mmode{T_\text{eff}}}
\newcommand{\Rsun}[0]{\mmode{\text{R}_{\odot}}}
\newcommand{\Msun}[0]{\mmode{\text{M}_{\odot}}}
\newcommand{\Lsun}[0]{\mmode{\text{L}_{\odot}}}

\newcommand{\Mjup}[0]{\mmode{\text{M}_{\text{Jup}}}}

\newcommand{\logg}[0]{\mmode{\log g}}
\newcommand{\Dnu}[0]{\mmode{\Delta\nu}}
\newcommand{\dnu}[1]{\mmode{\delta\nu_{#1}}}
\newcommand{\numax}[0]{\mmode{\nu_\text{max}}}
\newcommand{\Aosc}[0]{\mmode{A_\text{osc}}}

\newcommand{\muHz}[0]{\mmode{\mu{\rm Hz}}}
\newcommand{\Tobs}[0]{\mmode{T_{\rm obs}}}

\newcommand{\Yinit}[0]{\mmode{Y_{\rm init}}}

\newcommand{\amlt}[0]{\mmode{\alpha_{\rm MLT}}}

\newcommand{\mh}[0]{\mmode{{\rm [M/H]}}}
\newcommand{\feh}[0]{\mmode{{\rm [Fe/H]}}}

\newcommand{\logrphk}[0]{\mmode{\log R'_{\rm HK}}}
\newcommand{\Prot}[0]{\mmode{P_{\rm rot}}}

\newcommand{\Ro}[0]{\mmode{{\rm Ro}}}

\newcommand{\age}[0]{\mmode{\rm Age}}

\newcommand{\Lbol}{\mbox{$L_{bol}$}}

\newcommand{\cyan}[1]{\textcolor{cyan} }

\usepackage{CJKutf8}

\makeatletter
\newcommand\thefontsize[1]{{#1 The current font size is: \f@size pt\par}}
\makeatother

\begin{document}

\begin{CJK}{UTF8}{gbsn}

\title{A Test of Substellar Evolutionary Models with High-Precision Ages\\ from Asteroseismology and Gyrochronology for the Benchmark System HR 7672AB}

\newcommand{\PSUAA}{Department of Astronomy \& Astrophysics, 525 Davey Laboratory, The Pennsylvania State University, University Park, PA, 16802, USA}
\newcommand{\PSUCEHW}{Center for Exoplanets \& Habitable Worlds, 525 Davey Laboratory, The Pennsylvania State University, University Park, PA 16802, USA}
\newcommand{\PSETI}{Penn State Extraterrestrial Intelligence Center, 525 Davey Laboratory, The Pennsylvania State University, University Park, PA 16802, USA}
\newcommand{\UA}{Steward Observatory, The University of Arizona, 933 N.\ Cherry Ave, Tucson, AZ 85721, USA}
\newcommand{\Caltech}{Department of Astronomy, California Institute of Technology, Pasadena, CA 91125, USA}
\newcommand{\JHU}{Department of Physics \& Astronomy, Bloomberg Center, Johns Hopkins University, Baltimore, MD 21218, USA}
\newcommand{\Macquarie}{School of Mathematical and Physical Sciences, Macquarie University, Balaclava Road, North Ryde, NSW 2109, Australia}
\newcommand{\CUBoulder}{Department of Physics, 390 UCB, University of Colorado, Boulder, CO 80309, USA}
\newcommand{\JPL}{Jet Propulsion Laboratory, California Institute of Technology, 4800 Oak Grove Drive, Pasadena, CA 91109, USA}
\newcommand{\MITEAPS}{Department of Earth, Atmospheric, and Planetary Sciences, Massachusetts Institute of Technology, Cambridge, MA 02139, USA}
\newcommand{\MITKavli}{Kavli Institute for Astrophysics and Space Research, Massachusetts Institute of Technology, Cambridge, MA 02139, USA}
\newcommand{\UCI}{Department of Physics \& Astronomy, The University of California, Irvine, Irvine, CA 92697, USA}
\newcommand{\Carnegie}{Earth and Planets Laboratory, Carnegie Institution for Science, 5241 Broad Branch Road, NW, Washington, DC 20015, USA}
\newcommand{\PSUICS}{Institute for Computational and Data Sciences, The Pennsylvania State University, University Park, PA 16802, USA}
\newcommand{\PSUCASt}{Center for Astrostatistics, 525 Davey Laboratory, The Pennsylvania State University, University Park, PA 16802, USA}
\newcommand{\Princeton}{Department of Astrophysical Sciences, Princeton University, 4 Ivy Lane, Princeton, NJ 08540, USA}
\newcommand{\IAS}{Institute for Advance Study, 1 Einstein Drive, Princeton, NJ 08540, USA}
\newcommand{\Tsinghua}{Department of Astronomy, Tsinghua University, Beijing 100084, China}
\newcommand{\FlatironCCA}{Center for Computational Astrophysics, Flatiron Institute, 162 Fifth Avenue, New York, NY 10010, USA}
\newcommand{\ETH}{ETH Zurich, Institute for Particle Physics \& Astrophysics, Zurich, Switzerland}
\newcommand{\UCO}{UC Observatories, University of California, Santa Cruz, CA 95064, USA}
\newcommand{\SantaCruz}{University of California, Santa Cruz}
\newcommand{\WMKO}{W.\ M.\ Keck Observatory, 65-1120 Mamalahoa Hwy, Kamuela, HI 96743, USA}
\newcommand{\SSL}{Space Sciences Laboratory, University of California, Berkeley, CA 94720, USA}
\newcommand{\UH}{Institute for Astronomy, University of Hawai‘i, 2680 Woodlawn Drive, Honolulu, HI 96822, USA}
\newcommand{\UCB}{Department of Astronomy, 501 Campbell Hall, University of California, Berkeley, CA 94720, USA}
\newcommand{\UCLA}{Department of Physics \& Astronomy, University of California Los Angeles, Los Angeles, CA 90095, USA}
\newcommand{\nexsci}{NASA Exoplanet Science Institute/Caltech-IPAC, California Institute of Technology, Pasadena, CA
91125, USA}
\newcommand{\COO}{Caltech Optical Observatories, California Institute of Technology, Pasadena, CA 91125, USA}
\newcommand{\Sydney}{Sydney Institute for Astronomy (SIfA), School of Physics, University of Sydney, NSW 2006, Australia}
\newcommand{\Kansas}{Department of Physics and Astronomy, University of Kansas, Lawrence, KS, USA}
\newcommand{\Warwick}{Physics Department, University of Warwick, Coventry CV4 7AL, United Kingdom}
\newcommand{\Yale}{Department of Astronomy, Yale University, New Haven, CT 06511, USA}
\newcommand{\ICL}{Astrophysics Group, Department of Physics, Imperial College London, Prince Consort Rd, London SW7 2AZ, UK}
\newcommand{\Schmidt}{Astrophysics \& Space Institute, Schmidt Sciences, New York, NY 10011, USA}
\newcommand{\Amsterdam}{University of Amsterdam}
\newcommand{\ND}{Department of Physics and Astronomy, University of Notre Dame, Notre Dame, IN 46556, USA}
\newcommand{\NUS}{Department of Physics, National University of Singapore, 21 Lower Kent Ridge Road, 119077, Singapore}
\newcommand{\Geneva}{Observatoire Astronomique de l'Université de Genève, Chemin Pegasi 51, 1290 Versoix, Switzerland}
\newcommand{\PortoIAC}{Instituto de Astrof\'{i}sica e Ci\^{e}ncias do Espaço, Universidade do Porto, Rua das Estrelas, 4150-762 Porto, Portugal}
\newcommand{\Porto}{Departamento de F\'{i}sica e Astronomia, Faculdade de Ci\^{e}ncias da Universidade do Porto, Rua do Campo Alegre, s/n, 4169-007 Porto, Portugal}
\newcommand{\Aarhus}{Stellar Astrophysics Centre (SAC), Department of Physics and Astronomy, Aarhus University, Ny Munkegade 120, 8000 Aarhus C, Denmark}
\newcommand{\Edinburgh}{Institute for Astronomy, University of Edinburgh, Royal Observatory, Blackford Hill, Edinburgh EH9 3HJ, UK}
\newcommand{\Oslo}{Rosseland Centre for Solar Physics, University of Oslo, PO Box 1029 Blindern, 0315 Oslo, Norway}
\newcommand{\WDRC}{Center for Solar-Stellar Connections, WDRC, 9020 Brumm Trail, Golden, CO 80403, USA}


\author[0000-0003-3020-4437]{Yaguang~Li (李亚光)}
\affiliation{\UH}
\email[show]{yaguangl@hawaii.edu}

\author[orcid=0000-0003-2232-7664]{Michael C. Liu}
\email{mliu@ifa.hawaii.edu}
\affiliation{\UH}

\author[orcid=0000-0001-9823-1445]{Trent J. Dupuy}
\email{tdupuy@roe.ac.uk}
\affiliation{\Edinburgh}

\author[orcid=0000-0001-8832-4488]{Daniel Huber}
\email{huberd@hawaii.edu}
\affiliation{\UH}


\author[orcid=0000-0002-2696-2406]{Jingwen Zhang (张婧雯)}
\email{jingwen7@hawaii.edu}
\affiliation{\UH}
\affiliation{Department of Physics, University of California, Santa Barbara, CA 93106, USA}


\author[orcid=0000-0003-3244-5357]{Daniel Hey}
\email{dhey@hawaii.edu}
\affiliation{\UH}


\author[orcid=0009-0008-6039-6381]{R.R. Costa}
\email{up201909641@edu.fc.up.pt}
\affiliation{\PortoIAC}
\affiliation{\Porto}

\author[orcid=0009-0006-0423-2353]{Jens Reersted Larsen}
\email{jensrl@phys.au.dk}
\affiliation{\Aarhus}

\author[orcid=0000-0001-7664-648X]{J. M. Joel Ong (王加冕)}
\email{joelong@hawaii.edu}
\altaffiliation{NASA Hubble Fellow}
\affiliation{\UH}

\author[orcid=0000-0002-6163-3472]{Sarbani Basu}
\email{sarbani.basu@yale.edu}
\affiliation{\Yale}

\author[orcid=0000-0003-4034-0416]{Travis S. Metcalfe}
\email{travis@wdrc.org}
\affiliation{\WDRC}

\author[orcid=0000-0003-0817-6126]{Yixiao Zhou (周一啸)}
\email{yixiao.zhou@astro.uio.no}
\affiliation{\Oslo}
 

\author[orcid=0000-0002-4284-8638]{Jennifer van Saders}
\email{jlvs@hawaii.edu}
\affiliation{\UH}


\author[orcid=0000-0001-5222-4661]{Timothy R. Bedding}
\email{tim.bedding@sydney.edu.au}
\affiliation{\Sydney}

\author[orcid=0000-0003-2400-6960]{Marc Hon}
\email{mtyhon@mit.edu}
\affiliation{\NUS}

\author[orcid=0000-0002-9037-0018]{Hans Kjeldsen}
\email{hans@phys.au.dk}
\affiliation{\Aarhus}


\author[orcid=0000-0002-4588-5389]{Tiago L. Campante}
\email{Tiago.Campante@astro.up.pt}
\affiliation{\PortoIAC}
\affiliation{\Porto}

\author[orcid=0000-0003-0513-8116]{Mário J. P. F. G. Monteiro}
\email{mario.monteiro@astro.up.pt}
\affiliation{\PortoIAC}
\affiliation{\Porto}

\author[orcid=0000-0002-8661-2571]{Mia Sloth Lundkvist}
\email{lundkvist@phys.au.dk}
\affiliation{\Aarhus}

\author[orcid=0000-0003-1687-3271]{Mark Lykke Winther}
\email{mark@phys.au.dk}
\affiliation{\Aarhus}



\author[orcid=0000-0003-1125-2564]{Ashley Chontos}
\email{ashleychontos@astro.princeton.edu}
\affiliation{\Princeton}

\author[orcid=0000-0003-2657-3889]{Nicholas Saunders}
\email{saunders.nk@gmail.com}
\altaffiliation{NSF Graduate Research Fellow}
\affiliation{\UH}

\author[orcid=0000-0001-6416-1274]{Theron W. Carmichael}
\email{tcarmich@hawaii.edu}
\affiliation{\UH}


\author[orcid=0000-0001-9954-4398]{Antonin Bouchez}
\email{abouchez@keck.hawaii.edu}
\affiliation{\WMKO}

\author[orcid=0000-0003-0815-7953]{Carlos Alvarez}
\email{carlos04.alvarez@gmail.com}
\affiliation{\WMKO}

\author[orcid=0000-0001-7062-815X]{Samuel A. U. Walker}
\email{swalk@hawaii.edu}
\affiliation{\UH}






\author[orcid=0000-0002-8621-2682]{Aldo G. Sepulveda}
\email{aldogs@hawaii.edu}
\affiliation{\UH}

\author[orcid=0000-0002-0531-1073]{Howard Isaacson}
\email{hisaacson@berkeley.edu}
\affiliation{\UCB}

\author[orcid=0000-0001-8638-0320]{Andrew W. Howard}
\email{ahoward@caltech.edu}
\affiliation{\Caltech}


\author[orcid=0009-0004-4454-6053]{Steven R.\ Gibson}
\email{sgibson@caltech.edu}
\affiliation{\COO}

\author[orcid=0000-0003-1312-9391]{Samuel Halverson}
\email{samuel.halverson@jpl.nasa.gov}
\affiliation{\JPL}

\author{Kodi Rider}
\email{kodi.rider@ssl.berkeley.edu}
\affiliation{\SSL}

\author[orcid=0000-0001-8127-5775]{Arpita Roy}
\email{arpita308@gmail.com}
\affiliation{\Schmidt}

\author[orcid=0000-0002-6525-7013]{Ashley D.\ Baker}
\email{abaker@caltech.edu}
\affiliation{\COO}

\author[orcid=0009-0002-2419-8819]{Jerry Edelstein}
\email{jerrye@ssl.berkeley.edu}
\affiliation{\SSL}

\author{Chris Smith}
\email{christopher.smith@berkeley.edu}
\affiliation{\SSL}

\author[orcid=0000-0003-3504-5316]{Benjamin J.\ Fulton}
\email{bjfulton@ipac.caltech.edu}
\affiliation{\nexsci}

\author[orcid=0000-0002-6092-8295]{Josh Walawender}
\email{jwalawender@keck.hawaii.edu}
\affiliation{\WMKO}


\begin{abstract}
We present high-precision measurements for HR~7672AB, composed of a Sun-like (G0V) star and an L~dwarf companion.
Three nights of precise (70 cm/s) radial velocity (RV) asteroseismology with the Keck Planet Finder clearly detect 5-minute oscillations from the primary HR~7672A, and modeling of the frequency spectrum yields an asteroseismic age of $1.87\pm0.65$~Gyr. We also determine a gyrochronological age of $2.58\pm0.47$~Gyr, and we combine these two results for a final age of $2.26\pm0.40$~Gyr. 
In addition, we obtained new RVs for HR~7672A and new astrometry for the companion HR~7672B.
From a joint orbit fit, we measured a dynamical mass of $1.111\pm0.017$~\Msun{} for HR~7672A and $75.39\pm0.67$~\Mjup{} for  HR~7672B. 
This places the companion near the stellar/substellar boundary and thus particularly sensitive to differences in model predictions. 
The joint precision in host star age (18\% uncertainty) and companion mass (0.9\% uncertainty) makes HR~7672AB an exceptional substellar benchmark.  Combined with the companion's luminosity, we use these measurements to test predictions from six brown dwarf cooling models. 
The best agreement occurs with the Chabrier et al. (2023) models, which incorporate a new equation of state, resulting in predictions that agree within $<$0.3$\sigma$ with all the observations. The other 5 sets of models agree at the 1--3$\sigma$ level depending on the particular test, and some models struggle to predict a sufficient low luminosity for HR~7672B at any age given its dynamical mass.
We also detected a weak seismic signal in near-simultaneous TESS photometry of HR~7672A, with the resulting RV-to-photometry oscillation amplitude ratio consistent with solar values.
\end{abstract}


\section{Introduction} \label{sec:intro}

Brown dwarfs are objects with masses between the heaviest gas-giant planets and the lowest-mass stars ($\approx$13--75 \Mjup; e.g., \citealp{Chabrier2000, Dupuy2017}). They are not massive enough to sustain stable hydrogen fusion, which is the process that powers main-sequence stars. Instead, brown dwarfs can briefly fuse deuterium, and those above $\approx$65~\Mjup{} can fuse lithium early in their evolution. Without a long-lived fusion source, brown dwarfs cool and fade as they radiate away their residual heat. Their effective temperature and luminosity steadily decrease with age, causing them to evolve through later spectral types (from late-M to L to T and Y). This contrasts with main-sequence stars, which maintain stable luminosities for millions to trillions of years.

The cooling of brown dwarfs is predicted by evolutionary models. Early models used simplified treatments of interior and atmospheric physics, but the low temperatures of brown dwarfs require more sophisticated approaches. Modern models incorporate non-grey atmospheres \citep[e.g.,][]{Burrows1997, Chabrier1997, 2021ApJ...920...85M}, dust and cloud physics \citep[e.g.,][]{Chabrier2000, Baraffe2002, Saumon2008, 2012ApJ...756..172M}, disequilibrium chemistry \citep[e.g.,][]{2006ApJ...647..552S, 2020A&A...637A..38P}, vertical mixing \cite[e.g.,][]{2015ApJ...804L..17T}, detailed equations of state and degeneracy \citep[e.g.,][]{Chabrier2023}, and assumptions about initial conditions (e.g. hot vs.\ cold start; e.g., \citealp{2007ApJ...655..541M}).  

Evolutionary models provide predictions of brown dwarf luminosity, radius, and effective temperature at a given age and mass, often visualized as cooling curves (luminosity vs.\ age at fixed masses) and  isochrones (radius vs.\ mass at fixed ages).  When coupled to model atmospheres, such models also predict the mass- and age-dependence of an object's colors, magnitudes, and spectra. Thus, evolutionary models are at the heart of converting observations into the fundamental properties needed to understand both individual objects and the substellar population as a whole.   

Given the wide-spread use of such models, testing their predictions is imperative.  To do so, model-independent measurements of brown dwarf properties are needed, in particular the 3 key quantities of luminosity, age, and mass. 
Systems with at least 2 of these quantities measured are often referred to as ``benchmarks'' \citep[e.g.][]{liu08-2m1534orbit,Zhangzj2020}. Luminosities are relatively easy to measure using photometry and/or spectroscopy, supplemented by bolometric corrections or fitting model atmospheres \citep[e.g.][]{2023ApJ...959...63S}.  However, age and mass are not directly observable for most brown dwarfs, which are free-floating objects. For a minority fraction of known objects, ages can be established via their physical association with a star or stellar association \citep[e.g.][]{2006MNRAS.368.1281P}. And for an even smaller fraction of the known objects, dynamical masses can be measured via the visual/astrometric orbits of binary brown dwarfs \citep[e.g.,][]{Dupuy2017} and brown dwarfs that are companions to nearby stars \citep[e.g.][]{Brandt2019}.

Only a tiny fraction of known brown dwarfs have their luminosities, ages, and masses all measured.  Such systems provide the most stringent tests of evolutionary models.  For instance, using the models with the observed dynamical mass and luminosity produces a "cooling age" for the brown dwarf, which can be compared with the primary star's age.  This test was first done by \citet{2009ApJ...692..729D}, who found that the cooling age for the brown dwarf binary HD~130948BC (440$\pm$40~Myr) was inconsistent with the age of the solar-type host star HD~130948A as inferred from gyrochronology (790$\pm$150~Myr). Or put another way, the brown dwarfs appear to be 2--3$\times$ overluminous compared to model predictions given their masses and age. Evidence for a discrepancy between cooling ages and stellar ages has steadily grown in recent years, with now about half a dozen star+brown dwarf systems where the brown dwarf potentially shows either over- or under-luminosity relative to evolutionary model predictions (e.g., \citealp{2021AJ....162..301B} and references therein).
However, the discrepancy between model predictions and observation shown no clear pattern of behavior, limiting our ability to probe for the origin.   One key limiting factor is that the measured ages for these benchmarks are often imprecise.

This paper focuses on one notable benchmark system, HR~7672AB, composed of a Sun-like (G0V) star and an imaged L~dwarf companion with a 17~au semi-major axis. Studying the primary star HR 7672A (15 Sge; HD 190406) provides two critical measurements for testing evolutionary models: (1) a dynamical mass measurement for the companion based on its orbit, and (2) a stellar age estimate that applies to the system as a whole, under the conservative assumption of both components being coeval.

HR 7672B was discovered by \citet{Liu2002} with adaptive optics (AO) imaging at Keck and Gemini and shed light on the "brown dwarf desert," namely the scarcity of $\approx$15--75\Mjup{} companions in radial velocity surveys of solar-type stars. At $\approx$14~au (0.8") projected separation, HR 7672B demonstrated that brown dwarfs, while rare, can exist at planet-like separations. A decade later, \citet{Crepp2012} measured its orbit, finding a $\approx$73-year period, high eccentricity ($e\approx0.5$), and near edge-on inclination, and continued orbital monitoring by \citet{Bowler2023} showed no signs of spin-orbit misalignment.  Its eccentricity may reflect dynamical interactions (e.g., with a passing star or unseen companion) or simply the outcome of fragmentation-based formation, which naturally yields eccentric orbits (unlike the near-circular orbits expected for disk-formed gas-giant planets).

More recently, HR~7672B's atmosphere has been studied with AO-assisted high-spectral resolution measurements.  \citet{Wangj2022} used $K$-band spectroscopy with Keck/KPIC to measure carbon and oxygen abundances, key to CO, H$_2$O, and CH$_4$ chemistry.  The elemental abundances of the host star agree well with the host star, supporting a star-like formation scenario for the companion.  \citet{Kasagi2025} further obtained high-resolution $YJH$-band spectra with Subaru/REACH. Their atmospheric retrievals detected H$_2$O and FeH absorption and required an optically thick cloud layer to reproduce the data, consistent with L-dwarf cloud physics. They inferred cloud-top temperatures suggestive of condensates such as TiO$_2$, Al$_2$O$_3$, or Fe.

In this study, we derive the age of HR 7672A --- and hence the system --- using asteroseismology from three consecutive nights of Keck Planet Finder (KPF) radial velocity data and with gyrochronology based on stellar rotation (\S\ref{sec:obs}--\S\ref{sec:model}).  We also refine the dynamical mass of HR~7672B, using a new epoch of AO imaging combined with additional RV data for the primary. These new high-precision measurements of the system enable strong tests of current substellar evolutionary models (\S\ref{sec:cooling}). We also discuss the nearly-simultaneous KPF and TESS observations, which offer additional insights into stellar pulsations (\S\ref{sec:tess}).

\section{Observations} \label{sec:obs}

\begin{figure*}
    \centering
    \includegraphics{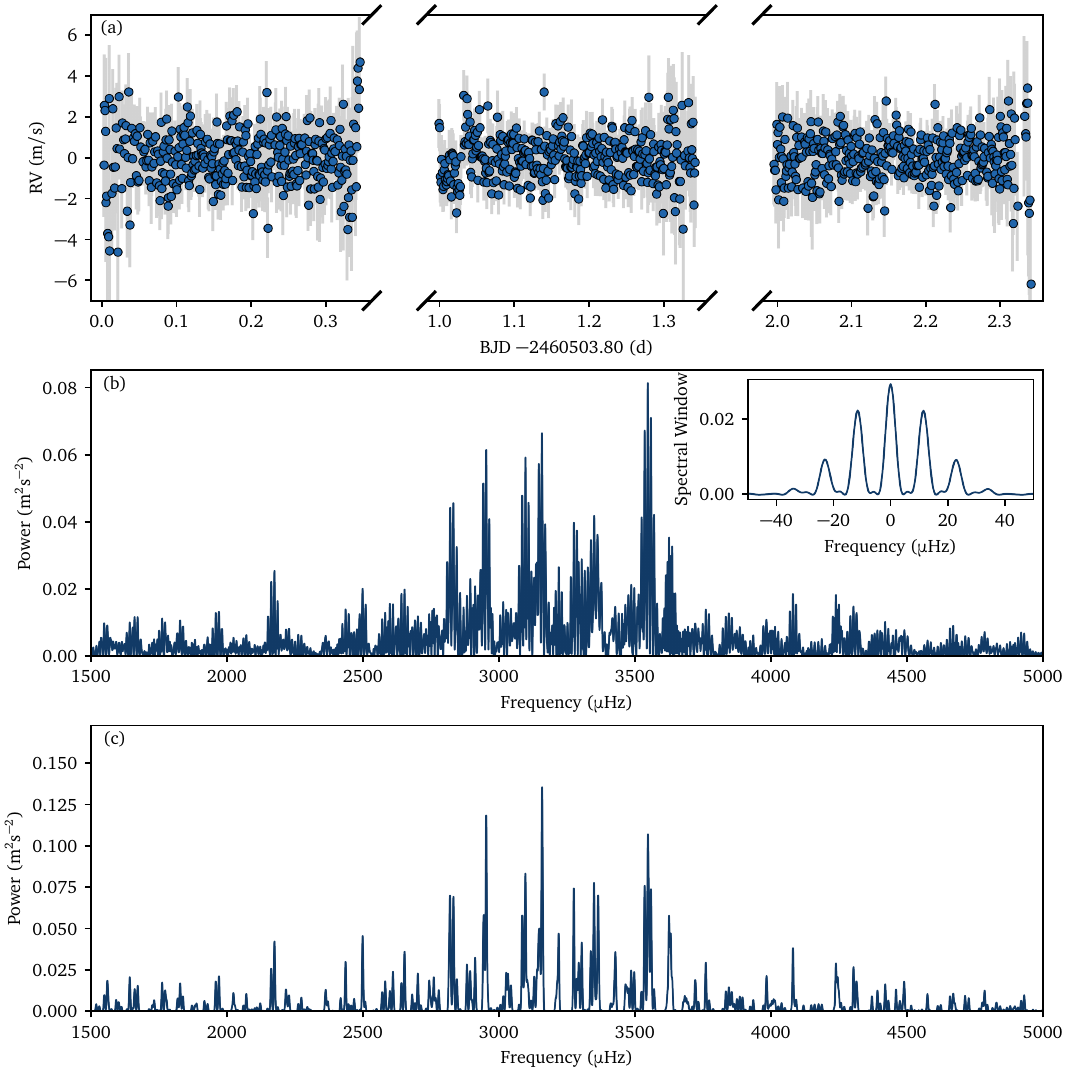}
    \caption{Radial velocity observations of HR 7672A over three consecutive nights using the Keck Planet Finder. (a) Radial-velocity time series after filtering out signals with periods longer than 1.2 hours. (b) Power spectrum of the RV time series, weighted by the reported RV uncertainties, displaying a clear power excess around 3000~\muHz{}. The inset shows the spectral window. (c) Power spectrum after deconvolution against the spectral window, followed by reconvolution with a Gaussian filter of width $1/\Tobs{}$ for clarity.}
    \label{fig:data}
\end{figure*}

\begin{figure}
    \centering
    \includegraphics{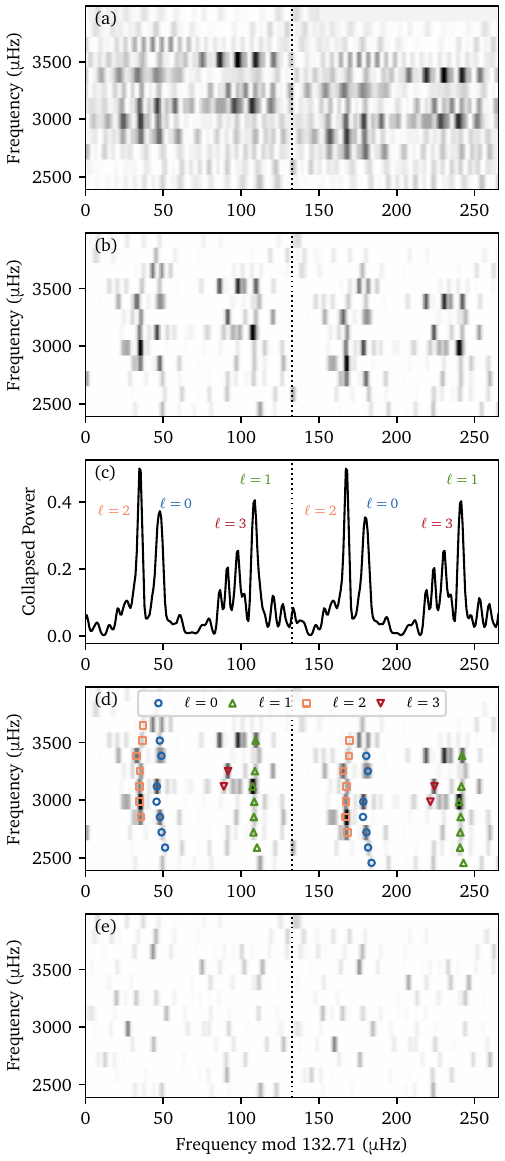}
    \caption{Replicated \'echelle diagrams showing structures of regular frequency spacings from KPF data. (a): \'echelle diagrams of the original power spectrum. (b): \'echelle diagram after deconvolving with the spectral window and then smoothing with a 1~$\mu$Hz-width Gaussian filter. (c): collapsed \'echelle diagram, by summing the power along the y-axis of panel b. (d): same as panel b, but highlighting the oscillation modes. (e): deconvolved power spectrum of the residual time series after subtracting oscillation modes.} 
    \label{fig:echelle}
\end{figure}

\subsection{Asteroseismology with Keck Planet Finder}\label{subsec:kpf}

\begin{deluxetable}{cccccccc}
\tabletypesize{\footnotesize}
\tablecolumns{3}
\tablewidth{\textwidth} 
\tablecaption{High-cadence radial-velocity data of HR~7672A measured from KPF spectra over three consecutive nights. \label{table:kpfrv}}
\tablehead{
\colhead{\hspace{1.0cm}BJD}\hspace{1.0cm} & \colhead{\hspace{1.0cm}RV }\hspace{1.0cm} & \colhead{\hspace{1.0cm}$\sigma({\rm RV})$}\hspace{1.0cm}  \\
d & km~s$^{-1}$ & km~s$^{-1}$ 
}
\startdata
2460503.80257248 & 5.6694026 & 0.0025670 \\
2460503.80334406 & 5.6723237 & 0.0024898 \\
2460503.80402596 & 5.6720894 & 0.0023852 \\
2460503.80475537 & 5.6711032 & 0.0035657 \\
2460503.80531227 & 5.6676321 & 0.0029298 \\
2460503.80612872 & 5.6679721 & 0.0030534 \\
2460503.80752271 & 5.6661874 & 0.0033515 \\
2460503.80882304 & 5.6660599 & 0.0050862 \\
... & ... & ... \\
\enddata
\tablecomments{Only the first 10 lines are shown. The full table can be accessed online.}
\end{deluxetable}

\begin{deluxetable}{cccccccc}
\tabletypesize{\footnotesize}
\tablecolumns{4}
\tablewidth{\textwidth} 
\tablecaption{ Oscillation frequencies of HR 7672, including corrections for line-of-sight Doppler shifts. \label{table:freqs}}
\tablehead{
\colhead{\hspace{0.6cm}$n$}\hspace{0.6cm} & \colhead{\hspace{0.6cm}$\ell$}\hspace{0.6cm} & \colhead{\hspace{0.6cm}$\nu_{n,\ell}$}\hspace{0.6cm} &  \colhead{\hspace{0.6cm}$\sigma(\nu_{n,\ell})$}\hspace{0.6cm} \\
 &  & \colhead{\muHz{}} & \colhead{\muHz{}}
}
\startdata
18 & 0 & 2572.67 & 1.75 \\
19 & 0 & 2703.14 & 1.74 \\
20 & 0 & 2834.84 & 1.25 \\
21 & 0 & 2965.31 & 1.66 \\
22 & 0 & 3098.25 & 0.93 \\
24 & 0 & 3366.58 & 1.26 \\
25 & 0 & 3498.28 & 1.79 \\
18 & 1 & 2631.75 & 1.62 \\
19 & 1 & 2762.22 & 1.61 \\
20 & 1 & 2895.16 & 1.35 \\
21 & 1 & 3028.09 & 1.32 \\
22 & 1 & 3159.79 & 0.74 \\
23 & 1 & 3293.96 & 1.15 \\
25 & 1 & 3559.82 & 0.84 \\
19 & 2 & 2822.53 & 1.27 \\
20 & 2 & 2954.24 & 0.59 \\
21 & 2 & 3087.17 & 1.33 \\
22 & 2 & 3220.10 & 0.84 \\
23 & 2 & 3350.58 & 0.90 \\
24 & 2 & 3487.20 & 1.39 \\
25 & 2 & 3620.14 & 1.53 \\
21 & 3 & 3141.33 & 1.40 \\
22 & 3 & 3276.72 & 1.42 \\
\enddata
\end{deluxetable}

Asteroseismology is the study of stellar oscillations, which probe internal structure and provide precise fundamental stellar properties \citep{Aerts2010,Basu2017}. Extremely Precise Radial Velocity (EPRV) techniques now achieve radial velocity measurements with precisions down to 30 cm/s for individual measurements \citep{Fischer2016, Wright2018}. At this precision, stars that oscillate at amplitudes too low to be detected from photometry like Kepler \citep{Gilliland2011} and TESS \citep{Ricker2015}, such as main-sequence K dwarfs, can now be routinely detected via the EPRV method. 
Oscillations have recently been detected in K-type main-sequence dwarfs with amplitudes of only 1--6 cm/s in $\epsilon$~Ind~A \citep{Campante2024,Lundkvist2024}, $\sigma$~Dra \citep{Hon2024-sig-dra} and HD~219134 \citep{Liyg2025}, as well as in the subgiant HD~118203 \citep{Zhangjw2024}.

The advantage of the EPRV method is particularly great for young, magnetically active stars \citep{Chaplin2011-B,Mathur2019}. Their oscillation amplitudes are, in general, smaller than those of older stars \citep{Huber2011,Campante2014,Sayeed2025}, so they can often be obscured by granulation or white noise in photometry \citep{Sulis2023}. EPRV is less sensitive to granulation at oscillation frequencies of interest, enabling higher signal-to-noise ratio (SNR) of oscillations in such targets. The star we are studying, HR 7672A, is a moderately young G-type dwarf that illustrates this capability.

We obtained approximately 8 hours of observations per night over 3 consecutive nights, from July 12 to 14, 2024 (Hawaii Standard Time), using the Keck Planet Finder (KPF) mounted on the Keck~I telescope \citep{gibson2016-kpf,gibson2018-kpf,gibson2020-kpf,gibson2024-kpf}. A total of 1,084 radial velocity measurements were collected. Observing conditions were moderately good, but with occasional gusts resulting in the poorest seeing reaching $2.0''$. To maintain the signal-to-noise ratio to reach adequate RV precision per spectra under the variable conditions, we used a mixture of 60-second and 90-second exposures, each followed by a 15-second readout, yielding cadences of 75 or 105 seconds, where the latter corresponds to a Nyquist frequency of 4800 \muHz{}. The average time-domain scatter measured using the power density at high frequency is 0.68~m/s. Table~\ref{table:kpfrv} lists these high-cadence radial velocity measurements obtained with the Keck Planet Finder. 

Figure~\ref{fig:data} shows the RV time series, along with the power spectrum calculated using the Lomb-Scargle method \citep{lomb1976,scargle1982}, using the reported RV uncertainties as weights \citep{Frandsen1995}. We see a clear power excess around 3000~\muHz{}. Based on a heavily smoothed power spectrum \citep{Kjeldsen2008}, we measured the frequency of maximum power as $\numax=3245\pm75$~\muHz{}. We note that this measured \numax{} does not translate to stellar properties accurately, because the value of \numax{} depends on the technique (photometric bandpass or selected spectroscopic lines; \citealt{Hon2024-sig-dra}), stellar-cycle variability \citep{Howe2020}, and time variability introduced by stochastic excitation \citep{Jiangc2023,Sreenivas2024}.

The oscillations observed around \numax{} are acoustic modes, and their frequencies can be well described by the asymptotic relation involving two quantum numbers, the radial order, $n$, and angular degree, $\ell$ \citep{Tassoul1980,Scherrer1983}:
\begin{equation}\label{eq:asymp}
    \nu_{n,\ell} \simeq \Delta\nu \left( n + \frac{\ell}{2} + \epsilon \right) - \delta\nu_{0,\ell}.
\end{equation}
The large frequency separation, \Dnu{}, depends primarily on the sound travel time across the star and, to first order, scales with the square root of the mean stellar density \citep{Ulrich1986}. The phase offset, $\epsilon$, generally falls between 0.8 and 1.6 in main-sequence stars \citep{White2012}, and depends on the turning points of modes \citep{Roxburgh2003}. The small separations, \dnu{0,\ell}, quantify the frequency offsets between modes of different $\ell$ but with adjacent $n$. In main-sequence stars, \dnu{0,\ell} is sensitive to the chemical gradient near the core \citep{Tassoul1994,Roxburgh1994}, making it a useful diagnostic for tracing hydrogen-burning processes and, therefore, for determining stellar ages on the main sequence \citep{jcd1984-02,White2011,Bellinger2019,Hon2024-flow}.

To identify and measure these oscillation modes, we followed the method outlined in \citet{Liyg2025}. We applied Gold deconvolution to the power spectrum with the spectral window to reveal regular structures of mode spacings (\citealt{gold1964,morhac2003-gold}; Ong et al. in preparation). Figure~\ref{fig:echelle}(a) shows the power spectrum displayed in an \'echelle format, which involves folding into segments of \Dnu{} and stacking them vertically (the \'echelle is duplicated for clarity). Figure~\ref{fig:echelle}(b) shows the deconvolved spectrum, revealing clear $\ell=0$--$3$ mode ridges, corresponding to the regular spacings dictated by Equation~\ref{eq:asymp} (see also Figure~\ref{fig:data}(c)). Figure~\ref{fig:echelle}(c) shows the collapsed power spectrum from summing power along the vertical direction in Figure~\ref{fig:echelle}(b). 
Because the small spacings \dnu{02} in main-sequence stars are close to 1~c/d, which is the spacing between sidelobes (see inset of Figure~\ref{fig:lc}), the sidelobes of the $l=0$ modes overlap with the main signal of the $l=2$ modes, and vice versa. Therefore it is very difficult to identify the true oscillation frequency from the original power spectrum (Figure~\ref{fig:echelle}a).  The Gold deconvolution provides a clean way to suppress power from the sidelobes and thus reveal the true mode structure.

From Figures~\ref{fig:echelle}(b) and (c), we identified the frequencies of peaks along each ridge, guided by the asymptotic relation (Equation~\ref{eq:asymp}), which predicts that modes of the same $\ell$ degree should align vertically. 
We then constructed a time-series model consisting of multiple sine waves, with frequencies fixed to those identified from the deconvolved spectrum and amplitudes $a$ and phases $\phi$ as free parameters. 
We fitted this model to the RV time series by minimizing RV-uncertainty-weighted $\chi^2$ statistic.
We estimated the average noise level $\sigma_{\rm noise}$ to be 0.03 m/s, based on the amplitude spectrum at high frequencies ($>5000$~\muHz{}).
Modes with amplitudes exceeding 3.5 times $\sigma_{\rm noise}$ were retained in the final list of modes. These are highlighted in Figure~\ref{fig:echelle}(d).
We then computed the power spectrum of the RV time series after subtracting the best-fitting model. The corresponding deconvolved spectrum is shown in Figure~\ref{fig:echelle}(e), and shows no clear remaining structure. The lack of any clear residual structure in Figure~\ref{fig:echelle}(e) indicates that treating the modes as coherent over the observing window is adequate.
We estimated the uncertainties of the frequencies using an analytical expression \citep{Montgomery1999,Kjeldsen2012}:
\begin{equation}
    \sigma(\nu) \approx 0.44 \sqrt{\pi/N} \sigma_{\rm noise} a^{-1} \sqrt{T_{\rm obs}^{-2} + \tau^{-2}},
\end{equation}
where $T_{\rm obs}$ is the effective observing duration and $\tau$ is the mode lifetime, which we assume to be 3~d, based on its dependency on effective temperature \citep{Chaplin2009}. 
Although the formal frequency resolution is set by $1/T_{\rm obs}$, the uncertainty of an extracted mode frequency can be much smaller. The resolution specifies the minimum spacing at which two unresolved peaks can be distinguished in the power spectrum, but it does not limit the precision with which the centroid of a single peak can be measured.

We also corrected the line-of-sight Doppler shifts for these frequencies following \citet{davies2014-los}, using a radial velocity of $5.6685$~km/s from an average value over three nights. The final list of frequencies are given in Table~\ref{table:freqs}. Fitting Equation~\ref{eq:asymp} to these frequencies yields the following asymptotic parameters:
$\Dnu{} = 132.798 \pm 0.122$~\muHz{},
$\epsilon = 1.336 \pm 0.022$,
$\dnu{0,1} = 5.43 \pm 0.69$~\muHz{},
$\dnu{0,2} = 12.45 \pm 0.64$~\muHz{}, and 
$\dnu{0,3} = 22.80 \pm 1.16$~\muHz{}. 

\subsection{Asteroseismology with TESS}\label{sec:tess}

\begin{figure*}
    \centering
    \includegraphics{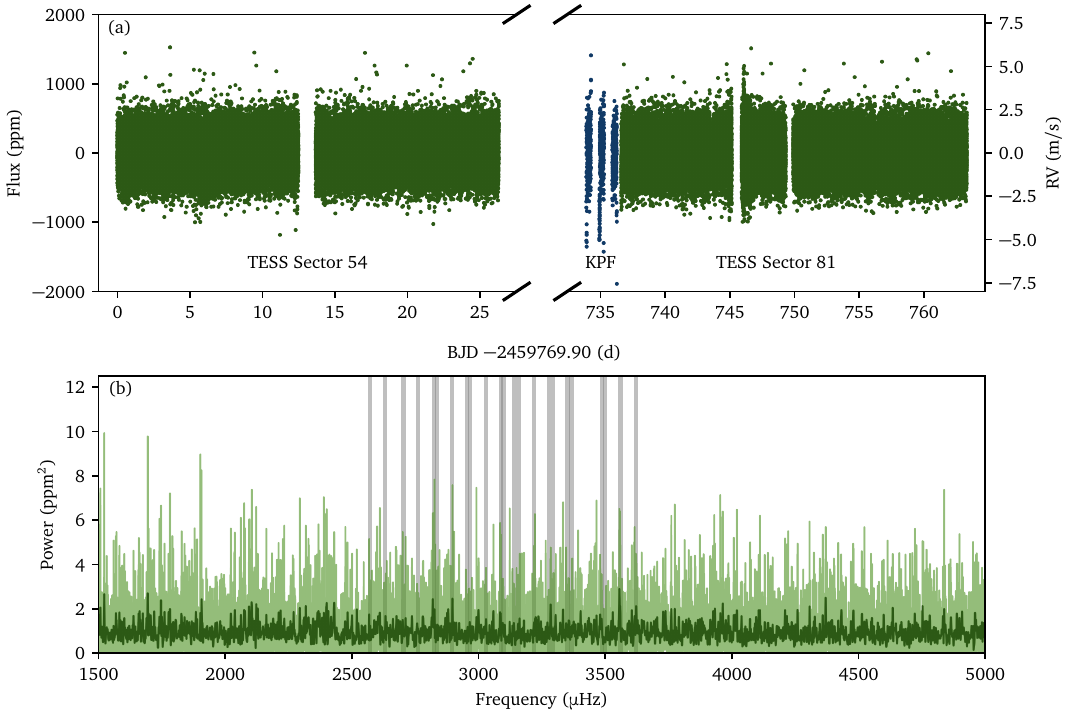}
    \caption{(a): TESS photometric observations of HR~7672 from Sectors 54 and 81. The KPF data were collected three days prior to the start of TESS Sector 81, resulting near-simultaneous coverage. (b): Power spectrum of the TESS light curve. The vertical lines highlight the oscillation frequencies derived from the KPF radial velocity data (Table~\ref{table:freqs}). }
    \label{fig:lc}
\end{figure*}

\begin{figure}
    \centering
    \includegraphics{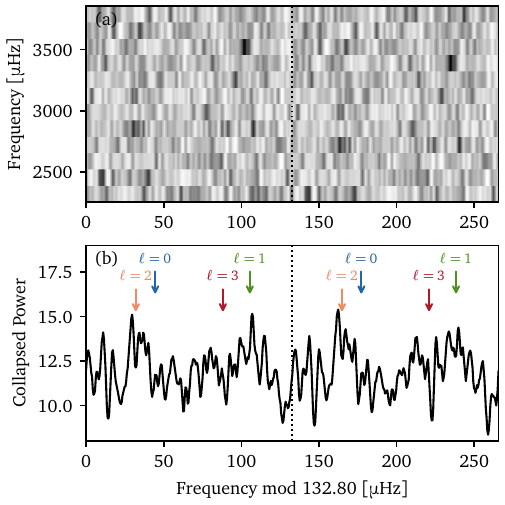}
    \caption{(a): Replicated \'echelle diagram of the TESS power spectrum. (b): Collapsed \'echelle diagram, by summing the power along the y-axis of panel (a). The expected mode positions, informed by the KPF data, are indicated by arrows.}
    \label{fig:lc-echelle}
\end{figure}

\begin{figure}
    \centering
    \includegraphics{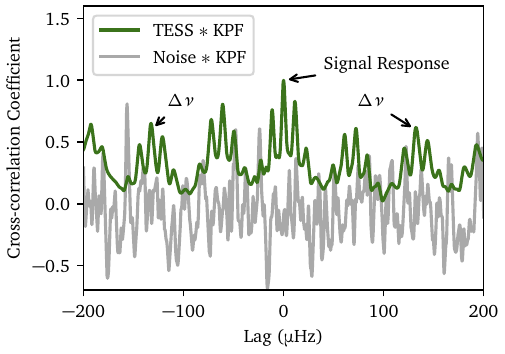}
    \caption{Cross-correlation coefficient between a template power spectrum, constructed using frequencies extracted from KPF RV data, and the TESS power spectrum. For comparison, the cross-correlation coefficient between the same template spectrum and a noise power spectrum generated from a random draw is also shown.}
    \label{fig:lc-correlate}
\end{figure}

\begin{figure}
    \centering
    \includegraphics{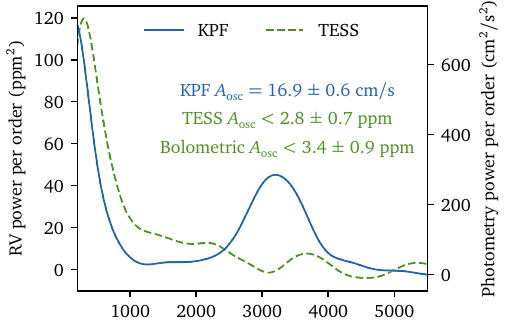}
    \caption{Power per radial order spectra computed from both KPF RV data and TESS photometric data.}
    \label{fig:amplitude-ratio}
\end{figure}

TESS observed HR~7672 in Sectors 54 and 81 with a 20-second cadence exposure. We analysed the SPOC light curves to search for oscillation signals in the TESS data using the Pre-search Data Conditioning SAP (PDCSAP) flux. Figure \ref{fig:lc}(a) shows the light curve after removing long-term trends from instrumental drifts as well as rotational modulation. The timestamps indicate that our KPF monitoring campaign began exactly three days before the start of Sector 81, yielding nearly contemporaneous RV and photometric coverage. Figure \ref{fig:lc}(b) displays the corresponding power spectrum of the TESS time series, with vertical lines marking the oscillation frequencies extracted from the KPF RVs. The oscillation signature is detectable but very weak. This is confirmed in Figure~\ref{fig:lc-echelle}: the ridges corresponding to the same $\ell$-degrees are only marginally discernible in both the \'{e}chelle diagram and the collapsed \'{e}chelle diagram.

To confirm these are real oscillation signals, we cross-correlated the TESS power spectrum with a template constructed from the KPF frequencies. This method of cross correlation should be more sensitive to regular structures in the spectrum. For each mode frequency listed in Table~\ref{table:freqs} (before applying the line-of-sight correction), we generated a unit-height, unit-width Lorentzian profile centered on that frequency and summed all these profiles to form the template. Figure~\ref{fig:lc-correlate} presents the cross-correlation coefficient as a function of lag between the template and TESS power spectra. The pronounced peak at zero lag confirms that the oscillation signals are indeed present in the TESS data. Additional peaks at integer multiples of \Dnu{} and at linear combinations of \Dnu{} and \dnu{0,l} further confirm the underlying comb structure. For comparison, we also include the cross-correlation spectrum between the template and a noise spectrum generated from normally distributed white noise with typical RV uncertainties. This noise cross-correlation shows no significant features.

Because oscillation frequencies are detected in both the TESS and KPF datasets, HR~7672 offers an opportunity to constrain photometry-to-RV amplitude ratios. These ratios have previously been measured in only a few stars with simultaneous data, including Procyon \citep{Huber2011-procyon}, HD~35833 \citep{Gupta2022}, and $\beta$~Aql \citep{Kjeldsen2025}. We follow the approach of \citet{Kjeldsen2008} to measure oscillation amplitudes. Each power density spectrum --- defined as the power spectrum multiplied by the effective observing duration --- was smoothed with a Gaussian kernel with a full width at half-maximum of $4\Dnu{}$ to produce a single power excess hump. The smoothed spectrum was then multiplied by $\Dnu{}/c$ to estimate the power per radial order, where $c$ accounts for the effective number of modes per order \citep{Ballot2011}. Given the star’s similarity to the Sun, we adopt solar values calculated by \citet{Kjeldsen2008}: $c=4.09$ for velocity data and $c=2.91$ for TESS photometry. Figure~\ref{fig:amplitude-ratio} shows the resulting spectra. The square root of the maximum near \numax{} yields the oscillation amplitude. Uncertainties in the oscillation amplitudes were estimated via bootstrapping: in each realization, the original spectrum was multiplied by a noise spectrum, and the oscillation amplitude was remeasured. The spread in amplitude estimates across all realizations was adopted as the uncertainty.

For the KPF data, we measured a RV amplitude of $16.9\pm0.6$ cm s$^{-1}$. In the TESS data, oscillations are not clearly visible as a distinct power excess due to dominant contributions from granulation and white noise. We therefore derive an upper limit of $2.8 \pm 0.7$ ppm for the photometric amplitude in the TESS bandpass, estimated from the maximum power around \numax{}. Applying a bolometric correction factor of 1.2 \citep{Lund2019}, the corresponding bolometric amplitude is $3.4 \pm 0.9$ ppm. These values yield an upper limit on the bolometric photometry-to-RV amplitude ratio of $20.1 \pm 7.6$ ppm (m s$^{-1}$)$^{-1}$. 
Both the amplitudes and their ratio are similar to the solar values. For comparison, the solar oscillation amplitude is approximately 18.7 cm s$^{-1}$ in RV \citep{Kjeldsen2008} and 3.6 ppm in photometry \citep{Michel2009}, corresponding to a ratio of about 19.2 ppm (m s$^{-1}$)$^{-1}$.

Further, both the amplitudes and their ratio are broadly in line with predictions from 3D hydrodynamic simulations \citep{zhouyx2021}. Repeating the methods of \citet{zhouyx2021}, we obtained a simulated solar amplitude ratio between bolometric intensity and fluid velocity of 20.12 ppm (m s$^{-1}$)$^{-1}$. For a hotter main-sequence model whose basic stellar parameters are close to those of HR 7672A ($\Teff{}=6000$~K and $\logg{}=4.5$~dex), the amplitude ratio is slightly lower at 19.27 ppm (m s$^{-1}$)$^{-1}$. The trend predicted from simulations is also consistent with the empirical amplitude scaling relation of \citet{kb95}. Assuming the conversion factor between fluid velocity and RV is invariant from the Sun to HR 7672A, which is reasonable given the similarity between the two stars, the expectation is that the actual photometry-to-RV amplitude ratio of the latter is slightly smaller than 19.2 ppm (m s$^{-1}$)$^{-1}$, in agreement with our measured upper limit.

Oscillation amplitudes are expected to be lower in stars with higher magnetic activity \citep{Campante2014,Sayeed2025}. \citet{Sayeed2025} quantified this dependency using the chromospheric activity levels derived from Ca II H\&K line emission, using the logarithmic flux ratio between the emission flux and the bolometric flux, \logrphk{}. \citet{BoroSaikia2018} reported \logrphk{} values between $-4.68$ and $-4.74$ for this star; for comparison, the solar value is around $-4.94$ \citep{Egeland2017}. According to Equation~10 of \citet{Sayeed2025}, this corresponds to an amplitude ratio between 0.81 and 0.85 relative to the Sun. Although our measurement is broadly consistent with this prediction, it does not provide conclusive evidence for a reduced amplitude in this star due to the large uncertainties.
The photometric-to-RV amplitude ratio, however, is likely less affected by magnetic activities, as it depends on the stellar structure near the photosphere to a first order \citep{zhouyx2021}. By comparing hydrodynamics and magnetohydrodynamics simulations of stellar atmospheres, \citet{2022A&A...663A.166B} show that changes in near-surface density, temperature, and pressure stratification caused by the small-scale dynamo are less than 1\% for G-type stars.

\subsection{Dynamical Orbit with NIRC2 and Radial Velocities}\label{subsec:dyn}

\begin{deluxetable}{cccccccc}
\tabletypesize{\footnotesize}
\tablecolumns{6}
\tablewidth{\textwidth} 
\tablecaption{Relative astrometry of HR~7672AB from direct imaging, including separations ($\rho$) and position angles (PA). \label{table:rel-astrometry}}

\tablehead{
\colhead{\hspace{0.2cm}Date}\hspace{0.2cm} & 
\colhead{\hspace{0.2cm}$\rho$}\hspace{0.2cm} & 
\colhead{\hspace{0.2cm}$\sigma({\rho})$}\hspace{0.2cm} & 
\colhead{\hspace{0.2cm}PA}\hspace{0.2cm} & 
\colhead{\hspace{0.2cm}$\sigma({\rm PA})$}\hspace{0.2cm} & 
\colhead{\hspace{0.2cm}Reference}\hspace{0.2cm} \\
 & arcsec & arcsec & deg & deg & 
}
\startdata
\midrule
2001.64 & 0.786 & 0.006 & 157.9 & 0.5 & \citet{Liu2002} \\
2001.94 & 0.794 & 0.005 & 157.3 & 0.6 & \citet{Liu2002} \\
2002.54 & 0.788 & 0.006 & 156.6 & 0.9 & \citet{Boccaletti2003} \\
2006.69 & 0.750 & 0.080 & 155.0 & 5.0 & \citet{Serabyn2009} \\
2007.73 & 0.742 & 0.035 & 151.8 & 2.9 & \citet{Crepp2012} \\
2011.37 & 0.519 & 0.006 & 147.1 & 0.5 & \citet{Crepp2012} \\
2024.78 & 0.852 & 0.007 & 329.5 & 1.1 & \S\ref{subsec:dyn} \\
\enddata
\end{deluxetable}

\begin{deluxetable}{cccccccc}
\tabletypesize{\footnotesize}
\tablecolumns{6}
\tablewidth{\textwidth} 
\tablecaption{Absolute astrometry of HR~7672A from Hipparcos and Gaia DR3, including proper motions in right ascension direction ($\mu_{\rm \alpha}$) and in declination direction ($\mu_{\delta}$). \label{table:abs-astrometry}}
\tablehead{
\colhead{Date} & 
\colhead{$\mu_{\rm \alpha}$} & 
\colhead{$\sigma(\mu_{\alpha})$} & 
\colhead{$\mu_{\delta}$} & 
\colhead{$\sigma(\mu_{\delta})$} & 
\colhead{Reference} \\
& mas/yr & mas/yr & mas/yr & mas/yr & 
}
\startdata
1991.25 & -394.07 & 0.63 & -406.42 & 0.64 & \citet{ESA1997} \\
2016.00 & -387.472 & 0.037 & -419.497 & 0.030 & \citet{GaiaCollaboration2022} \\
\enddata
\end{deluxetable}

\begin{deluxetable}{cccccccc}
\tabletypesize{\footnotesize}
\tablecolumns{4}
\tablewidth{\textwidth} 
\tablecaption{Long-baseline radial-velocity data of HR~7672A. \label{table:rv}}
\tablehead{
\colhead{\hspace{1.6cm}BJD}\hspace{1.6cm} & \colhead{\hspace{1.6cm}RV}\hspace{1.6cm} & \colhead{\hspace{1.6cm}$\sigma({\rm RV})$}\hspace{1.6cm} &
\colhead{\hspace{1.6cm}Instrument}\hspace{1.6cm} \\
d & km~s$^{-1}$ & km~s$^{-1}$ & 
}
\startdata
2447047.72690000 & 0.2516200 & 0.0089100 & Lick Observatory (Fischer) \\
2447373.79930000 & 0.2209600 & 0.0101700 & Lick Observatory (Fischer) \\
2447373.80310000 & 0.2141700 & 0.0100600 & Lick Observatory (Fischer) \\
2447373.80690000 & 0.2179700 & 0.0100500 & Lick Observatory (Fischer) \\
2447373.81110000 & 0.2180600 & 0.0100500 & Lick Observatory (Fischer) \\
2447373.81500000 & 0.1992400 & 0.0092300 & Lick Observatory (Fischer) \\
2447373.81900000 & 0.2152500 & 0.0116400 & Lick Observatory (Fischer) \\
2447373.93090000 & 0.2061600 & 0.0102500 & Lick Observatory (Fischer) \\
2447373.93700000 & 0.2005600 & 0.0102500 & Lick Observatory (Fischer) \\
2447374.83330000 & 0.1660100 & 0.0131700 & Lick Observatory (Fischer) \\
... & ... & ... & ... \\
\enddata
\tablecomments{Only the first 10 lines are shown. The full table can be accessed online.}
\end{deluxetable}

HR 7672A and B form a binary system, and their mutual orbit enables a dynamical determination of the masses of both the primary star and the brown dwarf. To model the orbit and constrain the component masses, we combined astrometry with long-baseline radial velocity (RV) measurements.

Relative astrometry of the system from direct imaging has been reported in previous studies \citep{Liu2002,Boccaletti2003,Serabyn2009,Crepp2012}. 
We also obtained a new epoch of direct imaging with the Keck~II facility imager NIRC2 on 2024~Oct~11~UT.   The seeing was $\approx$0.6--0.8\arcsec\ with thin ($<1$ mag) clouds.  We used its narrow camera mode ($9.971\pm0.004$\,mas\,pix$^{-1}$) with the standard MKO $K$-band filter \citep{Tokunaga2002} for deep images and the narrow-band Br$\gamma$ filter ($\lambda_{\rm cen} = 2.15762\,\mu$m) for shallow images. All images were obtained with the $256\times264$ subarray to achieve shorter exposure times. We performed standard bias subtraction and flat fielding corrections on all images, using 0.021-s matching subarray biases scaled to our actual exposure time of 0.05\,s and $K_s$-band dome flats for all images. 

In the deep images, the primary was saturated but the companion was well-detected, while in the shallow images, only the unsaturated primary was well-detected. We thus chose to use the shallow images, which bracketed the deep images in time, as a reference PSF image that we cross-correlated with both the primary and secondary in the deep images.  Before doing so, we masked the core of the saturated primary PSF, excluding any pixels within 3.4 pixels of the central pixel. We also attempted to remove any systematic effects of the primary's PSF wings by subtracting the median radial profile of the primary. To correct for NIRC2's geometric distortion, we placed our $(x,y)$ coordinates of the primary (i.e., its central pixel) and companion (i.e., the primary's coordinates with the offset computed from our cross-correlation analysis) in the full $1024\times1024$ NIRC2 coordinate system. We used the astrometric calibration of \citet{Service2016} and also applied corrections from differential aberration and atmospheric refraction using the same method as described by \citet{Dupuy2016}. We assumed a $K$-band effective wavelength of 2.18932\,$\mu$m for the primary and 2.19762\,$\mu$m for the secondary.

Our new measurement, along with all available measurements, are summarized in Table~\ref{table:rel-astrometry}. In addition, HR~7672 A has absolute astrometry from both Hipparcos \citep{ESA1997} and Gaia \citep{GaiaCollaboration2022}, allowing us to measure its proper motion acceleration over a 20-year baseline. These data are listed in Table~\ref{table:abs-astrometry}.
The California Planet Search (CPS) has acquired RVs of HR 7672A from 1970 to 2024 using Lick/APF and Keck/HIRES instruments \citep{Fischer2014,Rosenthal2021}. All RVs are compiled in Table~\ref{table:rv}.

We modeled the dynamical orbit by jointly fitting the Hipparcos and Gaia astrometry, direct imaging measurements, and RV data. We used open source package \texttt{orvara} \citep{Brandt2021b}, which performs a parallel-tempering Markov Chain Monte Carlo (MCMC) fitting. In total, our analysis used 16 free parameters. Two of them are the masses of the primary star ($M_{\rm{A}}$) and brown dwarf companion ($M_{\rm{B}}$). We adopted a log-uniform prior on $M_{\rm{A}}$. Six orbital parameters define the orbit of companion: semi-major axis ($a_B$), inclination ($i_{B}$), longitude of the ascending node ($\Omega_B$), mean longitude at a reference epoch ($t_{\rm ref}$) of 2455197.5 JD ($\lambda_{\rm ref}$), and the eccentricity ($e_B$) and the argument of periastron ($\omega_B$) in the form of $e_B\sin \omega_B$ and $e_B\cos \omega_B$. We also included four parameters to fit the zero-point for RV data from different instruments.  Finally, we have four parameters for the intrinsic jitter of RV data from different instruments. Because HIRES underwent an instrument upgrade in August 2014, we adopted separate radial velocity zero points and jitters for data taken before and after the upgrade. The likelihood was computed by comparing the observed separations, position angles, absolute astrometry, and radial velocities with those predicted from a synthetic orbit, under the assumption of Gaussian measurement errors. 

We used 100 walkers to sample our model with a total of $2.5\times10^{5}$ steps. The chain converged within the first 30\%, which we discarded as the burn-in portion. The best-fitting results to the astrometric and RV data are presented in \S\ref{sec:dyn}. In particular, we determined a primary-star mass of $1.11\pm0.02$~\Msun{}, and a companion mass of $75.4\pm0.7$~\Mjup{}. 
For comparison, \citet{Crepp2012} reported a companion mass of $68.7\pm2.8$\Mjup{}, \citet{Brandt2019} obtained $72.7\pm0.8$\Mjup{}, \citet{Feng2021} found $72.8\pm6.1$\Mjup{}, and \citet{An2025} derived $71.43\pm1.05$\Mjup{}. The difference primarily reflects the inclusion of updated radial-velocity and astrometric data. \citet{An2025} also adopted an informative prior on the host star (a normal prior with a small standard deviation), which can significantly shift the inferred companion mass. Because both component masses are independently constrained in this system, we recommend using uninformative priors, as in this work, \citet{Brandt2019}, and \citet{Feng2021}.

\subsection{Rotation and Activity}\label{subsec:rot}

\begin{figure}
    \centering
    \includegraphics[width=\columnwidth]{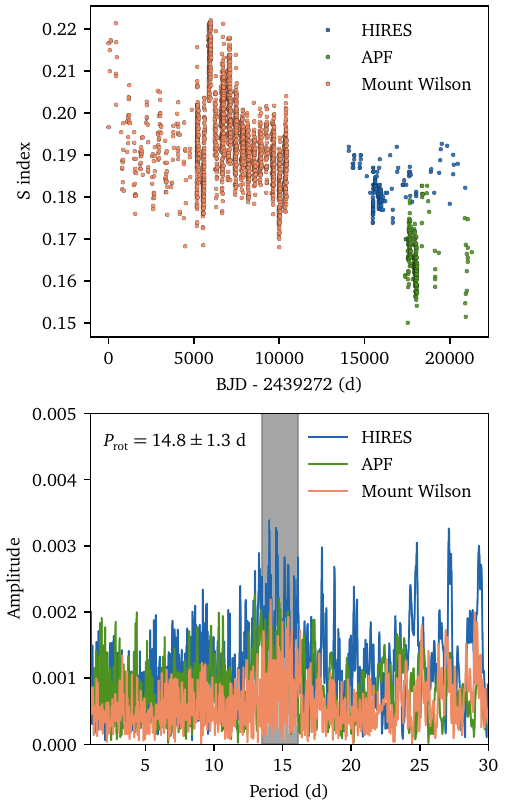}
    \caption{Top: Time series of S-index measured from Ca H \& K emissions. Bottom: Amplitude spectra of S-index measurements from various sites, showing clear signals around 14.8 days corresponding to rotation.}
    \label{fig:sindex}
\end{figure}

HR~7672A has a rather long coverage of S-index over a baseline of many years, from the Mount Wilson program \citep{Radick2018}, and later coverage by Lick/APF, and Keck/HIRES. Figure~\ref{fig:sindex} shows the time series of the S indices as well as the amplitude spectra, showing a clear signal around 15~d. The preprocessing of the time series involved removal of the long term ($>1000$ d) trend, subtraction of the median S index per instrument, and a 3$\sigma$ clipping. The power spectra of the S index for each instrument show a clear signal at approximately 15 days for both the HIRES and Mount Wilson data. We are unable to discern a signal above the noise level in the sparsely sampled APF data and so it is not included in the following rotation analysis. 

To obtain a rotation period from the S index, which is modulated by the star's rotation, we modeled the time series using a hierarchical model and Gaussian processes \citep{celerite2}. Each time series was assigned a single rotation period, with all periods assumed to follow a normal distribution with a common mean and a standard deviation, using the rotation kernel term in \textsc{celerite2}. This rotation kernel is a sum of two simple harmonic oscillator kernels (c.f., \citealt{celerite2}). A hierarchical model is well suited in this situation because of varying systematics, noise levels, and signal amplitudes across instruments. Using this model, we determined a rotation period of $\Prot=14.8\pm1.3$~d. This value agrees well with the rotation periods derived in previous work such as 13.94~d from \citet{Donahue1996} from Mount Wilson data alone, and 13.95~d from \citet{Messina2001} from V-band photometry, and 14.54~d from Keck/HIRES.

We also checked if there is any signature of rotation in TESS photometry, which has observed this star in Sector 54 and 81. We used the the Pre-search Data Conditioning SAP (PDCSAP) flux from the SPOC light curves. We found a cyclic modulation on the order of 6.27 d, which is about half of what S-indices suggests. This is expected in short photometric time series, where harmonics of the true rotation period are often seen as a signature of active regions emerging in different longitude.

The average S-index of 0.19 corresponds to a \logrphk{} value of $-4.74$ \citep{BoroSaikia2018}, placing the star firmly within an active regime where standard magnetic spin-down remains applicable. Weakened magnetic braking is only expected to commence when \logrphk{} reaches $-5.0$, or equivalently, at a Rossby number of $0.9 \Ro_{\odot}$ \citep{vanSaders2016-wmb,saunders2024-ro}.

Rotation lifts the degeneracy between modes with different $m$ but the same $n$ and $\ell$. In this case, the frequencies are given by $\nu_{n\ell m} \approx \nu_{n\ell} + m\nu_{\rm rot}$, where $\nu_{\rm rot}$ is the rotational frequency \citep{Ledoux1951}. The maximum possible splitting occurs between the $m = -\ell$ and $m = \ell$ components and is thus $2\ell \nu_{\rm rot}$, which is reached when the rotation axis is viewed edge-on \citep{Gizon2003}. Given the measured rotation period, the splittings are 1.56, 3.13, and 4.69~\muHz{} for $\ell = 1, 2, 3$ modes, respectively. Indeed, some broadening is observed in the ridges formed by non-radial modes in Figure~\ref{fig:echelle}(b), which could plausibly be attributed to this rotational splitting.

\subsection{Effective Temperature and Metallicity}\label{subsec:spec}

To determine the spectroscopic parameters, including effective temperature \Teff{} and metallicity \mh{}, we adopted the values from \citet{Rosenthal2021}, derived from Keck/HIRES spectra. They are 5932~K and 0.051~dex, respectively. We also compiled \Teff{} and \mh{} values in literature using the SIMBAD astronomical database. These values exhibit a spread of around 100~K in \Teff{} and 0.1~dex in \mh{}, which we adopt as the representative uncertainties. These are consistent with typical spectroscopic uncertainties for solar-type stars \citep{Serenelli2017,Tayar2022}. 

\subsection{Luminosity}\label{subsec:lum}

We used \texttt{isoclassify} \citep{Huber2017,Berger2020} to calculate the bolometric flux based on apparent magnitudes from the Tycho-2, Gaia, and 2MASS photometric systems. Combined with the Gaia DR3 distance \citep{bj2021}, this yields an estimate of the stellar luminosity. Depending on the band used, the derived luminosity ranges from 1.0 to 1.75~\Lsun{}, with typical formal uncertainties of approximately 0.04~\Lsun{}. The large spread reflects systematic issues such as saturation in bright-star photometry, uncertainties in bolometric corrections, and photometric zero points. The Gaia BP and RP bands are the most discrepant, likely because the systematic errors for bright stars are not yet well
calibrated for those Gaia bands \citep{Riello2021}. Among the available bands, the luminosity value based on the 2MASS $K_s$ band appears most consistent with that derived from asteroseismic modeling in \S\ref{subsec:astero-model}. However, we did not use it as a modeling constraint due to the inconsistency across photometric bands. All stellar parameters are summarized in Table~\ref{table:stp}.

\begin{deluxetable*}{cccccccccc}
\tabletypesize{\scriptsize}
\tablecolumns{7}
\tablecaption{Summary of the frequency modeling. \label{table:physics}}
\tablehead{
 & \colhead{Team 1} & \colhead{Team 2} & \colhead{Team 3} &  \colhead{Team 4} & \colhead{Team 5} & \colhead{Team 6}\\
}
\startdata
\midrule
\multicolumn{6}{c}{\textbf{Model Configuration}}\\
\midrule
Stellar evolution code 
& MESA r240301 
& MESA r12778 
& GARSTEC v2015
& YREC 
& GARSTEC v2020
& ASTEC\\
\midrule[0.1pt]
Pulsation code  
& GYRE v7.1 
& GYRE v5 
& ADIPLS v0.4 
& \citet{antia1994}
& GYRE v7.2
& ADIPLS\\
\midrule[0.1pt]
Opacities \& EOS 
& OPAL 
& OPAL 
& OPAL 
& OPAL
& OPAL
& OPAL\\
\midrule[0.1pt]
\makecell{Atmospheric\\boundary condition}
& \makecell{\citet{hauschildt1999a,hauschildt1999b} \\ \citet{castelli2003}}
& \citet{ks1966} 
& \citet{eddington1926} 
& \citet{eddington1926}
& \citet{trampedach2014}
& \citet{eddington1926} \\
\midrule[0.1pt]
Metal mixture 
& \citet{agss09} 
& \citet{agss09} 
& \citet{agss09} 
& \citet{gs98}
& \citet{gs98}
& \citet{gs98}\\
\midrule[0.1pt]
\makecell{Nuclear reaction rate}
& \makecell{\citet{jinareaclib2010} \\ \citet{nacre1999}} 
& \makecell{\citet{jinareaclib2010} \\ \citet{nacre1999}} 
& \makecell{\citet{nacre1999} \\ \citet{formicola2004} \\ \citet{hammer2005}} 
& \makecell{\citet{adelberger1998} \\ \citet{formicola2004}}
& \makecell{\citet{nacre1999} \\ \citet{formicola2004} \\ \citet{hammer2005}}
& \citet{nacre1999} \\ 
%
%
\midrule[0.1pt]
Mixing length formulation
& \citet{henyey1965} 
& \citet{cox1968} 
& \citet{bv1958} 
& \citet{bv1958}
& \citet{cox1968}
& \citet{bv1958} \\
\midrule[0.1pt]
\makecell{Mixing length parameter\\$\amlt$}
& Free 
& Solar-calibrated
& Free 
& Free 
& Varying vs. solar-calibrated
& Free \\
\midrule[0.1pt]
\makecell{$\alpha_{{\rm MLT},\odot}$}
& 1.96
& 1.71
& 1.79
& 1.84
& 1.83
& 2.12\\
\midrule[0.1pt]
Convective overshoot 
& \makecell{\citet{herwig2000}\\$f_{\rm ov, shell}=0.0174$} 
& None 
& None 
& None 
& None 
& None \\
\midrule[0.1pt]
Extra-mixing 
& None 
& \makecell{Gravitational settling\\Turbulent mixing}  
& \citet{thoul1994} 
& \citet{thoul1994}
& \makecell{\citet{thoul1994}\\Diffusive mixing}
& \citet{Michaud1993} \\
\midrule[0.1pt]
Surface correction 
& \makecell{\citet{bg14} two-term\\+\citet{liyg2023} ensemble correction}  
& \citet{sonoi2015} 
& \makecell{\citet{bg14}\\cubic-term}
& \makecell{\citet{bg14}\\two-term}
& \makecell{\citet{bg14} two-term\\+ \citet{Roxburgh2016} $\epsilon$-matching }
& \citet{Kjeldsen2008} \\
\midrule[0.1pt]
Grid free parameters 
& $M$, [M/H], \amlt{}, \Yinit{}
& $M$, [M/H], \Yinit{} 
& $M$, [M/H], \amlt{}, \Yinit{}
& $M$, [M/H], \amlt{}, \Yinit{}
& $M$, [M/H], \Yinit{} 
& $M$, [M/H], \amlt{}, \Yinit{} \\
\midrule\midrule
\multicolumn{6}{c}{\textbf{Derived Stellar Parameters}}\\
\midrule
Mass $M_\star$ (\Msun{})
& $1.114 \pm 0.013$
& $1.111 \pm 0.017$
& $1.110 \pm 0.015$
& $1.106 \pm 0.016$
& $1.110 \pm 0.016$
& $1.110 \pm 0.030$
\\
Radius $R_\star$ (\Rsun{})
& $1.051 \pm 0.004$
& $1.047 \pm 0.006$
& $1.049 \pm 0.006$
& $1.076 \pm 0.007$
& $1.050 \pm 0.007$
& $1.048 \pm 0.009$
\\
Age $t_\star$ (Gyr)
& $1.869 \pm 0.560$
& $1.376 \pm 0.493$
& $1.750 \pm 0.945$
& $1.726 \pm 0.281$
& $1.160 \pm 0.450$
& $1.000 \pm 0.500$
\\
Density $\rho_\star$ ($\rho_\odot$)
& $0.960 \pm 0.002$
& $0.966 \pm 0.006$
& $0.962 \pm 0.006$
& $0.892 \pm 0.021$
& $0.957 \pm 0.013$
& $0.964 \pm 0.004$
\\
Surface gravity $\logg_\star$ (dex)
& $4.442 \pm 0.002$
& $4.443 \pm 0.003$
& $4.442 \pm 0.002$
& $4.416 \pm 0.004$
& $4.440 \pm 0.004$
& $4.443 \pm 0.003$
\\
\Yinit{} 
& $0.246 \pm 0.017$
& $0.276 \pm 0.016$
& $0.258 \pm 0.016$
& $0.274 \pm 0.009$
& $0.267 \pm 0.011$
& $0.269 \pm 0.016$
\\
$\amlt{}/\amlt_{,\odot}$ 
& $1.046 \pm 0.053$
& --- 
& $1.004 \pm 0.072$
& $0.903 \pm 0.047$
& --- 
& $0.953 \pm 0.066$
\\
\enddata
\vspace{-0.8cm}
\end{deluxetable*}

\begin{deluxetable*}{lccccccc}
\tabletypesize{\footnotesize}
\tablecolumns{6}
\tablecaption{Stellar properties of HR 7672A (15 Sge; HD 190406; HIP 98819).\label{table:stp}} 
\tablehead{Property & Value & Reference }
\startdata
\multicolumn{3}{c}{\textbf{Photometry}}  \\
\midrule
$K_s$-band magnitude & 4.388 $\pm$ 0.027 & \citet{Cutri2003} \\
\midrule
\multicolumn{3}{c}{\textbf{Dynamical Orbit}}  \\
\midrule
Mass (\Msun{}) & 1.111 $\pm$ 0.017 $^{1, 2}$ & \S\ref{subsec:dyn} \\
Mass (\Mjup{}; HR 7672 B) & 75.39 $\pm$ 0.67 & \S\ref{subsec:dyn} \\
\midrule
\multicolumn{3}{c}{\textbf{Astrometry}}  \\
\midrule
Distance (pc) & 11.758 $\pm$ 0.014 & \citet{bj2021} \\
Luminosity $L_\star$ (\Lsun{}) & 1.224 $\pm$ 0.043 & \S\ref{subsec:lum} \\
\midrule
\multicolumn{3}{c}{\textbf{Spectroscopy}} \\ 
\midrule
Effective temperature \Teff{} (K) & 5932 $\pm$ 81 $^{1, 2}$ & \citet{Rosenthal2021} \\
Metallicity \feh{} (dex) & 0.051 $\pm$ 0.057 $^{1, 2}$ & \citet{Rosenthal2021} \\
\midrule
\multicolumn{3}{c}{\textbf{Asteroseismology}} \\ 
\midrule
$\nu_{n,\ell}$ (\muHz{}) &  Table~\ref{table:freqs} $^{1}$  & \S\ref{subsec:kpf} \\
$\epsilon$ & $1.335 \pm 0.023$ & \S\ref{subsec:kpf} \\
\Dnu{} (\muHz{}) & $132.806 \pm 0.126$  & \S\ref{subsec:kpf} \\
\dnu{0,1} (\muHz{}) & $5.47 \pm 0.69$ & \S\ref{subsec:kpf} \\
\dnu{0,2} (\muHz{}) & $12.53 \pm 0.67$ & \S\ref{subsec:kpf} \\
\dnu{0,3} (\muHz{}) & $22.87 \pm 1.17$ & \S\ref{subsec:kpf} \\
\numax{} (\muHz{}) & $3245 \pm 75$ & \S\ref{subsec:kpf} \\
\Aosc{} (cm~s$^{-1}$) & $16.9 \pm 0.6$ & \S\ref{sec:tess} \\
\Aosc{} (ppm) & $< 3.4 \pm 0.9$ & \S\ref{sec:tess} \\
Mass ($M_\odot$) & $1.114$ $\pm$ $0.013$ (stat) $\pm$ $0.002$ (sys) & \S\ref{subsec:astero-model} \\
Radius ($R_\odot$) & $1.051$ $\pm$ $0.004$ (stat) $\pm$ $0.010$ (sys) & \S\ref{subsec:astero-model} \\
Luminosity ($L_\odot$) & $1.25$ $\pm$ $0.05$ (stat) $\pm$ $0.01$ (sys) & \S\ref{subsec:astero-model} \\
Asteroseismic age (Gyr) & $1.87$ $\pm$ $0.56$ (stat) $\pm$ $0.32$ (sys) & \S\ref{subsec:astero-model} \\
Density ($\rho_\odot$) & $0.960$ $\pm$ $0.002$ (stat) $\pm$ $0.026$ (sys) & \S\ref{subsec:astero-model} \\
$\log g$ (cgs; dex) & $4.442$ $\pm$ $0.002$ (stat) $\pm$ $0.010$ (sys) & \S\ref{subsec:astero-model} \\
\midrule 
\multicolumn{3}{c}{\textbf{Rotation}} \\ 
\midrule
Rotation period \Prot{} (d) & 14.8 $\pm$ 1.3 $^{2}$ & \S\ref{subsec:rot} \\
Rotational age (Gyr) & 2.58 $\pm$ 0.47 & \S\ref{subsec:rot-model} \\
\midrule 
\multicolumn{3}{c}{\textbf{Asteroseismology + Rotation}} \\ 
\midrule
Asteroseismic + rotational joint age (Gyr) & 2.26 $\pm$ 0.40 & \S\ref{subsec:joint-mod} \\
\enddata
\tablecomments{1. Used for asteroseismic modeling. 2. Used for rotational modeling. }
\end{deluxetable*}

\section{Stellar Ages}\label{sec:model}

\begin{figure}
    \centering
    \includegraphics{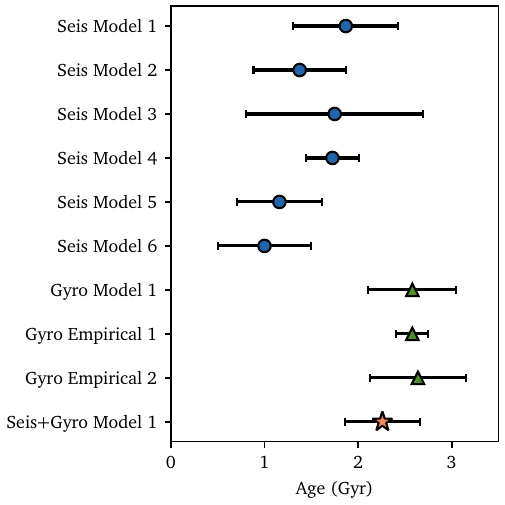}
    \caption{Estimates of stellar ages derived from various methods, including six asteroseismic modeling ages, one rotational modeling age, two rotational empirical ages, and one asteroseismic and rotational joint modeling age.}
    \label{fig:ages}
\end{figure}

\begin{figure}
    \centering
    \includegraphics{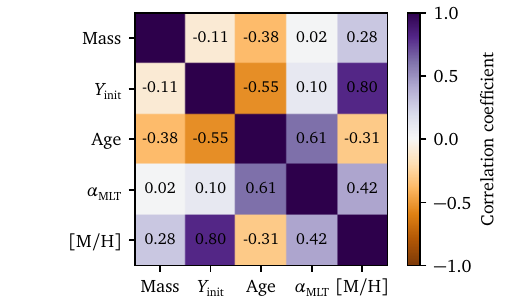}
    \caption{Correlation coefficients of stellar model input parameters ${\mathbf x} = (M, \Yinit{}, \age, \amlt{}, \mh{})$ constrained using asteroseismology, estimated from models implemented by Team 1. }
    \label{fig:corr}
\end{figure}

\subsection{Asteroseismic Modeling}\label{subsec:astero-model}

To estimate the stellar age from oscillation frequencies, we used six independent modeling pipelines. This approach allows us to assess systematic uncertainties associated with differences in input physics and modeling codes.

Stellar evolution codes used by the six teams included MESA \citep{mesa2011,mesa2013,mesa2015,mesa2018,mesa2019,mesa2023,Moedas2024}, GARSTEC \citep{garstec2008}, YREC \citep{yrec2008}, and ASTEC \citep{ChristensenDalsgaard2008a}. 
Pulsation codes used for calculating oscillation frequencies included GYRE \citep{gyre2013}, ADIPLS \citep{adipls2008}, and the code used by \citet{antia1994}. 

The input physics used by the six teams also differed.
Treatments of atmospheric boundary conditions included the Eddington \citep{eddington1926} and Krishna-Swamy \citep{ks1966} $T$-$\tau$ relations, and pre-computed photosphere tables \citep{hauschildt1999a,hauschildt1999b,castelli2003,trampedach2014}.
Choices of nuclear reaction rates varied from \citet{jinareaclib2010}, \citet{nacre1999}, \citet{formicola2004}, \citet{hammer2005}, and \citet{adelberger1998}.
The mixing length formulations included \citet{henyey1965}, \citet{cox1968} and \citet{bv1958}, with some teams using solar-calibrated mixing length parameters, others treating it as a free variable, and still others using calibrated prescriptions for varying it. 
Two main metal mixtures were used: those of \citet{agss09} and of \citet{gs98}.
Mass, metallicity, and initial helium abundance were all treated as free parameters by all teams.

Different teams employed different correction procedures for near-surface modeling errors. Several of the teams used empirical correction formulae \citep{Kjeldsen2008,bg14}, where an offset is applied to the mode frequencies to account for the effects of the surface term, before the corrected mode frequencies are used to construct a conventional likelihood function. These empirical formulae admit additional free parameters, which may either be fitted against the data, or potentially be externally calibrated against 3D models \citep{sonoi2015} or from an empirical ensemble \citep{liyg2023}. Other teams employed the nonparametric $\epsilon$-matching technique \citep{Roxburgh2016}, where combinations of mode frequencies are computed from both the model and observed set, chosen so that the likelihood function evaluated from these combinations is insensitive to the near-surface structure of the star \citep{Ong2021-surferr}. Table~\ref{table:physics} lists the detailed configurations as well as the derived stellar parameters by each team.

Overall, the derived stellar properties for HR~7672A show consistency across modeling teams, despite intentional variations in input physics. The standard deviation of median values from each team is comparable to or smaller than the average uncertainties reported by each team, indicating that model uncertainties arising from differences in input physics and stellar evolution codes are generally smaller than the formal uncertainties reported.

We report our stellar properties of HR~7672A as follows: we adopted the median values and the statistical uncertainties from Team 1's median and formal uncertainty, and adopted the standard deviation of the median values reported by the different teams as an estimate of the systematic uncertainties. 
The results are reported in Table~\ref{table:stp}. 
We find a radius of $R_\star=1.051$ $\pm$ $0.004$ (stat) $\pm$ $0.010$ (sys)~\Rsun{},
a luminosity of $L_\star=1.25$ $\pm$ $0.05$ (stat) $\pm$ $0.01$ (sys)~\Lsun{}, and
an age of $t_\star=1.87$ $\pm$ $0.56$ (stat) $\pm$ $0.32$ (sys)~Gyr, based on asteroseismology.

\subsection{Rotational Modeling}\label{subsec:rot-model}
To model the star based on its rotation period using gyrochronology, we require rotation period predictions from stellar models. For this purpose, we used Team 1’s stellar models in combination with the \texttt{rotevol} code \citep{vanSaders2013-sg, somers2017}. Angular momentum loss was modeled using a calibrated prescription, based on open cluster data and asteroseismic field stars, following the exact configuration described in \citet{Liyg2025}.

We briefly summarize the calibration procedure used in the angular momentum loss modeling. Stellar models were initialized with a rotation period of $P_{\rm disk} = 4$~days and a disk-locking timescale of $\tau_{\rm disk} = 10$~Myr, following the setup of \citet{chiti2024-wd}. 
We note that these initial conditions does not influence the rotation–age relation once convergence has been achieved. Angular momentum loss was modeled using the prescription from \citet{vanSaders2016-wmb}, which relates the loss rate to the stellar rotation rate and convective overturn timescale, scaled by a braking efficiency parameter $f_K$. When the Rossby number exceeds a critical threshold $\Ro_{\rm crit}$, the star enters a weakened magnetic braking regime, and angular momentum loss is no longer applied. This is less relevant in our case because the star has not entered into that phase. Stellar rotation was modeled using a two-zone framework in which the radiative core and convective envelope rotate as solid bodies but exchange angular momentum over a coupling timescale $\tau_{\rm ce}$. This timescale is assumed to scale with stellar mass as $\tau_{\rm ce}/\tau_{\rm ce,\odot} = (M/M_{\odot})^{-\alpha_{\rm ce}}$, where $\tau_{\rm ce,\odot} \approx 22$~Myr \citep{sl20}. The free parameters in the model --- $f_K$, $\Ro_{\rm crit}$, and $\alpha_{\rm ce}$ --- were calibrated to reproduce the rotation periods of stars in a set of well-characterized open clusters ranging from the 120~Myr Pleiades to the 4~Gyr M67 \citep{rebull2016-pleiades,douglas2017-praesepe,douglas2019-praesepe,curtis2019-6811,meibom2015-6819,curtis2020-147,barnes2016-m67,dungee2022-m67,long2023-cl}, asteroseismic field stars \citep{hall2021,silvaaguirre2015-kages,silvaaguirre2017-legacy}, and the Sun \citep{Bahcall1995} at their respective ages. The \Teff{} and rotation period of HR~7672A place it in a regime where gyrochronology is well calibrated.

The models calculated by Team 1 are described by the input parameters ${\mathbf x} = (M, \Yinit{}, \age, \amlt{}, \mh{})$. Following calibration, each model can be assigned with a rotation period. Then we fitted these models with observational constraints $\mathbf{D}$. We constructed a $\chi^2$ statistic using stellar mass, rotation period, effective temperature, and metallicity, leading to a posterior probability of the form:
\begin{equation}
\begin{aligned}
    \ln p(\mathbf{x} \mid \mathbf{D}) & \propto 
    - \frac{1}{2} \left[ \left(M_{\rm mod} - M_{\rm obs}\right)^2/\sigma_{M}^2 \right.\\ 
    & + \left(\Teff_{,\rm mod} - \Teff_{,\rm obs}\right)^2/\sigma_{\Teff}^2 \\ 
    & + \left(\mh_{\rm mod} - \mh_{\rm obs}\right)^2/\sigma_{\mh}^2 \\ 
    & + \left. \left(\Prot_{,\rm mod} - \Prot_{,\rm obs}\right)^2/\sigma_\Prot^2 \right] .\\ 
\end{aligned}
\end{equation}
Any other model parameter $\theta$ to be estimated can be obtained by marginalization:
\begin{equation}
    p(\theta \mid \mathbf{D})  = \int \delta (\theta-\theta(\mathbf{x})) \ p({\mathbf x} \mid \mathbf{D}) \ {\text d} \mathbf{x},
\end{equation}
where $\delta$ is the Kronecker delta function.
We report the result using the median and the 16th and 84th percentile credible intervals. From this, we determined a rotation-based age of $\tau_{\star} = 2.58 \pm 0.47$ Gyr.

In addition to this physics-based approach, which models stellar rotation using an angular momentum loss prescription, we also consider empirical rotation-age-temperature relations, which are more data-driven. We adopted two empirical models. The \texttt{gpgyro} model \citep{Lu2024} calibrates the spin-down relation using stellar clusters and kinematic ages. For this star, it predicts an age of $2.58 \pm 0.17$ Gyr, using \Prot{}, \Teff{}, and the absolute Gaia G-band magnitude as inputs. The \texttt{gyrointerp} model \citep{Bouma2024} is calibrated on stellar clusters and predicts an age of $2.64 \pm 0.51$ Gyr, based on \Prot{} and \Teff{}. The larger uncertainty in the latter estimate likely reflects a more realistic treatment of the intrinsic spread in gyrochrones, with considerations for the empirical limits of gyrochronology.

These rotation-based age estimates are listed in Table~\ref{table:stp}. A comparison of all estimates is shown in Figure~\ref{fig:ages}, where we find that the results are broadly consistent across methods. However, the rotation-based ages tend to be systematically higher than those derived from asteroseismology, although the difference is not statistically significant. This offset could arise from systematic effects in either method. For example, latitudinal differential rotation can limit the accuracy of rotation-based estimates \citep{Epstein2014}, while uncertainties in the initial helium abundance can affect the seismic diagnostics of the sound-speed profile. To mitigate such systematics, a more robust approach is to perform a joint modeling of both methods, which we explore next.

\subsection{Asteroseismic and Rotational Joint Modeling}\label{subsec:joint-mod}

Since Team 1’s models include both oscillation frequencies and rotation periods as outputs, we use this opportunity to jointly model the star using both asteroseismic and rotational constraints. We also aim to account for systematic uncertainties of ages arising from uncertainties in the underlying stellar physics adopted by a single modeling pipeline. 

For this purpose, we introduce a workflow that incorporates systematic uncertainties in age estimation during the modeling process, while still combining both asteroseismic and rotational constraints. In the asteroseismic analysis, under constraints on \Teff{}, \mh{}, mass, and oscillation frequencies, we can approximate the posterior distributions of the model input parameters ${\mathbf x}$ 
using a multivariate Gaussian distribution with a mean of $\mu$ and covariance $\Sigma$, such that $p({\mathbf x}) \sim \mathcal{N} (\mu, \Sigma)$. In Figure~\ref{fig:corr}, we show the standardized correlation coefficient matrix, which reveals strong correlations among several model parameters. The diagonal elements of the covariance matrix, in particular, quantify the uncertainties in each parameter within the context of Team 1’s model grid. To incorporate systematic uncertainty in age, we computed the spread in seismic ages across all modeling teams and inflate the variance of stellar age (i.e., the diagonal element corresponding to age) by adding this systematic component in quadrature.

To perform the joint fit using both asteroseismic and rotational constraints, we treated the multivariate Gaussian distribution of the model input parameters $\mathbf x$ from the asteroseismic analysis, $\mathcal{N} (\mu, \Sigma)$, as a prior, and adopted the observed rotation period as an observational constraint to include in the likelihood function. The resulting posterior probability for the model parameter is:
\begin{equation}
\begin{aligned}
     \ln p(\mathbf{x} \mid \mathbf{D}) & \propto \ln p(\mathbf{x}) + \ln  p(\mathbf{D} \mid \mathbf{x})  \\
    & = -\frac{1}{2} (\mathbf{x}-\mathbf{\mu})^{T} \mathbf{\Sigma}^{-1} (\mathbf{x}-\mathbf{\mu}) \\
    & = - \frac{1}{2}\left(\Prot_{,\rm mod} - \Prot_{,\rm obs}\right)^2/\sigma_\Prot^2  .\\ 
\end{aligned}
\end{equation}
In this construction, we treat each class of observational constraint independently, so we do not expect any overconstraining of the age due to double-counting of information. We obtained a final age estimate of $t_{\star} = 2.26\pm 0.40$ Gyr. The joint estimate falls between the asteroseismic and rotational ages, with an uncertainty of 18\%. These results are summarized in Table~\ref{table:stp} and shown in Figure~\ref{fig:ages}. 
For comparison, \citet{Crepp2012} estimated the system age to be $2.5\pm1.8$~Gyr from isochrone fitting and $2.5\pm0.7$~Gyr from gyrochronology relations using \citet{mam08-ages}. The combination of asteroseismology and gyrochronology employed in this work relies on independent diagnostics, making the resulting age scale more accurate and less affected by method-specific systematics. We also find an improvement in precision compared to previous estimates.

\section{Testing Brown Dwarf Cooling Models}\label{sec:cooling}

Our new high-precision measurements of the HR~7672B dynamical mass and the age of the HR~7627AB system, combined with the existing HR~7672B luminosity measurement, enable this benchmark system to test substellar evolutionary models with a degree of rigor previously not possible.  We consider six widely-used model grids: \citet[Burrows1997]{Burrows1997}, \citet[SM2008]{Saumon2008}, \citet[BHAC2015]{Baraffe2015}, \citet[ATMO2020]{Phillips2020}, \citet[CBPD2023]{Chabrier2023}, and \citet[Diamondback]{Morley2024}. 
The different models employ different interior physics (e.g., equations of state) and boundary conditions (e.g., atmosphere models). Brown-dwarf evolution grids have progressed from early non-gray, baseline cooling tracks to modern models that fold in updated H/He equations of state (EOS), opacities, cloud physics, chemistry, and vertical mixing. Burrows1997 \citep{Burrows1997} sets the foundation with non-gray atmospheres and coupled spectra/cooling. SM2008 \citep{Saumon2008} introduced a hybrid cloudy/clear prescription to capture the L/T transition. BHAC2015 \citep{Baraffe2015} refreshed pre-main-sequence/young-BD tracks with boundary conditions based on BT-Settl model atmospheres. ATMO2020 \citep{Phillips2020} added a new H-He EOS and expanded opacities and chemistry. CBPD2023 \citep{Chabrier2023} then isolated interior updates with a newer EOS, shifting hydrogen-burning limits and late-time cooling. Diamondback delivered a modern cloudy grid for warm L/T objects that systematically varies metallicity and cloud sedimentation.   We use the solar-metallicity versions of all these models, in accord with the metallicity of HR~7672A.

Along our new dynamical mass ($M=75.39\pm0.67$~\Mjup) and  host star age ($t=2.26\pm0.40$~Gyr), the third observation of HR~7672B needed to test models is the bolometric luminosity.   We adopt $\log(L/\Lsun) = -4.19 \pm 0.04$~dex from \citet{Brandt2019}, which represents the weighted average of values derived from the companion's $H$-band  ($-4.23\pm0.05$~dex) and $K_S$-band ($-4.14\pm0.06$~dex) photometry measured by \citet{Boccaletti2003} and then combined with a bolometric correction from \citet{Dupuy2017}.  We note that the \citet{Boccaletti2003} photometry leads to a color of $(H-K_S) = 1.00\pm0.17$~mag, which is somewhat redder compare to colors for mid-L dwarfs, e.g., $(H-K_S) = 0.64_{-0.20}^{+0.11}$~mag from the L4--L4.9 template of \citet{2018ApJS..234....1B}.

\begin{figure}
    \centering
    \includegraphics[width=\linewidth]{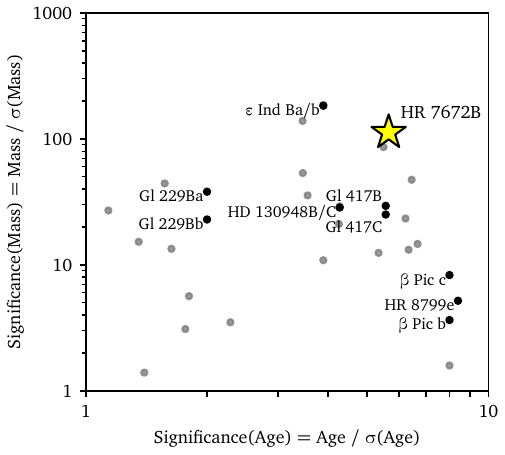}
    \caption{Summary of all known substellar companions with dynamical masses, plotting the significance of their measured dynamical masses and host star ages, defined as the ratio between the measured value and its uncertainty (Table~\ref{tab:benchmarks}). HR~7672B's high precision in both quantities (0.9\% in mass and 18\% in age) makes it an exceptional benchmark for testing brown dwarf evolutionary models. }
    \label{fig:snr}
\end{figure}

\begin{figure}
    \centering
    \includegraphics[width=\columnwidth]{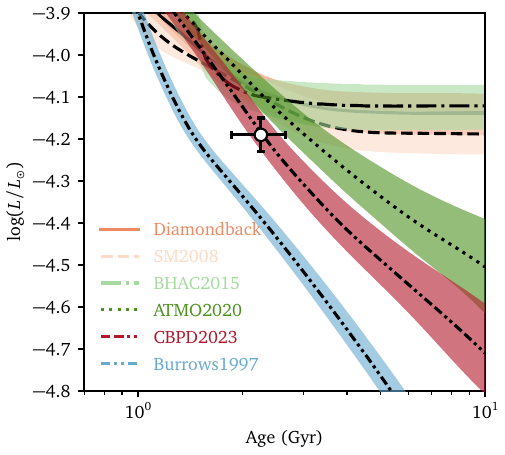}
    \caption{Brown dwarf cooling tracks from 6 evolutionary model grids for HR~7672B given its measured dynamical mass and 1-$\sigma$ uncertainty. The circle indicates the observed values.}
    \label{fig:cooling}
\end{figure}

Figure~\ref{fig:snr} and Table~\ref{tab:benchmarks} summarize the properties of all known substellar benchmarks with both substellar mass and system age determinations, showing how HR~7672 is distinguished by its combination of high precision in both parameters.
Figure~\ref{fig:cooling} provides an illustration of the evolutionary models that we consider here, constructed using the measured dynamical mass of HR~7672B --- as is apparent, our new measurements are in tension with a number of the models, which we examine quantitatively below.

\newcommand{\tn}[1]{\tablenotemark{\scriptsize #1}}

{\catcode`\&=11
  \gdef\Boucsixteencite{\citet{2016A&A...585A..46B}}
  \gdef\Cevatwentyfivecite{\citet{2025A&A...701A..78C}}
  \gdef\Cheeeighteenbcite{\citet{2018A&A...614A..16C}}
  \gdef\DeRotwentythreecite{\citet{2023A&A...672A..94D}}
  \gdef\Deloseventeenbcite{\citet{2017A&A...608A..79D}}
  \gdef\Hinktwentythreecite{\citet{2023A&A...671L...5H}}
  \gdef\Kingtencite{\citet{2010A&A...510A..99K}}
  \gdef\Mairtwentyacite{\citet{2020A&A...633L...2M}}
  \gdef\Mairtwentybcite{\citet{2020A&A...639A..47M}}
  \gdef\Mairtwentyfourcite{\citet{2024A&A...691A.263M}}
  \gdef\Mesatwentythreecite{\citet{2023A&A...672A..93M}}
  \gdef\Millseventeencite{\citet{2017A&A...597L...2M}}
  \gdef\Nasetwentyfourcite{\citet{2024A&A...687A.298N}}
  \gdef\Nowatwentycite{\citet{2020A&A...642L...2N}}
  \gdef\Perenineteencite{\citet{2019A&A...631A.107P}}
  \gdef\Ricknineteencite{\citet{2019A&A...625A..71R}}
  \gdef\Ricktwentycite{\citet{2020A&A...635A.203R}}
  \gdef\Ricktwentyfourcite{\citet{2024A&A...684A..88R}}
  \gdef\Schothreecite{\citet{2003A&A...398L..29S}}
  \gdef\Vigasixteencite{\citet{2016A&A...587A..55V}}
  \gdef\Winttwentyfourcite{\citet{2024A&A...688A..44W}}
  \gdef\Winttwentyfivecite{\citet{2025A&A...700A...4W}}
  \gdef\Zurlsixteencite{\citet{2016A&A...587A..57Z}}
}

\begin{longrotatetable}
\begin{deluxetable*}{lcccccccl}
\tablecaption{Imaged Ultracool Companions with Dynamical Mass Measurements\label{tab:benchmarks}}
\tablewidth{0pt}
\tabletypesize{\footnotesize}
\setlength{\tabcolsep}{2pt}   
\tablehead{
   \colhead{Object} &
   \colhead{RA, Dec} &
   \colhead{SpT} &
   \colhead{Host SpT} &
   \colhead{Separation} &
   \colhead{Mass} &
   \colhead{$\log(\Lbol/\Lsun)$} &
   \colhead{Host Age} &
   \colhead{Refs:} \\
   [-6pt]%
   \colhead{} &
   \colhead{(J2000)} &
   \colhead{} &
   \colhead{} &
   \colhead{} &
   \colhead{(\Mjup)} &
   \colhead{(dex)} &
   \colhead{(Gyr)} &
   \colhead{Discovery; SpT; Mass; \Lbol; Age}
}
\decimalcolnumbers
\startdata                 
HD 984 B      &    3.5427, $-$7.1991   &  M6                 &  F7V     &  0.2\arcsec\   (9.1 au)      &  $61\pm4$               &  $-2.815\pm0.024$            &  30$-$200                   &  Mesh15b; Mesh15b; Fran22; Mesh15b; Mesh15b                  \\
HD 4113 C     &   10.8025, $-$37.9826  &  T9                 &  G5V     &  0.54\arcsec\  (22.5 au)     &  $66_{-4}^{+5}$         &  $-6.3\pm0.22$               &  $5_{-0.8}^{+0.7}$          &  Chee18b; Chee18b; Chee18b; Bran21b; Bran21b                 \\
HD 4747 B     &   12.3615, $-$23.2125  &  L9                 &  G8/K0V  &  0.61\arcsec\  (11.5 au)     &  $68.7_{-1.4}^{+1.5}$   &  $-4.55\pm0.08$              &  $2.9_{-0.5}^{+0.4}$        &  Crep16; Pere19; An25; Bran21b; Bran21b                  \\
HD 13724 B    &   33.0862, $-$46.8164  &  T4                 &  G3/5V   &  0.18\arcsec\  (7.8 au)      &  $36.2_{-1.5}^{+1.6}$   &  $-4.78\pm0.07$              &  $2.8_{-0.4}^{+0.5}$        &  Rick19;Rick20; Rick20; Bran21b; Bran21b; Bran21b            \\
HD 17155 B    &   40.8926, $-$46.4549  &  M9$_{-1.5}^{+1.5}$ &  K4V     &  0.04\arcsec\  (1.2 au)      &  $80\pm4$               &  $-3.57\pm0.1$               &  \nodata                    &  Wint24; Kamm25; Wint24; Kamm25; --                          \\
HD 19467B     &   46.8270, $-$13.7620  &  T6:                &  G3V     &  1.65\arcsec\  (52.9 au)     &  $65.4_{-4.6}^{+5.9}$   &  $-5.16\pm0.08$              &  7$-$10                     &  Crep15; Mesa20; Bran21b; Bran21b; Mair20b                   \\
HIP 21152 B   &   68.0200,   +5.4100   &  T0                 &  F5V     &  0.37\arcsec\  (16.0 au)     &  $24_{-4}^{+6}$         &  $-4.57\pm0.07$              &  $0.75\pm0.1$               &  Kuzu22, Bona22, Fran23b; Fran23b; Fran23b; Fran23b; Bran15  \\
HD 33632 Ab   &   78.3227,   +37.3373  &  L9.5$_{-3}^{+1}$   &  F8V     &  0.22\arcsec\  (5.8 au)      &  $52.8_{-2.4}^{+2.6}$   &  $-4.69\pm0.07$              &  $1.7\pm0.4$                &  Curr20; Curr20; ElMo25; Bran21b; Bran21b                    \\
AF Lep b      &   81.7698, $-$11.9010  &  T2                 &  F8V     &  0.34\arcsec\  (9.1 au)      &  $3.75\pm0.5$           &  $-5.2_{-0.2}^{+0.1}$        &  $0.024\pm0.003$            &  DeRo23;Fran23;Mesa23; Kamm25; Balm25; Balm24; Bell15        \\
$\beta$ Pic b &   86.8212, $-$51.0665  &  L2+/-1             &  A6V     &  0.41\arcsec\  (8.1 au)      &  $9.3_{-2.5}^{+2.6}$    &  $-3.74\pm0.03$\tn{a}              &  $0.024\pm0.003$            &  Lagr10; Chil17; Bran21c; Kamm24; Bell15                     \\
$\beta$ Pic c &   86.8212, $-$51.0665  &  \nodata            &  A6V     &  0.13\arcsec\  (2.6 au)      &  $8.3\pm1$              &  $-4.43\pm0.10$\tn{j}              &  $0.024\pm0.003$            &  Lagr19;Nowa20; --; Bran21c; Nowa20, Li26; Bell15        \\
Gl 229 Ba     &   92.6450, $-$21.8667  &  T7(phot)           &  M1V     &  7.8\arcsec\   (44.9 au)     &  $38.1\pm1$             &  $-5.36\pm0.07$              &  $4\pm2$                    &  Naka95b; Xuan24; Xuan24; Xuan24; Bran20                     \\
Gl 229 Bb     &   92.6450, $-$21.8667  &  T8(phot)           &  M1V     &  7.8\arcsec\   (44.9 au)     &  $34.4\pm1.5$           &  $-5.56\pm0.07$              &  $4\pm2$                    &  Naka95b; Xuan24; Xuan24; Xuan24; Bran20                     \\
HD 63754 B\tn{d}    &  117.4386, $-$20.2076  &  \nodata               &  G0V     &  0.47\arcsec\  (23.6 au)     &  $81.9_{-5.8}^{+6.4}$   &  $-4.55\pm0.08$              &  2.4$-$10\tn{d}                   &  Li24; --; Li24; Li24; Li24                  \\
HIP 39017 b   &  119.7664, $-$4.3324   &  T2                 &  F9/A0   &  0.4\arcsec\   (26.4 au)     &  $30_{-12}^{+31}$       &  $-4.5\pm0.1$                &  0.116$-$0.7                &  Tobi24; Tobi24; Tobi24; Tobi24; Tobi24                      \\
HD 72946 B    &  128.9636,   +6.6228   &  L5+/-1.5           &  G5V     &  0.24\arcsec\  (6.2 au)      &  $69.5\pm0.5$           &  $-4.12\pm0.02$            &  $1.9_{-0.6}^{+0.5}$        &  Bouc16;Mair20a; Mair20a; Balm23; Bran21b; Bran21b           \\
HIP 54515 B   &  167.3069, $-$1.3333   &  L0+/-2             &  A5V     &  0.14\arcsec\  (11.6 au)     &  $17.7_{-4.9}^{+7.6}$   &  $-3.52\pm0.03$              &  $0.115_{-0.092}^{+0.085}$  &  Curr25; Curr25; Curr25; Curr25; Curr25                      \\
Gl 417B       &  168.1068,   +35.8035  &  L4.5:              &  G2V     &  90\arcsec\    (2,133.0 au)  &  $58.2\pm1.5$\tn{c}     &  $-4.132_{-0.031}^{+0.029}$  &  $0.75_{-0.12}^{+0.15}$     &  Kirk00; Dupu12; Dupu17, Li26; Dupu17; Dupu14                      \\
Gl 417C       &  168.1068,   +35.8035  &  L6:                &  G2V     &  90\arcsec\    (2,133.0 au)  &  $51.8\pm2.4$\tn{c}     &  $-4.22\pm0.03$              &  $0.75_{-0.12}^{+0.15}$     &  Kirk00; Dupu12; Dupu17, Li26; Dupu17; Dupu14                      \\
HD 112863 B   &  194.9396, $-$4.4303   &  L3                 &  K1V     &  0.11\arcsec\  (4.1 au)      &  $77.1_{-2.8}^{+2.9}$   &  $-3.76\pm0.035$\tn{b}             &  $3.31\pm2.91$              &  Rick24; Ceva25; Rick24; Ceva25; Rick24                      \\
HD 130948 B   &  222.5667,   +23.9116  &  L4:                &  F9IV-V  &  2.64\arcsec\  (48.0 au)     &  $59.4\pm0.9$\tn{c}     &  $-3.85_{-0.06}^{+0.07}$     &  $0.79_{-0.15}^{+0.22}$\tn{e}     &  Pott02; Goto02; Dupu17, Li26; Dupu17; Dupu09a           \\
HD 130948 C   &  222.5667,   +23.9116  &  L4:                &  F9IV-V  &  2.64\arcsec\  (48.0 au)     &  $56.5\pm1.8$\tn{c}     &  $-3.96\pm0.06$              &  $0.79_{-0.15}^{+0.22}$\tn{e}     &  Pott02; Goto02; Dupu17, Li26; Dupu17; Dupu09a           \\
Gl 569 Ba     &  223.6226,   +16.1024  &  M8.5               &  M3V     &  5\arcsec\     (49.7 au)     &  $80.1_{-8}^{+8.6}$\tn{c}     &  $-3.44\pm0.03$              &  \nodata                    &  Skru87b; Henr90;Lane01; Kono10, Dupu17; Dupu17; --          \\
Gl 569 Bb     &  223.6226,   +16.1024  &  M9.0               &  M3V     &  5\arcsec\     (49.7 au)     &  $57.5_{-8.5}^{+6.8}$\tn{c}   &  $-3.67\pm0.05$              &  \nodata                    &  Skru87b; Henr90;Lane01; Kono10, Dupu17; Dupu17; --          \\
HIP 75056 Ab\tn{f}  &  230.0558, $-$34.9254  &  M7$_{-0.5}^{+1.5}$ &  A2V     &  0.15\arcsec\  (18.3 au)     &  $35\pm10$              &  $-2.8\pm0.3$                &  $0.016\pm0.007$\tn{f}            &  Wagn20; Kamm25; Balm24; Balm24; Peca16                      \\
14 Her c      &  242.6013,   +43.8176  &  \nodata               &  K0V     &  1.11\arcsec\  (19.9 au)     &  $7.9_{-1.2}^{+1.6}$    &  \nodata                     &  $4.6_{-1.3}^{+3.8}$        &  Bard25; --; Bard25; --; Bard21                              \\
HD 167665 B   &  274.3497, $-$28.2896  &  T4$_{-2}^{+1}$     &  F9V     &  0.18\arcsec\  (5.6 au)      &  $60.3\pm0.7$           &  $-4.892_{-0.028}^{+0.024}$  &  $6.2\pm1.13$               &  Mair24; Mair24; Mair24; Mair24; Mair24                      \\
PZ Tel B      &  283.2745, $-$50.1805  &  M6                 &  G9IV    &  0.33\arcsec\  (15.6 au)     &  $27_{-9}^{+25}$        &  $-2.59\pm0.08$              &  $0.024\pm0.003$            &  Bill10; Kamm25; Fran23c; Fran23c; Bell15                    \\
HD 176535 B   &  285.3312, $-$13.6907  &  T6                 &  F9V     &  0.35\arcsec\  (12.9 au)     &  $65.9_{-1.7}^{+2}$     &  $-5.26\pm0.07$              &  $3.59_{-1.15}^{+0.87}$     &  Li23; Li23; Li23; Li23; Li23            \\
GJ 758 B      &  290.8917,   +33.2220  &  T8                 &  G8V     &  1.9\arcsec\   (29.7 au)     &  $35.9_{-0.65}^{+0.69}$ &  $-6.07\pm0.03$              &  $8.3_{-2.1}^{+2.7}$        &  Thal09; Viga16; An25; Bowl18; Bran21b                   \\
HR 7672 B     &  301.0259,   +17.0702  &  L4.5:              &  G0V     &  0.79\arcsec\  (14.0 au)     &  $75.39\pm0.67$         &  $-4.19\pm0.04$              &  $2.26\pm0.4$               &  Liu02b; Liu02b; Li26; Bran19; Li26              \\
HIP 99770 b   &  303.6335,   +36.8063  &  L8                 &  A2V     &  0.45\arcsec\  (18.3 au)     &  $17_{-5}^{+6}$         &  $-4.53\pm0.02$              &  0.115$-$0.415              &  Curr23; Wint25; Wint25; Curr23; Curr23                      \\
Gl 802 B      &  310.8303,   +55.3481  &  \nodata               &  M5Ve    &  0.09\arcsec\  (1.5 au)      &  $66\pm5$               &  $-4.43\pm0.06$\tn{g}              &  8$-$11                     &  Prav05; --; Irel08; Li26; Belo18, Helm18                \\
WT 766 B      &  315.2809, $-$49.1236  &  L1                 &  M4.5    &  0.09\arcsec\  (1.2 au)      &  $77_{-1.3}^{+1.4}$     &  $-3.58_{-0.08}^{+0.06}$     &  \nodata                    &  Wint24; Kamm25; Wint24; Kamm25; --                          \\
HD 206893 B   &  326.3413, $-$12.7834  &  L6+/-2             &  F5V     &  0.27\arcsec\  (11.0 au)     &  $28_{-2.1}^{+2.2}$     &  $-4.24_{-0.09}^{+0.1}$\tn{h}     &  $0.25_{-0.2}^{+0.45}$\tn{h}       &  Mill17; Ward21; Hink23; Kamm25; Delo17b                     \\
HD 206893 c   &  326.3413, $-$12.7834  &  \nodata               &  F5V     &  0.11\arcsec\  (4.5 au)      &  $12.7_{-1}^{+1.2}$     &  $-4.42_{-0.1}^{+0.12}$      &  $0.25_{-0.2}^{+0.45}$\tn{h}      &  Hink23; --; Hink23; Kamm25; Delo17b                         \\
HD 206505 B   &  327.1291, $-$78.4332  &  L2                 &  K1V     &  0.28\arcsec\  (12.3 au)     &  $79.8\pm1.8$           &  $-3.689\pm0.028$\tn{b}            &  $3.94\pm2.51$              &  Rick24; Ceva25; Rick24; Ceva25; Rick24                      \\
eps Indi Ab   &  330.8724, $-$56.7973  &  \nodata               &  K5V     &  4.1\arcsec\   (14.9 au)     &  $6.31_{-0.56}^{+0.6}$  &  \nodata                     &  $3.5_{-1}^{+0.8}$          &  Matt24; --; Matt24; --; Chen22                              \\
eps Indi Ba   &  331.0442, $-$56.7828  &  T1                 &  K5V     &  402\arcsec\   (1,485.0 au)  &  $66.92\pm0.36$         &  $-4.691\pm0.017$            &  $3.5_{-1}^{+0.8}$          &  Scho03; King10; Chen22; King10, Chen22; Chen22              \\
eps Indi Bb   &  331.0442, $-$56.7828  &  T6                 &  K5V     &  402\arcsec\   (1,485.0 au)  &  $53.25\pm0.29$         &  $-5.224\pm0.02$             &  $3.5_{-1}^{+0.8}$          &  Scho03; King10; Chen22; King10, Chen22; Chen22              \\
HR 8799 e     &  346.8697,   +21.1343  &  L7                 &  A5V     &  0.37\arcsec\  (15.1 au)     &  [$9.6_{-1.8}^{+1.9}$]\tn{i}  &  $-4.72\pm0.06$\tn{i}              &  $0.042_{-0.004}^{+0.006}$  &  Maro10b; Zurl16; Bran21a; Nase24; Bell15                    \\
\enddata
\tablecomments{In addition, \citet{2021ApJ...913L..26B} discovered a faint companion to the old ($\approx$7--10~Gyr) G5V star HD~417127, with a dynamical mass of 67.5--177~\Mjup\ (95\% confidence limits) measured by combining RV, imaging, and Hipparcos+Gaia astrometry.  The companion's $H$-band absolute magnitude is too faint relative to this measured mass, suggesting the companion may be an unresolved binary.}
%

\tablenotetext{a}{Computed from the average and RMS of their fitting results using 3 sets of model atmospheres (BT-Settl, DRIFT-PHOENIX, and Exo-REM) applied to all the data.}

\tablenotetext{b}{Computed from the average and RMS of their fitting results using 2 sets of model atmospheres (BT-Settl and Sonora Diamondback).}

\tablenotetext{c}{Individual component masses are computed from the total dynamical mass of the binary, the flux ratio, and an evolutionary model.  While strictly speaking these results are model-dependent, the near-equal flux ratios suggest that systematics from this process are small.  For HD~130948BC and GJ~417BC, the results from \citet{Dupuy2017} have been updated to account for the change in parallax from Hipparcos to Gaia DR3, using the mass ratios estimated in \citet{2009ApJ...692..729D} and \citet{2014ApJ...790..133D}, respectively.}

\tablenotetext{d}{(1) Object: SIMBAD indicates this companion is improperly designated as HD~63754~B in \citet{2024MNRAS.533.3501L}, since the Washington Double Star catalog \citep{2001AJ....122.3466M} notes a faint ($\Delta{\rm mag}\approx7$) companion at $\approx$5\arcsec\ separation (known as \text{**}SEE 87B), first observed by Thomas J.\ J.\ See in 1897.  \citet{2014AJ....147...86T} note that the companion does not share the same proper motion as the primary (a.k.a.\ HIP~38216) based on an observation in 2016.  However, Gaia DR3 \citep{GaiaCollaboration2023} show the companion shares a common parallax with the primary, and thus the designation for the ultracool companion should be changed to "C".  
(2)~Host Age: The age is reported by Li24 as $>$3.4 Gyr at 68\% CL and 2.4 Gyr at $>$95\% CL.  We report this here as 2.4-10~Gyr, but this is not a particularly good description of the posterior (see Figure~1 of \citealp{2024MNRAS.533.3501L}).}

\tablenotetext{e}{Computed from gyrochronology, where the posterior of the inferred age distribution is normal in $\log$(age).}

\tablenotetext{f}{(1) Object: a.k.a.\ HD 136164 Ab \citep{2024AJ....167...64B}.  
(2)~Host Age: The host star is a member of the Upper Centaurus Lupus (UCL) region, and the the uncertainty in the age listed here comes from the intrinsic age spread of UCL reported in Table~12 of \citet{2016MNRAS.461..794P}, which is larger than simply quoting their $\pm$2~Myr uncertainty on the region's representation age.}

\tablenotetext{g}{Luminosity computed from $M(K)$ in \citet{2008ApJ...678..463I} and the $M(K)-\Lbol$ relation in \citet{2017ApJS..231...15D}.}

\tablenotetext{h}{(1) Log(\Lbol): \Hinktwentythreecite\ report a more precise value of $-4.23\pm0.01$~dex, but this comes from fitting only a single set of model atmospheres, whereas Kamm25 fit multiple sets and also add in quadrature a 10\% systematic uncertainty due to flux calibration errors.  
(2)~Host Age: The reported error bars on the Delo17b ages "should be considered as very conservative (max allowed range)".}

\tablenotetext{i}{Mass: Bran21a measured a dynamical mass using Hipparcos+Gaia acceleration and imaging data, which also required adopting light priors on the mass ratios of the 4~planets, though the resulting mass for planet~e is not strongly dependent on the prior.
(2) Log(\Lbol): Bayesian Model Average of retrieval results.}

\tablenotetext{j}{Using the \citet{2023ApJ...959...63S} relation between $M(K_{MKO})$ and $\log(\Lbol)$ for young objects, we  compute the difference in $\log(\Lbol)$ between $\beta$~Pic~b and $\beta$~Pic~c, and then use the value for the former from Kamm25 to compute the latter.  Photometry errors are propagated in the calculation, as is the intrinsic 0.057~dex scatter in the Sanghi et al. relation.}

\end{deluxetable*}
\end{longrotatetable}

  {\footnotesize
  \noindent{\bf References ---} 
  An25: \citet{2025ApJS..280...61A}; 
  Balm23: \citet{2023ApJ...956...99B}; 
  Balm24: \citet{2024AJ....167...64B}; 
  Balm25: \citet{2025AJ....169...30B}; 
  Bard21: \citet{2021ApJ...922L..43B}; 
  Bard25: \citet{2025ApJ...988L..18B}; 
  Bell15: \citet{2015MNRAS.454..593B}; 
  Belo18: \citet{2018MNRAS.478..611B}; 
  Bill10: \citet{2010ApJ...720L..82B}; 
  Bona22: \citet{2022MNRAS.513.5588B}; 
  Bouc16: \Boucsixteencite; 
  Bowl18: \citet{2018AJ....155..159B}; 
  Bran15: \citet{2015ApJ...807...58B}; 
  Bran19: \citet{2019AJ....158..140B}; 
  Bran20: \citet{2020AJ....160..196B}; 
  Bran21a: \citet{2021ApJ...915L..16B}; 
  Bran21b: \citet{2021AJ....162..301B}; 
  Bran21c: \citet{2021AJ....161..179B}; 
  Ceva25: \Cevatwentyfivecite; 
  Chee18b: \Cheeeighteenbcite; 
  Chen22: \citet{2022AJ....163..288C}; 
  Chil17: \citet{2017AJ....153..182C}; 
  Crep15: \citet{2015ApJ...798L..43C}; 
  Crep16: \citet{2016ApJ...831..136C}; 
  Curr20: \citet{2020ApJ...904L..25C}; 
  Curr23: \citet{2023Sci...380..198C}; 
  Curr25: \citet{2025arXiv251202159C};
  DeRo23: \DeRotwentythreecite; 
  Delo17b: \Deloseventeenbcite; 
  Dupu09a: \citet{2009ApJ...692..729D}; 
  Dupu12: \citet{2012ApJS..201...19D}; 
  Dupu14: \citet{2014ApJ...790..133D}; 
  Dupu17: \citet{2017ApJS..231...15D}; 
  ElMo25: \citet{2025ApJ...981...20E}; 
  Fran22: \citet{2022AJ....163...50F}; 
  Fran23: \citet{2023ApJ...950L..19F}; 
  Fran23b: \citet{2023AJ....165...39F}; 
  Fran23c: \citet{2023AJ....165..246F}; 
  Goto02: \citet{2002ApJ...567L..59G}; 
  Helm18: \citet{2018Natur.563...85H}; 
  Henr90: \citet{1990ApJ...354L..29H}; 
  Hink23: \Hinktwentythreecite; 
  Irel08: \citet{2008ApJ...678..463I}; 
  Kamm24: \citet{2024AJ....168...51K}; 
  Kamm25: \citet{2025arXiv251008691K}; 
  King10: \Kingtencite; 
  Kirk00: \citet{2000AJ....120..447K}; 
  Kono10: \citet{2010ApJ...711.1087K}; 
  Kuzu22: \citet{2022ApJ...934L..18K}; 
  Lagr10: \citet{2010Sci...329...57L}; 
  Lagr19: \citet{2019NatAs...3.1135L}; 
  Lane01: \citet{2001ApJ...560..390L}; 
  Li23: \citet{2023MNRAS.522.5622L}; 
  Li24: \citet{2024MNRAS.533.3501L}; 
  Li26: this work; 
  Liu02b: \citet{2002ApJ...571..519L}; 
  Mair20a: \Mairtwentyacite; 
  Mair20b: \Mairtwentybcite; 
  Mair24: \Mairtwentyfourcite; 
  Maro10b: \citet{2010Natur.468.1080M}; 
  Matt24: \citet{2024Natur.633..789M}; 
  Mesa20: \citet{2020MNRAS.495.4279M}; 
  Mesa23: \Mesatwentythreecite; 
  Mesh15b: \citet{2015MNRAS.453.2378M}; 
  Mill17: \Millseventeencite; 
  Naka95b: \citet{1995Natur.378..463N}; 
  Nase24: \Nasetwentyfourcite; 
  Nowa20: \Nowatwentycite; 
  Peca16: \citet{2016MNRAS.461..794P}; 
  Pere19: \Perenineteencite; 
  Pott02: \citet{2002ApJ...567L.133P}; 
  Prav05: \citet{2005ApJ...630..528P}; 
  Rick19: \Ricknineteencite; 
  Rick20: \Ricktwentycite; 
  Rick24: \Ricktwentyfourcite; 
  Scho03: \Schothreecite; 
  Skru87b: \citet{1987BAAS...19.1128S}; 
  Thal09: \citet{2009ApJ...707L.123T}; 
  Tobi24: \citet{2024AJ....167..205T}; 
  Viga16: \Vigasixteencite; 
  Wagn20: \citet{2020ApJ...902L...6W}; 
  Ward21: \citet{2021AJ....161....5W}; 
  Wint24: \Winttwentyfourcite; 
  Wint25: \Winttwentyfivecite; 
  Xuan24: \citet{2024Natur.634.1070X}; 
  Zurl16: \Zurlsixteencite
  }

\subsection{Consistency tests among mass, age, and luminosity}

Independent measurements of mass, age, and luminosity allow any two quantities to predict the third using an evolutionary model, enabling a direct comparison between the model-predicted and observed values for the third quantity.  We performed this test using the rejection-sampling procedure described in \citet{Dupuy2017} and \citet{Dupuy2023}. To obtain the model-predicted distribution of $\log L$ based on mass and age, we drew normally distributed random values for mass and age according to their observational uncertainties and calculated the corresponding luminosity for each mass-age pair through bicubic interpolation on the model grids.

Conversely, to derive the model-predicted distributions of age (or mass) given the observed distributions of $\log L$ and mass (or age), we applied a rejection-sampling method. We drew normally distributed random values for mass (or age) based on their measured uncertainties and uniformly sampled age (or mass). For each mass–age pair, we computed $\log L$ using the evolutionary model. A random number $u$, uniformly distributed between 0 and 1, was then generated, and the rejection probability was defined as $p \propto \exp[-(X_{\rm obs} - X_{\rm mod})^2 / \sigma_X^2]$, where $X = \log L$. Samples with $p < u$ were rejected, and the accepted samples were used to construct the final age (or mass) distributions.

\begin{figure}
    \centering
    \includegraphics[width=\columnwidth]{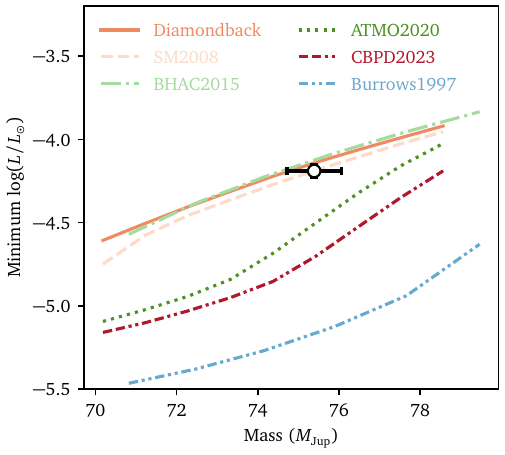}
    \caption{The minimum luminosity reached along the brown dwarf cooling tracks is shown as a function of mass. This minimum luminosity corresponds to the luminosity at the 10~Gyr age in the cooling tracks. Diamondback and SM2008 go beyond 10~Gyr, but we truncate at 10~Gyr for our tests. The evolutionary models populate only the region above the plotted lines, and the circle indicates the observed values for HR~7672B. The minimum luminosities of the Diamondback and BHAC2015 models are only marginally consistent with the observations (model-coverage fractions of 16\% and 24\%, respectively) and the SM2008 models also may be in tension (50\% model-coverage fraction), suggesting that the stellar/substellar boundary in these models is too low in mass.}
    \label{fig:minlogl}
\end{figure}

When using $\log L$ and $M$ to derive the model age distribution, we encountered a limitation: some models do not reach luminosities as low as the observations. As shown in Figure~\ref{fig:cooling}, for Diamondback, SM2008, and BHAC2015, the minimum luminosity
remains higher than the observed value, though still somewhat/marginally consistent when considering the errors in the observed $L$ and the mass. 
Consequently, many samples cannot find valid solutions during rejection sampling due to the restricted model space. Figure~\ref{fig:minlogl} illustrates this effect by showing the minimum luminosity reached by each evolutionary model as a function of mass. Each model's tracks occupy the parameter space above these lines. For Diamondback, SM2008, and BHAC2015, these boundaries lie above the observed luminosity, implying that many randomly drawn samples cannot find acceptable model counterparts. To quantify the impact, we computed a ``model-coverage fraction,'' defined as the proportion of sampled points that fall within the bounds of physical models at 10~Gyr, based on normally sampled luminosity and mass values.

\begin{figure}
    \centering
    \includegraphics[width=\columnwidth]{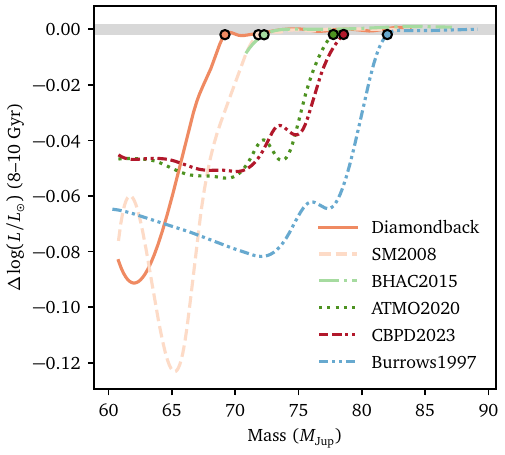}
    \caption{The luminosity differences between 8 and 10 Gyr, $\Delta \log (L/\Lsun)$, shown as a function of mass. The measured stellar/substellar boundaries are shown in circles.}
    \label{fig:boundary}
\end{figure}

\begin{figure*}
    \centering
    \includegraphics[width=\linewidth]{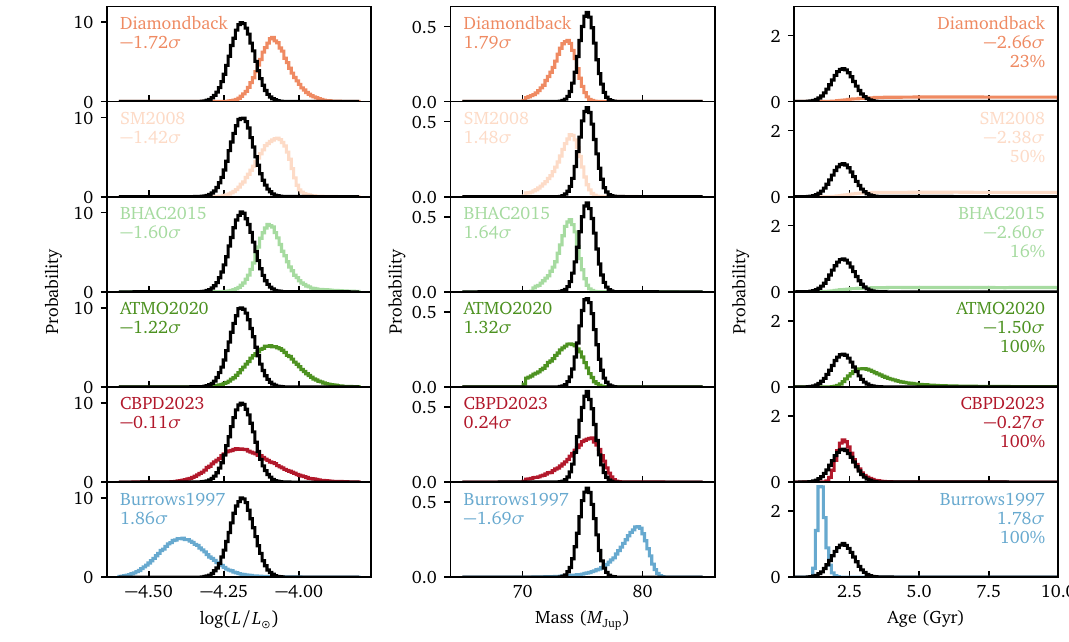}
    \caption{Comparisons between the observed and model-predicted properties of HR~7672B, where the predicted value of each property is calculated based on the other two. The black lines represent the observed distributions, and the colored lines show the model predictions. The numbers in each panel indicate the significance level of the discrepancy between the two distributions. In the rightmost column (age), the model-coverage fraction (the proportion of sampled points that lie within the bounds of physical models at 10~Gyr) is also shown.}
    \label{fig:prob}
\end{figure*}

Figure~\ref{fig:prob} presents the resulting model-predicted probability distributions compared to the observed distributions for $\log L$, mass, and age. 
Table~\ref{table:sigma} lists the significance levels of the discrepancies between the observed and modeled values as well as the model-coverage fractions. The significance levels were derived using a nonparametric sign test. We first computed the residuals between the observational and modeled samples, then determined the fractions of positive and negative residuals. The smaller of the two fractions represents the tail probability from the side with fewer samples, which was multiplied by two to obtain the two-sided tail probability $p$. This probability $p$ was then converted into an equivalent $\sigma$ value using the inverse survival function of a normal distribution. For the age test, to handle models that only partly overlap the observations, we scaled the generate posterior using the model-coverage fraction, by adding a corresponding set of samples with infinite ages to produce the final posterior.

\begin{table*}
\centering
\caption{Significance levels of the discrepancies between the model-predicted and measured properties of HR~7672B. The signs in the first three columns indicate the direction of the discrepancy, expressed as O$-$C (observed$-$modeled). \label{table:sigma}}
\begin{tabular}{lccccc}
\hline\hline
Model &
 $L_{\rm obs}$ vs. $L_{\rm mod}$$(M_{\rm obs},t_{\rm obs})$ &
$M_{\rm obs}$ vs. $M_{\rm mod}$$(L_{\rm obs},t_{\rm obs})$ &
$t_{\rm obs}$ vs. $t_{\rm mod}$$(M_{\rm obs},L_{\rm obs})$ &
$D^2$\\
 &
 &
 &
 (model-coverage fraction) &
\\
\hline
Diamondback & $-1.71\sigma$ & $1.78\sigma$ & $-2.66\sigma$ (0.226) & $1.63\sigma$ \\
SM2008 & $-1.42\sigma$ & $1.48\sigma$ & $-2.38\sigma$ (0.509) & $1.31\sigma$ \\
BHAC2015 & $-1.61\sigma$ & $1.62\sigma$ & $-2.58\sigma$ (0.170) & $1.54\sigma$ \\
ATMO2020 & $-1.22\sigma$ & $1.32\sigma$ & $-1.51\sigma$ (0.998) & $1.15\sigma$ \\
CBPD2023 & $-0.11\sigma$ & $0.25\sigma$ & $-0.27\sigma$ (1.000) & $0.06\sigma$ \\
Burrows1997 & $1.86\sigma$ & $-1.68\sigma$ & $1.78\sigma$ (1.000) & $1.92\sigma$ \\
\hline
\end{tabular}
\end{table*}

\begin{table}
\centering
\caption{Predicted stellar/substellar boundary from each evolutionary model. For reference, HR~7672B's mass is $75.39\pm0.67$~\Mjup{}. \label{table:boundary}}
\begin{tabular}{lccccc}
\hline\hline
Model &
 Boundary Mass (\Mjup{})\\
\hline
Diamondback & $69.2$ \\
SM2008 & $71.8$ \\
BHAC2015 & $72.3$ \\
ATMO2020 & $77.7$ \\
CBPD2023 & $78.6$ \\
Burrows1997 & $82.0$ \\
\hline
\end{tabular}
\end{table}

From these comparisons, we found that the Burrows1997 models predict the lowest luminosities and a clearly substellar nature for HR~7672B. The ATMO2020 model lies near the transition between the bottom of the main sequence and the brown dwarf regime, while its successor CBPD2023 shows the best agreement with our measurements and remains below the threshold for sustained hydrogen fusion. The three models that include cloud treatment through the L/T transition (BHAC2015, SM2008, and Diamondback) predict slightly higher luminosities than observed, indicating a stellar outcome for HR~7672B at the low-mass end of the main sequence given its measured dynamical mass. These models suggest that the object would reach the zero-age main sequence in approximately 1--2~Gyr, producing the long tail in the age distribution in Figure~\ref{fig:prob} and the flattening in the cooling tracks in Figure~\ref{fig:cooling}. The implication for these three models is that their predicted stellar/substellar boundary is too low in mass. For reference, we list the stellar/substellar boundary in Table~\ref{table:boundary}. To compute these values, we examined the brown dwarf cooling sequences and evaluated the change in $\log L$ between 8 and 10~Gyr. We denote the stellar/substellar boundary as the mass above which the $\log L$ change remains below 0.002~dex. Figure~\ref{fig:boundary} illustrates this procedure. 

Despite the differing evolutionary pathways predicted by the models for HR~7672B’s mass, these predictions have not yet diverged significantly at the age of the object. As summarized in Table~\ref{table:sigma}, all models remain consistent with the measurements within $\approx$2.7$\sigma$.

Finally, we consider which observed quantities could most improve model discrimination (Figure~\ref{fig:prob}). The age uncertainty (18\%) is relatively large, and further improvements in the host star age will be challenging, given the substantial observing time needed to measure the current asteroseismic age.  Moreover, an improved age is unlikely to enhance model separation, as the system lies in a region where models predict similar luminosities. The narrow substellar mass range already provides some discriminatory power, and upcoming data from Gaia DR4 can further refine the mass and improve model differentiation. A reduced luminosity uncertainty (currently 10\%) would likely provide the strongest improvement, as it could distinguish between the higher-mass substellar boundary predicted by ATMO2020 and the lower-mass boundary suggested by models such as SM2008, which perform well across the L/T transition \citep{2015ApJ...805...56D, Dupuy2017, Chenmh2022}.

\subsection{Goodness-of-fit test for mass, age, and luminosity}

In the three-dimensional parameter space defined by mass, age, and luminosity, the evolutionary models form two-dimensional manifolds, whereas the observations correspond to a single point with associated uncertainties. We assess the goodness of fit by evaluating how close the observed point lies to each model manifold. To quantify this, we compute the Mahalanobis distance $D^2$, which reduces to a $\chi^2$ metric in the absence of correlations among variables, which is a reasonable assumption in our case. The observed luminosity and dynamical mass both depend on the measured parallax to the host star, but its uncertainty is an irrelevantly small contribution to the total uncertainties for these two quantities. The asteroseismic analysis also did not use the parallax.

We determined the distance $D^2$ between the observed point and the nearest point on each model manifold. The distance $D^2$ follows a chi-squared distribution with 1 degree of freedom, corresponding to the single direction normal to the two-dimensional manifold in three-dimensional parameter space (3–2). We calculated the two-sided tail probability $p$, defined as the probability that a random draw from this distribution exceeds the observed $D^2$. The probability $p$ is then converted to an equivalent $\sigma$ value using the inverse survival function of a normal distribution. 
The results are reported in Table~\ref{table:sigma}. 

We see that the overall trends are consistent with those from the tests performed earlier, with CBPD2023 showing the best agreement with the observations. The $\sigma$ values reported are also generally lower than those reported from the consistency tests because the current method allows the model to move along the manifold surface to find the closest point to the observations. This always yields the smallest possible residual compatible with the model. In contrast, the consistency test fixes two coordinates at their observed values and evaluates the residual in the third dimension. It does not allow movement along the surface, so the resulting residual is measured along a coordinate direction rather than normal to the surface, and is therefore typically larger.

\section{Conclusions}\label{sec:conc}

In this work, we obtained three nights of extremely precise RV data for HR 7672A with the Keck Planet Finder and determined the stellar age using RV asteroseismology and activity-based gyrochronology.   We also obtained a new epoch of astrometry for HR~7672B and, combined with new long-term RV monitoring of the host star, resulted in a more precise dynamical mass for the system.  The resulting measurements enable a direct high-precision test of brown dwarf evolutionary models at the stellar/substellar boundary. Our main findings are as follows:

\begin{enumerate}

\item Asteroseismic modeling of HR 7672A yields an age of $1.87\pm0.65$ Gyr; gyrochronology gives $2.39\pm0.48$ Gyr; and a joint analysis provides a final estimate of $2.26\pm0.40$ Gyr, corresponding to a 18\% fractional uncertainty for this young Sun-like star.

\item Modeling of the relative astrometry, RV, and Hipparcos-Gaia accelerations data produced the the dynamical orbit of the system, resulting in revised dynamical masses of $1.111\pm0.017$~\Msun{} for HR 7672A and $75.39\pm0.67$~\Mjup{} for HR 7672B.

\item We performed consistency tests on six sets of brown dwarf evolutionary models against the observed mass, age, and luminosity of the companion. Each test used two of these quantities as inputs and compared the model predictions for the third.  The CBPD2023 models \citep{Chabrier2023}, which incorporates an updated equation of state, showing excellent agreement, within $<0.3\sigma$ of all the observations.  The other 5 sets of models agree within 1--3$\sigma$ of the observations depending on the test. Also, three of them (Diamondback, SM2008, and BHAC2015) struggle to predict luminosities as low as the observed luminosity, a discrepancy that is independent of the system’s age (Figure~\ref{fig:minlogl}). 

\item We also conducted goodness-of-fit tests by measuring the distance of the observed properties from the two-dimensional manifold defined by each evolutionary model in the three-dimensional (mass, age, luminosity) space. The CBPD2023 models again show the best overall agreement, while the other models are consistent within 1--2$\sigma$.

\item TESS observed the system nearly simultaneously with our KPF monitoring. We detected a weak seismic signal in the TESS data and constrained the photometry-to-RV oscillation amplitude ratio to be close to the solar value, consistent with predictions from 3D hydrodynamical simulations.
\end{enumerate}

Future measurements of HR~7672B's bolometric luminosity, for example with JWST photometry, could significantly improve its precision. Combined with a more precise astrometric solution from Gaia DR4, which will benefit from a baseline roughly twice long as Gaia DR3, the dynamical mass constraints are expected to improve further. As shown in Figure~\ref{fig:cooling}, these improvements could enable much stronger discrimination among cooling models. This is especially valuable for testing their performance near the stellar/substellar boundary, where HR 7672B resides. Some evolutionary models already predict that hydrogen fusion may occur in the future of this object. With these new measurements, HR 7672B will continue to serve as one of the most robust benchmarks for brown dwarf evolution and cooling physics. Moreover, the differences between models in this mass range increases for older ages, so the discovery of older analogs to the HR~7672AB system from, e.g., Gaia DR4, will be exceptionally valuable to test our understanding of the evolution of low-mass stars and brown dwarfs.

\section*{Acknowledgements}
The authors wish to recognize and acknowledge the very significant cultural role and reverence that the summit of Maunakea has always had within the Native Hawaiian community. We are most fortunate to have the opportunity to conduct observations from this mountain.
This work has made use of data from the European Space Agency (ESA) mission {\it Gaia} (\url{https://www.cosmos.esa.int/gaia}), processed by the {\it Gaia}
Data Processing and Analysis Consortium (DPAC, \url{https://www.cosmos.esa.int/web/gaia/dpac/consortium}). Funding for the DPAC has been provided by national institutions, in particular the institutions participating in the {\it Gaia} Multilateral Agreement.
This work has benefited from The UltracoolSheet at \url{http://bit.ly/UltracoolSheet}, maintained by Will Best, Trent Dupuy, Michael Liu, Aniket Sanghi, Rob Siverd, and Zhoujian Zhang, and developed from compilations by \citet{2012ApJS..201...19D, 2013Sci...341.1492D, 2014ApJ...792..119D, 2016ApJ...833...96L, 2018ApJS..234....1B, 2021AJ....161...42B, 2023ApJ...959...63S, 2023AJ....166..103S}.
We thank Will Balmer, Brendan Bowler, Kyle Franson, and Jerry Xuan for discussions about the census of substellar companions with dynamical masses, and Adam Burrows for comments on the manuscript.
Y.L. acknowledges support from the National Aeronautics and Space Administration (80NSSC25K7904) and the Beatrice Watson Parrent Fellowship.  
M.L. acknowledges past support from the Parrent Fellowship 24~years ago that enabled the discovery of HR~7672B and current support from the Gordon and Betty Moore Foundation through grant GBMF8550.
T.S.M. acknowledges support from NASA grant 80NSSC25K7898.
T.R.B. acknowledges support from the Australian Research Council (FL220100117). 
T.L.C. is supported by Funda\c c\~ao para a Ci\^encia e a Tecnologia (FCT) in the form of a work contract (\href{https://doi.org/10.54499/2023.08117.CEECIND/CP2839/CT0004}{2023.08117.CEECIND/CP2839/CT0004}).
J.R.L. and M.S.L. acknowledge support from VILLUM FONDEN (42101) and The Independent Research Fund Denmark’s Inge Lehmann program (1131-00014B).

\bibliography{references/yaguang,references/others,references/mliu}{}

\begin{thebibliography}{}
\expandafter\ifx\csname natexlab\endcsname\relax\def\natexlab#1{#1}\fi
\providecommand{\url}[1]{\href{#1}{#1}}
\providecommand{\mhref}[2]{\href{#1}{\color{teal}#2}}
\providecommand{\dodoi}[1]{doi:~\href{http://doi.org/#1}{\nolinkurl{#1}}}
\providecommand{\doeprint}[1]{\href{http://ascl.net/#1}{\nolinkurl{http://ascl.net/#1}}}
\providecommand{\doarXiv}[1]{\href{https://arxiv.org/abs/#1}{\nolinkurl{https://arxiv.org/abs/#1}}}

\bibitem[{E.~G. {Adelberger} {et~al.}(1998){Adelberger}, {Austin}, {Bahcall},
  {Balantekin}, {Bogaert}, {Brown}, {Buchmann}, {Cecil}, {Champagne}, {de
  Braeckeleer}, {Duba}, {Elliott}, {Freedman}, {Gai}, {Goldring}, {Gould},
  {Gruzinov}, {Haxton}, {Heeger}, {Henley}, {Johnson}, {Kamionkowski},
  {Kavanagh}, {Koonin}, {Kubodera}, {Langanke}, {Motobayashi}, {Pandharipande},
  {Parker}, {Robertson}, {Rolfs}, {Sawyer}, {Shaviv}, {Shoppa}, {Snover},
  {Swanson}, {Tribble}, {Turck-Chi{\`e}ze}, \& {Wilkerson}}]{adelberger1998}
{Adelberger}, E.~G., {Austin}, S.~M., {Bahcall}, J.~N., {et~al.} 1998,
  {\mhref{http://doi.org/10.1103/RevModPhys.70.1265}{Reviews of Modern
  Physics}},
  {\href{https://ui.adsabs.harvard.edu/abs/1998RvMP...70.1265A}{70}}{\href{https://ui.adsabs.harvard.edu/abs/1998RvMP...70.1265A}{,
  1265}}

\bibitem[{C.~{Aerts} {et~al.}(2010){Aerts}, {Christensen-Dalsgaard}, \&
  {Kurtz}}]{Aerts2010}
{Aerts}, C., {Christensen-Dalsgaard}, J., \& {Kurtz}, D.~W. 2010,
  {Asteroseismology}, \dodoi{10.1007/978-1-4020-5803-5}

\bibitem[{Q.~{An} {et~al.}(2025{\natexlab{a}}){An}, {Brandt}, {Brandt}, \&
  {Venner}}]{An2025}
{An}, Q., {Brandt}, T.~D., {Brandt}, G.~M., \& {Venner}, A. 2025{\natexlab{a}},
  {\mhref{http://doi.org/10.3847/1538-4365/adfa99}{\apjs}},
  {\href{https://ui.adsabs.harvard.edu/abs/2025ApJS..280...61A}{280}}{\href{https://ui.adsabs.harvard.edu/abs/2025ApJS..280...61A}{,
  61}}

\bibitem[{Q.~{An} {et~al.}(2025{\natexlab{b}}){An}, {Brandt}, {Brandt}, \&
  {Venner}}]{2025ApJS..280...61A}
---. 2025{\natexlab{b}},
  {\mhref{http://doi.org/10.3847/1538-4365/adfa99}{\apjs}},
  {\href{https://ui.adsabs.harvard.edu/abs/2025ApJS..280...61A}{280}}{\href{https://ui.adsabs.harvard.edu/abs/2025ApJS..280...61A}{,
  61}}

\bibitem[{C.~{Angulo} {et~al.}(1999){Angulo}, {Arnould}, {Rayet},
  {Descouvemont}, {Baye}, {Leclercq-Willain}, {Coc}, {Barhoumi}, {Aguer},
  {Rolfs}, {Kunz}, {Hammer}, {Mayer}, {Paradellis}, {Kossionides}, {Chronidou},
  {Spyrou}, {degl'Innocenti}, {Fiorentini}, {Ricci}, {Zavatarelli},
  {Providencia}, {Wolters}, {Soares}, {Grama}, {Rahighi}, {Shotter}, \& {Lamehi
  Rachti}}]{nacre1999}
{Angulo}, C., {Arnould}, M., {Rayet}, M., {et~al.} 1999,
  {\mhref{http://doi.org/10.1016/S0375-9474(99)00030-5}{\nphysa}},
  {\href{https://ui.adsabs.harvard.edu/abs/1999NuPhA.656....3A}{656}}{\href{https://ui.adsabs.harvard.edu/abs/1999NuPhA.656....3A}{,
  3}}

\bibitem[{H.~M. {Antia} \& S.~{Basu}(1994){Antia} \& {Basu}}]{antia1994}
{Antia}, H.~M., \& {Basu}, S. 1994, \aaps,
  {\href{https://ui.adsabs.harvard.edu/abs/1994A&AS..107..421A}{107}}{\href{https://ui.adsabs.harvard.edu/abs/1994A&AS..107..421A}{,
  421}}

\bibitem[{M.~{Asplund} {et~al.}(2009){Asplund}, {Grevesse}, {Sauval}, \&
  {Scott}}]{agss09}
{Asplund}, M., {Grevesse}, N., {Sauval}, A.~J., \& {Scott}, P. 2009,
  {\mhref{http://doi.org/10.1146/annurev.astro.46.060407.145222}{\araa}},
  {\href{https://ui.adsabs.harvard.edu/abs/2009ARA&A..47..481A}{47}}{\href{https://ui.adsabs.harvard.edu/abs/2009ARA&A..47..481A}{,
  481}}

\bibitem[{J.~N. {Bahcall} {et~al.}(1995){Bahcall}, {Pinsonneault}, \&
  {Wasserburg}}]{Bahcall1995}
{Bahcall}, J.~N., {Pinsonneault}, M.~H., \& {Wasserburg}, G.~J. 1995,
  {\mhref{http://doi.org/10.1103/RevModPhys.67.781}{Reviews of Modern
  Physics}},
  {\href{https://ui.adsabs.harvard.edu/abs/1995RvMP...67..781B}{67}}{\href{https://ui.adsabs.harvard.edu/abs/1995RvMP...67..781B}{,
  781}}

\bibitem[{C.~A.~L. {Bailer-Jones} {et~al.}(2021){Bailer-Jones}, {Rybizki},
  {Fouesneau}, {Demleitner}, \& {Andrae}}]{bj2021}
{Bailer-Jones}, C.~A.~L., {Rybizki}, J., {Fouesneau}, M., {Demleitner}, M., \&
  {Andrae}, R. 2021, {\mhref{http://doi.org/10.3847/1538-3881/abd806}{\aj}},
  {\href{https://ui.adsabs.harvard.edu/abs/2021AJ....161..147B}{161}}{\href{https://ui.adsabs.harvard.edu/abs/2021AJ....161..147B}{,
  147}}

\bibitem[{W.~H. {Ball} \& L.~{Gizon}(2014){Ball} \& {Gizon}}]{bg14}
{Ball}, W.~H., \& {Gizon}, L. 2014,
  {\mhref{http://doi.org/10.1051/0004-6361/201424325}{\aap}},
  {\href{https://ui.adsabs.harvard.edu/abs/2014A&A...568A.123B}{568}}{\href{https://ui.adsabs.harvard.edu/abs/2014A&A...568A.123B}{,
  A123}}

\bibitem[{J.~{Ballot} {et~al.}(2011){Ballot}, {Barban}, \& {van't
  Veer-Menneret}}]{Ballot2011}
{Ballot}, J., {Barban}, C., \& {van't Veer-Menneret}, C. 2011,
  {\mhref{http://doi.org/10.1051/0004-6361/201016230}{\aap}},
  {\href{https://ui.adsabs.harvard.edu/abs/2011A&A...531A.124B}{531}}{\href{https://ui.adsabs.harvard.edu/abs/2011A&A...531A.124B}{,
  A124}}

\bibitem[{W.~O. {Balmer} {et~al.}(2023){Balmer}, {Pueyo}, {Stolker},
  {Reggiani}, {Maire}, {Lacour}, {Molli{`e}re}, {Nowak}, {Sing}, {Pourr{'e}},
  {Blunt}, {Wang}, {Rickman}, {Kammerer}, {Henning}, {Ward-Duong}, {Abuter},
  {Amorim}, {Asensio-Torres}, {Benisty}, {Berger}, {Beust}, {Boccaletti},
  {Bohn}, {Bonnefoy}, {Bonnet}, {Bourdarot}, {Brandner}, {Cantalloube},
  {Caselli}, {Charnay}, {Chauvin}, {Chavez}, {Choquet}, {Christiaens},
  {Cl{'e}net}, {Coud{'e} Du Foresto}, {Cridland}, {Dembet}, {Dexter},
  {Drescher}, {Duvert}, {Eckart}, {Eisenhauer}, {Gao}, {Garcia}, {Garcia
  Lopez}, {Gendron}, {Genzel}, {Gillessen}, {Girard}, {Haubois}, {Hei{ss}el},
  {Hinkley}, {Hippler}, {Horrobin}, {Houll{'e}}, {Hubert}, {Jocou}, {Keppler},
  {Kervella}, {Kreidberg}, {Lagrange}, {Lapeyr{`e}re}, {Le Bouquin}, {L{'e}na},
  {Lutz}, {Monnier}, {Mouillet}, {Nasedkin}, {Ott}, {Otten}, {Paladini},
  {Paumard}, {Perraut}, {Perrin}, {Pfuhl}, {Rameau}, {Rodet}, {Rousset},
  {Rustamkulov}, {Shangguan}, {Shimizu}, {Stadler}, {Straub}, {Straubmeier},
  {Sturm}, {Tacconi}, {van Dishoeck}, {Vigan}, {Vincent}, {von Fellenberg},
  {Widmann}, {Wieprecht}, {Wiezorrek}, {Winterhalder}, {Woillez}, {Yazici},
  {Young}, \& {Gravity Collaboration}}]{2023ApJ...956...99B}
{Balmer}, W.~O., {Pueyo}, L., {Stolker}, T., {et~al.} 2023,
  {\mhref{http://doi.org/10.3847/1538-4357/acf761}{\apj}},
  {\href{https://ui.adsabs.harvard.edu/abs/2023ApJ...956...99B}{956}}{\href{https://ui.adsabs.harvard.edu/abs/2023ApJ...956...99B}{,
  99}}

\bibitem[{W.~O. {Balmer} {et~al.}(2024){Balmer}, {Pueyo}, {Lacour}, {Wang},
  {Stolker}, {Kammerer}, {Pourr{\'e}}, {Nowak}, {Rickman}, {Blunt},
  {Sivaramakrishnan}, {Sing}, {Wagner}, {Marleau}, {Lagrange}, {Abuter},
  {Amorim}, {Asensio-Torres}, {Berger}, {Beust}, {Boccaletti}, {Bohn},
  {Bonnefoy}, {Bonnet}, {Bordoni}, {Bourdarot}, {Brandner}, {Cantalloube},
  {Caselli}, {Charnay}, {Chauvin}, {Chavez}, {Choquet}, {Christiaens},
  {Cl{\'e}net}, {Coud{\'e} Du Foresto}, {Cridland}, {Davies}, {Dembet},
  {Drescher}, {Duvert}, {Eckart}, {Eisenhauer}, {Schreiber}, {Garcia}, {Garcia
  Lopez}, {Gendron}, {Genzel}, {Gillessen}, {Girard}, {Grant}, {Haubois},
  {Hei{\ss}el}, {Henning}, {Hinkley}, {Hippler}, {Houll{\'e}}, {Hubert},
  {Jocou}, {Keppler}, {Kervella}, {Kreidberg}, {Kurtovic}, {Lapeyr{\`e}re}, {Le
  Bouquin}, {L{\'e}na}, {Lutz}, {Maire}, {Mang}, {M{\'e}rand}, {Molli{\`e}re},
  {Mordasini}, {Mouillet}, {Nasedkin}, {Ott}, {Otten}, {Paladini}, {Paumard},
  {Perraut}, {Perrin}, {Pfuhl}, {Ribeiro}, {Rodet}, {Rustamkulov}, {Shangguan},
  {Shimizu}, {Straubmeier}, {Sturm}, {Tacconi}, {Vigan}, {Vincent},
  {Ward-Duong}, {Widmann}, {Winterhalder}, {Woillez}, {Yazici}, \& {Gravity
  Collaboration}}]{2024AJ....167...64B}
{Balmer}, W.~O., {Pueyo}, L., {Lacour}, S., {et~al.} 2024,
  {\mhref{http://doi.org/10.3847/1538-3881/ad1689}{\aj}},
  {\href{https://ui.adsabs.harvard.edu/abs/2024AJ....167...64B}{167}}{\href{https://ui.adsabs.harvard.edu/abs/2024AJ....167...64B}{,
  64}}

\bibitem[{W.~O. {Balmer} {et~al.}(2025){Balmer}, {Franson}, {Chomez}, {Pueyo},
  {Stolker}, {Lacour}, {Nowak}, {Nasedkin}, {Bonse}, {Thorngren},
  {Palma-Bifani}, {Molli{`e}re}, {Wang}, {Zhang}, {Chavez}, {Kammerer},
  {Blunt}, {Bowler}, {Bonnefoy}, {Brandner}, {Charnay}, {Chauvin}, {Henning},
  {Lagrange}, {Pourr{'e}}, {Rickman}, {De Rosa}, {Vigan}, \&
  {Winterhalder}}]{2025AJ....169...30B}
{Balmer}, W.~O., {Franson}, K., {Chomez}, A., {et~al.} 2025,
  {\mhref{http://doi.org/10.3847/1538-3881/ad9265}{\aj}},
  {\href{https://ui.adsabs.harvard.edu/abs/2025AJ....169...30B}{169}}{\href{https://ui.adsabs.harvard.edu/abs/2025AJ....169...30B}{,
  30}}

\bibitem[{I.~{Baraffe} {et~al.}(2002){Baraffe}, {Chabrier}, {Allard}, \&
  {Hauschildt}}]{Baraffe2002}
{Baraffe}, I., {Chabrier}, G., {Allard}, F., \& {Hauschildt}, P.~H. 2002,
  {\mhref{http://doi.org/10.1051/0004-6361:20011638}{\aap}},
  {\href{https://ui.adsabs.harvard.edu/abs/2002A&A...382..563B}{382}}{\href{https://ui.adsabs.harvard.edu/abs/2002A&A...382..563B}{,
  563}}

\bibitem[{I.~{Baraffe} {et~al.}(2015){Baraffe}, {Homeier}, {Allard}, \&
  {Chabrier}}]{Baraffe2015}
{Baraffe}, I., {Homeier}, D., {Allard}, F., \& {Chabrier}, G. 2015,
  {\mhref{http://doi.org/10.1051/0004-6361/201425481}{\aap}},
  {\href{https://ui.adsabs.harvard.edu/abs/2015A&A...577A..42B}{577}}{\href{https://ui.adsabs.harvard.edu/abs/2015A&A...577A..42B}{,
  A42}}

\bibitem[{D.~C. {Bardalez Gagliuffi} {et~al.}(2021){Bardalez Gagliuffi},
  {Faherty}, {Li}, {Brandt}, {Williams}, {Brandt}, \&
  {Gelino}}]{2021ApJ...922L..43B}
{Bardalez Gagliuffi}, D.~C., {Faherty}, J.~K., {Li}, Y., {et~al.} 2021,
  {\mhref{http://doi.org/10.3847/2041-8213/ac382c}{\apjl}},
  {\href{https://ui.adsabs.harvard.edu/abs/2021ApJ...922L..43B}{922}}{\href{https://ui.adsabs.harvard.edu/abs/2021ApJ...922L..43B}{,
  L43}}

\bibitem[{D.~C. {Bardalez Gagliuffi} {et~al.}(2025){Bardalez Gagliuffi},
  {Balmer}, {Pueyo}, {Brandt}, {Giovinazzi}, {Millholland}, {Black}, {Lu},
  {Rice}, {Mang}, {Morley}, {Lacy}, {Girard}, {Matthews}, {Carter}, {Bowler},
  {Faherty}, {Fontanive}, \& {Rickman}}]{2025ApJ...988L..18B}
{Bardalez Gagliuffi}, D.~C., {Balmer}, W.~O., {Pueyo}, L., {et~al.} 2025,
  {\mhref{http://doi.org/10.3847/2041-8213/ade30f}{\apjl}},
  {\href{https://ui.adsabs.harvard.edu/abs/2025ApJ...988L..18B}{988}}{\href{https://ui.adsabs.harvard.edu/abs/2025ApJ...988L..18B}{,
  L18}}

\bibitem[{S.~A. {Barnes} {et~al.}(2016){Barnes}, {Weingrill}, {Fritzewski},
  {Strassmeier}, \& {Platais}}]{barnes2016-m67}
{Barnes}, S.~A., {Weingrill}, J., {Fritzewski}, D., {Strassmeier}, K.~G., \&
  {Platais}, I. 2016,
  {\mhref{http://doi.org/10.3847/0004-637X/823/1/16}{\apj}},
  {\href{https://ui.adsabs.harvard.edu/abs/2016ApJ...823...16B}{823}}{\href{https://ui.adsabs.harvard.edu/abs/2016ApJ...823...16B}{,
  16}}

\bibitem[{S.~{Basu} \& W.~J. {Chaplin}(2017){Basu} \& {Chaplin}}]{Basu2017}
{Basu}, S., \& {Chaplin}, W.~J. 2017, {Asteroseismic Data Analysis: Foundations
  and Techniques}

\bibitem[{C.~P.~M. {Bell} {et~al.}(2015){Bell}, {Mamajek}, \&
  {Naylor}}]{2015MNRAS.454..593B}
{Bell}, C.~P.~M., {Mamajek}, E.~E., \& {Naylor}, T. 2015,
  {\mhref{http://doi.org/10.1093/mnras/stv1981}{\mnras}},
  {\href{http://adsabs.harvard.edu/abs/2015MNRAS.454..593B}{454}}{\href{http://adsabs.harvard.edu/abs/2015MNRAS.454..593B}{,
  593}}

\bibitem[{E.~P. {Bellinger}(2019){Bellinger}}]{Bellinger2019}
{Bellinger}, E.~P. 2019, {\mhref{http://doi.org/10.1093/mnras/stz714}{\mnras}},
  {\href{https://ui.adsabs.harvard.edu/abs/2019MNRAS.486.4612B}{486}}{\href{https://ui.adsabs.harvard.edu/abs/2019MNRAS.486.4612B}{,
  4612}}

\bibitem[{V.~{Belokurov} {et~al.}(2018){Belokurov}, {Erkal}, {Evans},
  {Koposov}, \& {Deason}}]{2018MNRAS.478..611B}
{Belokurov}, V., {Erkal}, D., {Evans}, N.~W., {Koposov}, S.~E., \& {Deason},
  A.~J. 2018, {\mhref{http://doi.org/10.1093/mnras/sty982}{\mnras}},
  {\href{https://ui.adsabs.harvard.edu/abs/2018MNRAS.478..611B}{478}}{\href{https://ui.adsabs.harvard.edu/abs/2018MNRAS.478..611B}{,
  611}}

\bibitem[{T.~A. {Berger} {et~al.}(2020){Berger}, {Huber}, {van Saders},
  {Gaidos}, {Tayar}, \& {Kraus}}]{Berger2020}
{Berger}, T.~A., {Huber}, D., {van Saders}, J.~L., {et~al.} 2020,
  {\mhref{http://doi.org/10.3847/1538-3881/159/6/280}{\aj}},
  {\href{https://ui.adsabs.harvard.edu/abs/2020AJ....159..280B}{159}}{\href{https://ui.adsabs.harvard.edu/abs/2020AJ....159..280B}{,
  280}}

\bibitem[{W.~M.~J. {Best} {et~al.}(2021){Best}, {Liu}, {Magnier}, \&
  {Dupuy}}]{2021AJ....161...42B}
{Best}, W. M.~J., {Liu}, M.~C., {Magnier}, E.~A., \& {Dupuy}, T.~J. 2021,
  {\mhref{http://doi.org/10.3847/1538-3881/abc893}{\aj}},
  {\href{https://ui.adsabs.harvard.edu/abs/2021AJ....161...42B}{161}}{\href{https://ui.adsabs.harvard.edu/abs/2021AJ....161...42B}{,
  42}}

\bibitem[{W.~M.~J. {Best} {et~al.}(2018){Best}, {Magnier}, {Liu}, {Aller},
  {Zhang}, {Burgett}, {Chambers}, {Draper}, {Flewelling}, {Kaiser},
  {Kudritzki}, {Metcalfe}, {Tonry}, {Wainscoat}, \&
  {Waters}}]{2018ApJS..234....1B}
{Best}, W. M.~J., {Magnier}, E.~A., {Liu}, M.~C., {et~al.} 2018,
  {\mhref{http://doi.org/10.3847/1538-4365/aa9982}{\apjs}},
  {\href{https://ui.adsabs.harvard.edu/abs/2018ApJS..234....1B}{234}}{\href{https://ui.adsabs.harvard.edu/abs/2018ApJS..234....1B}{,
  1}}

\bibitem[{T.~S. {Bhatia} {et~al.}(2022){Bhatia}, {Cameron}, {Solanki}, {Peter},
  {Przybylski}, {Witzke}, \& {Shapiro}}]{2022A&A...663A.166B}
{Bhatia}, T.~S., {Cameron}, R.~H., {Solanki}, S.~K., {et~al.} 2022,
  {\mhref{http://doi.org/10.1051/0004-6361/202243607}{\aap}},
  {\href{https://ui.adsabs.harvard.edu/abs/2022A&A...663A.166B}{663}}{\href{https://ui.adsabs.harvard.edu/abs/2022A&A...663A.166B}{,
  A166}}

\bibitem[{B.~A. {Biller} {et~al.}(2010){Biller}, {Liu}, {Wahhaj}, {Nielsen},
  {Close}, {Dupuy}, {Hayward}, {Burrows}, {Chun}, {Ftaclas}, {Clarke},
  {Hartung}, {Males}, {Reid}, {Shkolnik}, {Skemer}, {Tecza}, {Thatte},
  {Alencar}, {Artymowicz}, {Boss}, {de Gouveia Dal Pino}, {Gregorio-Hetem},
  {Ida}, {Kuchner}, {Lin}, \& {Toomey}}]{2010ApJ...720L..82B}
{Biller}, B.~A., {Liu}, M.~C., {Wahhaj}, Z., {et~al.} 2010,
  {\mhref{http://doi.org/10.1088/2041-8205/720/1/L82}{\apjl}},
  {\href{http://adsabs.harvard.edu/abs/2010ApJ...720L..82B}{720}}{\href{http://adsabs.harvard.edu/abs/2010ApJ...720L..82B}{,
  L82}}

\bibitem[{A.~{Boccaletti} {et~al.}(2003){Boccaletti}, {Chauvin}, {Lagrange}, \&
  {Marchis}}]{Boccaletti2003}
{Boccaletti}, A., {Chauvin}, G., {Lagrange}, A.~M., \& {Marchis}, F. 2003,
  {\mhref{http://doi.org/10.1051/0004-6361:20031216}{\aap}},
  {\href{https://ui.adsabs.harvard.edu/abs/2003A&A...410..283B}{410}}{\href{https://ui.adsabs.harvard.edu/abs/2003A&A...410..283B}{,
  283}}

\bibitem[{E.~{B{\"o}hm-Vitense}(1958){B{\"o}hm-Vitense}}]{bv1958}
{B{\"o}hm-Vitense}, E. 1958, \zap,
  {\href{https://ui.adsabs.harvard.edu/abs/1958ZA.....46..108B}{46}}{\href{https://ui.adsabs.harvard.edu/abs/1958ZA.....46..108B}{,
  108}}

\bibitem[{M.~{Bonavita} {et~al.}(2022){Bonavita}, {Fontanive}, {Gratton},
  {Mu{v{z}}i{'c}}, {Desidera}, {Mesa}, {Biller}, {Scholz}, {Sozzetti}, \&
  {Squicciarini}}]{2022MNRAS.513.5588B}
{Bonavita}, M., {Fontanive}, C., {Gratton}, R., {et~al.} 2022,
  {\mhref{http://doi.org/10.1093/mnras/stac1250}{\mnras}},
  {\href{https://ui.adsabs.harvard.edu/abs/2022MNRAS.513.5588B}{513}}{\href{https://ui.adsabs.harvard.edu/abs/2022MNRAS.513.5588B}{,
  5588}}

\bibitem[{S.~{Boro Saikia} {et~al.}(2018){Boro Saikia}, {Marvin}, {Jeffers},
  {Reiners}, {Cameron}, {Marsden}, {Petit}, {Warnecke}, \&
  {Yadav}}]{BoroSaikia2018}
{Boro Saikia}, S., {Marvin}, C.~J., {Jeffers}, S.~V., {et~al.} 2018,
  {\mhref{http://doi.org/10.1051/0004-6361/201629518}{\aap}},
  {\href{https://ui.adsabs.harvard.edu/abs/2018A&A...616A.108B}{616}}{\href{https://ui.adsabs.harvard.edu/abs/2018A&A...616A.108B}{,
  A108}}

\bibitem[{F.~{Bouchy} {et~al.}(2016){Bouchy}, {S{'e}gransan}, {D{'i}az},
  {Forveille}, {Boisse}, {Arnold}, {Astudillo-Defru}, {Beuzit}, {Bonfils},
  {Borgniet}, {Bourrier}, {Courcol}, {Delfosse}, {Demangeon}, {Delorme},
  {Ehrenreich}, {H{'e}brard}, {Lagrange}, {Mayor}, {Montagnier}, {Moutou},
  {Naef}, {Pepe}, {Perrier}, {Queloz}, {Rey}, {Sahlmann}, {Santerne}, {Santos},
  {Sivan}, {Udry}, \& {Wilson}}]{2016A&A...585A..46B}
{Bouchy}, F., {S{'e}gransan}, D., {D{'i}az}, R.~F., {et~al.} 2016,
  {\mhref{http://doi.org/10.1051/0004-6361/201526347}{\aap}},
  {\href{https://ui.adsabs.harvard.edu/abs/2016A&A...585A..46B}{585}}{\href{https://ui.adsabs.harvard.edu/abs/2016A&A...585A..46B}{,
  A46}}

\bibitem[{L.~G. {Bouma} {et~al.}(2024){Bouma}, {Hillenbrand}, {Howard},
  {Isaacson}, {Masuda}, \& {Palumbo}}]{Bouma2024}
{Bouma}, L.~G., {Hillenbrand}, L.~A., {Howard}, A.~W., {et~al.} 2024,
  {\mhref{http://doi.org/10.3847/1538-4357/ad855f}{\apj}},
  {\href{https://ui.adsabs.harvard.edu/abs/2024ApJ...976..234B}{976}}{\href{https://ui.adsabs.harvard.edu/abs/2024ApJ...976..234B}{,
  234}}

\bibitem[{B.~P. {Bowler} {et~al.}(2018){Bowler}, {Dupuy}, {Endl}, {Cochran},
  {MacQueen}, {Fulton}, {Petigura}, {Howard}, {Hirsch}, {Kratter}, {Crepp},
  {Biller}, {Johnson}, \& {Wittenmyer}}]{2018AJ....155..159B}
{Bowler}, B.~P., {Dupuy}, T.~J., {Endl}, M., {et~al.} 2018,
  {\mhref{http://doi.org/10.3847/1538-3881/aab2a6}{\aj}},
  {\href{https://ui.adsabs.harvard.edu/abs/2018AJ....155..159B}{155}}{\href{https://ui.adsabs.harvard.edu/abs/2018AJ....155..159B}{,
  159}}

\bibitem[{B.~P. {Bowler} {et~al.}(2021){Bowler}, {Endl}, {Cochran}, {MacQueen},
  {Crepp}, {Doppmann}, {Dulz}, {Brandt}, {Mirek Brandt}, {Li}, {Dupuy},
  {Franson}, {Kratter}, {Morley}, \& {Zhou}}]{2021ApJ...913L..26B}
{Bowler}, B.~P., {Endl}, M., {Cochran}, W.~D., {et~al.} 2021,
  {\mhref{http://doi.org/10.3847/2041-8213/abfec8}{\apjl}},
  {\href{https://ui.adsabs.harvard.edu/abs/2021ApJ...913L..26B}{913}}{\href{https://ui.adsabs.harvard.edu/abs/2021ApJ...913L..26B}{,
  L26}}

\bibitem[{B.~P. {Bowler} {et~al.}(2023){Bowler}, {Tran}, {Zhang}, {Morgan},
  {Ashok}, {Blunt}, {Bryan}, {Evans}, {Franson}, {Huber}, {Nagpal}, {Wu}, \&
  {Zhou}}]{Bowler2023}
{Bowler}, B.~P., {Tran}, Q.~H., {Zhang}, Z., {et~al.} 2023,
  {\mhref{http://doi.org/10.3847/1538-3881/acbd34}{\aj}},
  {\href{https://ui.adsabs.harvard.edu/abs/2023AJ....165..164B}{165}}{\href{https://ui.adsabs.harvard.edu/abs/2023AJ....165..164B}{,
  164}}

\bibitem[{G.~M. {Brandt} {et~al.}(2021{\natexlab{a}}){Brandt}, {Brandt},
  {Dupuy}, {Li}, \& {Michalik}}]{2021AJ....161..179B}
{Brandt}, G.~M., {Brandt}, T.~D., {Dupuy}, T.~J., {Li}, Y., \& {Michalik}, D.
  2021{\natexlab{a}}, {\mhref{http://doi.org/10.3847/1538-3881/abdc2e}{\aj}},
  {\href{https://ui.adsabs.harvard.edu/abs/2021AJ....161..179B}{161}}{\href{https://ui.adsabs.harvard.edu/abs/2021AJ....161..179B}{,
  179}}

\bibitem[{G.~M. {Brandt} {et~al.}(2021{\natexlab{b}}){Brandt}, {Brandt},
  {Dupuy}, {Michalik}, \& {Marleau}}]{2021ApJ...915L..16B}
{Brandt}, G.~M., {Brandt}, T.~D., {Dupuy}, T.~J., {Michalik}, D., \& {Marleau},
  G.-D. 2021{\natexlab{b}},
  {\mhref{http://doi.org/10.3847/2041-8213/ac0540}{\apjl}},
  {\href{https://ui.adsabs.harvard.edu/abs/2021ApJ...915L..16B}{915}}{\href{https://ui.adsabs.harvard.edu/abs/2021ApJ...915L..16B}{,
  L16}}

\bibitem[{G.~M. {Brandt} {et~al.}(2021{\natexlab{c}}){Brandt}, {Dupuy}, {Li},
  {Chen}, {Brandt}, {Wong}, {Currie}, {Bowler}, {Liu}, {Best}, \&
  {Phillips}}]{2021AJ....162..301B}
{Brandt}, G.~M., {Dupuy}, T.~J., {Li}, Y., {et~al.} 2021{\natexlab{c}},
  {\mhref{http://doi.org/10.3847/1538-3881/ac273e}{\aj}},
  {\href{https://ui.adsabs.harvard.edu/abs/2021AJ....162..301B}{162}}{\href{https://ui.adsabs.harvard.edu/abs/2021AJ....162..301B}{,
  301}}

\bibitem[{T.~D. {Brandt} {et~al.}(2019{\natexlab{a}}){Brandt}, {Dupuy}, \&
  {Bowler}}]{Brandt2019}
{Brandt}, T.~D., {Dupuy}, T.~J., \& {Bowler}, B.~P. 2019{\natexlab{a}},
  {\mhref{http://doi.org/10.3847/1538-3881/ab04a8}{\aj}},
  {\href{https://ui.adsabs.harvard.edu/abs/2019AJ....158..140B}{158}}{\href{https://ui.adsabs.harvard.edu/abs/2019AJ....158..140B}{,
  140}}

\bibitem[{T.~D. {Brandt} {et~al.}(2019{\natexlab{b}}){Brandt}, {Dupuy}, \&
  {Bowler}}]{2019AJ....158..140B}
---. 2019{\natexlab{b}},
  {\mhref{http://doi.org/10.3847/1538-3881/ab04a8}{\aj}},
  {\href{https://ui.adsabs.harvard.edu/abs/2019AJ....158..140B}{158}}{\href{https://ui.adsabs.harvard.edu/abs/2019AJ....158..140B}{,
  140}}

\bibitem[{T.~D. {Brandt} {et~al.}(2020){Brandt}, {Dupuy}, {Bowler}, {Bardalez
  Gagliuffi}, {Faherty}, {Brandt}, \& {Michalik}}]{2020AJ....160..196B}
{Brandt}, T.~D., {Dupuy}, T.~J., {Bowler}, B.~P., {et~al.} 2020,
  {\mhref{http://doi.org/10.3847/1538-3881/abb45e}{\aj}},
  {\href{https://ui.adsabs.harvard.edu/abs/2020AJ....160..196B}{160}}{\href{https://ui.adsabs.harvard.edu/abs/2020AJ....160..196B}{,
  196}}

\bibitem[{T.~D. {Brandt} {et~al.}(2021){Brandt}, {Dupuy}, {Li}, {Brandt},
  {Zeng}, {Michalik}, {Bardalez Gagliuffi}, \& {Raposo-Pulido}}]{Brandt2021b}
{Brandt}, T.~D., {Dupuy}, T.~J., {Li}, Y., {et~al.} 2021, {orvara: Orbits from
  Radial Velocity, Absolute, and/or Relative Astrometry}.
\newblock \doeprint{2105.012}

\bibitem[{T.~D. {Brandt} \& C.~X. {Huang}(2015){Brandt} \&
  {Huang}}]{2015ApJ...807...58B}
{Brandt}, T.~D., \& {Huang}, C.~X. 2015,
  {\mhref{http://doi.org/10.1088/0004-637X/807/1/58}{\apj}},
  {\href{http://adsabs.harvard.edu/abs/2015ApJ...807...58B}{807}}{\href{http://adsabs.harvard.edu/abs/2015ApJ...807...58B}{,
  58}}

\bibitem[{A.~{Burrows} {et~al.}(1997){Burrows}, {Marley}, {Hubbard}, {Lunine},
  {Guillot}, {Saumon}, {Freedman}, {Sudarsky}, \& {Sharp}}]{Burrows1997}
{Burrows}, A., {Marley}, M., {Hubbard}, W.~B., {et~al.} 1997,
  {\mhref{http://doi.org/10.1086/305002}{\apj}},
  {\href{https://ui.adsabs.harvard.edu/abs/1997ApJ...491..856B}{491}}{\href{https://ui.adsabs.harvard.edu/abs/1997ApJ...491..856B}{,
  856}}

\bibitem[{T.~L. {Campante} {et~al.}(2014){Campante}, {Chaplin}, {Lund},
  {Huber}, {Hekker}, {Garc{\'\i}a}, {Corsaro}, {Handberg}, {Miglio},
  {Arentoft}, {Basu}, {Bedding}, {Christensen-Dalsgaard}, {Davies}, {Elsworth},
  {Gilliland}, {Karoff}, {Kawaler}, {Kjeldsen}, {Lundkvist}, {Metcalfe}, {Silva
  Aguirre}, \& {Stello}}]{Campante2014}
{Campante}, T.~L., {Chaplin}, W.~J., {Lund}, M.~N., {et~al.} 2014,
  {\mhref{http://doi.org/10.1088/0004-637X/783/2/123}{\apj}},
  {\href{https://ui.adsabs.harvard.edu/abs/2014ApJ...783..123C}{783}}{\href{https://ui.adsabs.harvard.edu/abs/2014ApJ...783..123C}{,
  123}}

\bibitem[{T.~L. {Campante} {et~al.}(2024){Campante}, {Kjeldsen}, {Li}, {Lund},
  {Silva}, {Corsaro}, {Gomes da Silva}, {Martins}, {Adibekyan}, {Azevedo
  Silva}, {Bedding}, {Bossini}, {Buzasi}, {Chaplin}, {Costa}, {Cunha},
  {Cristo}, {Faria}, {Garc{\'\i}a}, {Huber}, {Lundkvist}, {Metcalfe},
  {Monteiro}, {Neitzel}, {Nielsen}, {Poretti}, {Santos}, \&
  {Sousa}}]{Campante2024}
{Campante}, T.~L., {Kjeldsen}, H., {Li}, Y., {et~al.} 2024,
  {\mhref{http://doi.org/10.1051/0004-6361/202449197}{\aap}},
  {\href{https://ui.adsabs.harvard.edu/abs/2024A&A...683L..16C}{683}}{\href{https://ui.adsabs.harvard.edu/abs/2024A&A...683L..16C}{,
  L16}}

\bibitem[{F.~{Castelli} \& R.~L. {Kurucz}(2003){Castelli} \&
  {Kurucz}}]{castelli2003}
{Castelli}, F., \& {Kurucz}, R.~L. 2003, in IAU Symposium, Vol. 210, Modelling
  of Stellar Atmospheres, ed. N.~{Piskunov}, W.~W. {Weiss}, \& D.~F. {Gray},
  {\href{https://ui.adsabs.harvard.edu/abs/2003IAUS..210P.A20C}{A20}}

\bibitem[{W.~{Ceva} {et~al.}(2025){Ceva}, {Matthews}, {Rickman},
  {S{'e}gransan}, {Vigan}, {Bowler}, {Forveille}, {Franson}, {Hagelberg}, \&
  {Udry}}]{2025A&A...701A..78C}
{Ceva}, W., {Matthews}, E.~C., {Rickman}, E.~L., {et~al.} 2025,
  {\mhref{http://doi.org/10.1051/0004-6361/202555310}{\aap}},
  {\href{https://ui.adsabs.harvard.edu/abs/2025A&A...701A..78C}{701}}{\href{https://ui.adsabs.harvard.edu/abs/2025A&A...701A..78C}{,
  A78}}

\bibitem[{G.~{Chabrier} \& I.~{Baraffe}(1997){Chabrier} \&
  {Baraffe}}]{Chabrier1997}
{Chabrier}, G., \& {Baraffe}, I. 1997,
  {\mhref{http://doi.org/10.48550/arXiv.astro-ph/9704118}{\aap}},
  {\href{https://ui.adsabs.harvard.edu/abs/1997A&A...327.1039C}{327}}{\href{https://ui.adsabs.harvard.edu/abs/1997A&A...327.1039C}{,
  1039}}

\bibitem[{G.~{Chabrier} \& I.~{Baraffe}(2000){Chabrier} \&
  {Baraffe}}]{Chabrier2000}
---. 2000, {\mhref{http://doi.org/10.1146/annurev.astro.38.1.337}{\araa}},
  {\href{https://ui.adsabs.harvard.edu/abs/2000ARA&A..38..337C}{38}}{\href{https://ui.adsabs.harvard.edu/abs/2000ARA&A..38..337C}{,
  337}}

\bibitem[{G.~{Chabrier} {et~al.}(2023){Chabrier}, {Baraffe}, {Phillips}, \&
  {Debras}}]{Chabrier2023}
{Chabrier}, G., {Baraffe}, I., {Phillips}, M., \& {Debras}, F. 2023,
  {\mhref{http://doi.org/10.1051/0004-6361/202243832}{\aap}},
  {\href{https://ui.adsabs.harvard.edu/abs/2023A&A...671A.119C}{671}}{\href{https://ui.adsabs.harvard.edu/abs/2023A&A...671A.119C}{,
  A119}}

\bibitem[{W.~J. {Chaplin} {et~al.}(2009){Chaplin}, {Houdek}, {Karoff},
  {Elsworth}, \& {New}}]{Chaplin2009}
{Chaplin}, W.~J., {Houdek}, G., {Karoff}, C., {Elsworth}, Y., \& {New}, R.
  2009, {\mhref{http://doi.org/10.1051/0004-6361/200911952}{\aap}},
  {\href{https://ui.adsabs.harvard.edu/abs/2009A&A...500L..21C}{500}}{\href{https://ui.adsabs.harvard.edu/abs/2009A&A...500L..21C}{,
  L21}}

\bibitem[{W.~J. {Chaplin} {et~al.}(2011){Chaplin}, {Bedding}, {Bonanno},
  {Broomhall}, {Garc{\'\i}a}, {Hekker}, {Huber}, {Verner}, {Basu}, {Elsworth},
  {Houdek}, {Mathur}, {Mosser}, {New}, {Stevens}, {Appourchaux}, {Karoff},
  {Metcalfe}, {Molenda-{\.Z}akowicz}, {Monteiro}, {Thompson},
  {Christensen-Dalsgaard}, {Gilliland}, {Kawaler}, {Kjeldsen}, {Ballot},
  {Benomar}, {Corsaro}, {Campante}, {Gaulme}, {Hale}, {Handberg}, {Jarvis},
  {R{\'e}gulo}, {Roxburgh}, {Salabert}, {Stello}, {Mullally}, {Li}, \&
  {Wohler}}]{Chaplin2011-B}
{Chaplin}, W.~J., {Bedding}, T.~R., {Bonanno}, A., {et~al.} 2011,
  {\mhref{http://doi.org/10.1088/2041-8205/732/1/L5}{\apjl}},
  {\href{https://ui.adsabs.harvard.edu/abs/2011ApJ...732L...5C}{732}}{\href{https://ui.adsabs.harvard.edu/abs/2011ApJ...732L...5C}{,
  L5}}

\bibitem[{A.~{Cheetham} {et~al.}(2018){Cheetham}, {S{'e}gransan}, {Peretti},
  {Delisle}, {Hagelberg}, {Beuzit}, {Forveille}, {Marmier}, {Udry}, \&
  {Wildi}}]{2018A&A...614A..16C}
{Cheetham}, A., {S{'e}gransan}, D., {Peretti}, S., {et~al.} 2018,
  {\mhref{http://doi.org/10.1051/0004-6361/201630136}{\aap}},
  {\href{https://ui.adsabs.harvard.edu/abs/2018A&A...614A..16C}{614}}{\href{https://ui.adsabs.harvard.edu/abs/2018A&A...614A..16C}{,
  A16}}

\bibitem[{M.~{Chen} {et~al.}(2022{\natexlab{a}}){Chen}, {Li}, {Brandt},
  {Dupuy}, {Cardoso}, \& {McCaughrean}}]{2022AJ....163..288C}
{Chen}, M., {Li}, Y., {Brandt}, T.~D., {et~al.} 2022{\natexlab{a}},
  {\mhref{http://doi.org/10.3847/1538-3881/ac66d2}{\aj}},
  {\href{https://ui.adsabs.harvard.edu/abs/2022AJ....163..288C}{163}}{\href{https://ui.adsabs.harvard.edu/abs/2022AJ....163..288C}{,
  288}}

\bibitem[{M.~{Chen} {et~al.}(2022{\natexlab{b}}){Chen}, {Li}, {Brandt},
  {Dupuy}, {Cardoso}, \& {McCaughrean}}]{Chenmh2022}
---. 2022{\natexlab{b}},
  {\mhref{http://doi.org/10.3847/1538-3881/ac66d2}{\aj}},
  {\href{https://ui.adsabs.harvard.edu/abs/2022AJ....163..288C}{163}}{\href{https://ui.adsabs.harvard.edu/abs/2022AJ....163..288C}{,
  288}}

\bibitem[{J.~{Chilcote} {et~al.}(2017){Chilcote}, {Pueyo}, {De Rosa}, {Vargas},
  {Macintosh}, {Bailey}, {Barman}, {Bauman}, {Bruzzone}, {Bulger}, {Burrows},
  {Cardwell}, {Chen}, {Cotten}, {Dillon}, {Doyon}, {Draper}, {Duch{\^e}ne},
  {Dunn}, {Erikson}, {Fitzgerald}, {Follette}, {Gavel}, {Goodsell}, {Graham},
  {Greenbaum}, {Hartung}, {Hibon}, {Hung}, {Ingraham}, {Kalas}, {Konopacky},
  {Larkin}, {Maire}, {Marchis}, {Marley}, {Marois}, {Metchev},
  {Millar-Blanchaer}, {Morzinski}, {Nielsen}, {Norton}, {Oppenheimer},
  {Palmer}, {Patience}, {Perrin}, {Poyneer}, {Rajan}, {Rameau},
  {Rantakyr{\"o}}, {Sadakuni}, {Saddlemyer}, {Savransky}, {Schneider}, {Serio},
  {Sivaramakrishnan}, {Song}, {Soummer}, {Thomas}, {Wallace}, {Wang},
  {Ward-Duong}, {Wiktorowicz}, \& {Wolff}}]{2017AJ....153..182C}
{Chilcote}, J., {Pueyo}, L., {De Rosa}, R.~J., {et~al.} 2017,
  {\mhref{http://doi.org/10.3847/1538-3881/aa63e9}{\aj}},
  {\href{http://adsabs.harvard.edu/abs/2017AJ....153..182C}{153}}{\href{http://adsabs.harvard.edu/abs/2017AJ....153..182C}{,
  182}}

\bibitem[{F.~{Chiti} {et~al.}(2024){Chiti}, {van Saders}, {Heintz}, {Hermes},
  {Ong}, {Hey}, {Ramirez-Weinhouse}, \& {Dugas}}]{chiti2024-wd}
{Chiti}, F., {van Saders}, J.~L., {Heintz}, T.~M., {et~al.} 2024,
  {\mhref{http://doi.org/10.48550/arXiv.2403.12129}{arXiv
  e-prints}}{\href{https://ui.adsabs.harvard.edu/abs/2024arXiv240312129C}{,
  arXiv:2403.12129}}

\bibitem[{J.~{Christensen-Dalsgaard}(1984){Christensen-Dalsgaard}}]{jcd1984-02}
{Christensen-Dalsgaard}, J. 1984, in Space Research in Stellar Activity and
  Variability, ed. A.~{Mangeney} \& F.~{Praderie},
  {\href{https://ui.adsabs.harvard.edu/abs/1984srps.conf...11C}{11}}

\bibitem[{J.~{Christensen-Dalsgaard}(2008{\natexlab{a}}){Christensen-Dalsgaard}}]{ChristensenDalsgaard2008a}
{Christensen-Dalsgaard}, J. 2008{\natexlab{a}},
  {\mhref{http://doi.org/10.1007/s10509-007-9675-5}{\apss}},
  {\href{https://ui.adsabs.harvard.edu/abs/2008Ap&SS.316...13C}{316}}{\href{https://ui.adsabs.harvard.edu/abs/2008Ap&SS.316...13C}{,
  13}}

\bibitem[{J.~{Christensen-Dalsgaard}(2008{\natexlab{b}}){Christensen-Dalsgaard}}]{adipls2008}
---. 2008{\natexlab{b}},
  {\mhref{http://doi.org/10.1007/s10509-007-9689-z}{\apss}},
  {\href{https://ui.adsabs.harvard.edu/abs/2008Ap&SS.316..113C}{316}}{\href{https://ui.adsabs.harvard.edu/abs/2008Ap&SS.316..113C}{,
  113}}

\bibitem[{J.~P. {Cox} \& R.~T. {Giuli}(1968){Cox} \& {Giuli}}]{cox1968}
{Cox}, J.~P., \& {Giuli}, R.~T. 1968, {Principles of stellar structure}

\bibitem[{J.~R. {Crepp} {et~al.}(2016){Crepp}, {Gonzales}, {Bechter}, {Montet},
  {Johnson}, {Piskorz}, {Howard}, \& {Isaacson}}]{2016ApJ...831..136C}
{Crepp}, J.~R., {Gonzales}, E.~J., {Bechter}, E.~B., {et~al.} 2016,
  {\mhref{http://doi.org/10.3847/0004-637X/831/2/136}{\apj}},
  {\href{https://ui.adsabs.harvard.edu/abs/2016ApJ...831..136C}{831}}{\href{https://ui.adsabs.harvard.edu/abs/2016ApJ...831..136C}{,
  136}}

\bibitem[{J.~R. {Crepp} {et~al.}(2012){Crepp}, {Johnson}, {Fischer}, {Howard},
  {Marcy}, {Wright}, {Isaacson}, {Boyajian}, {von Braun}, {Hillenbrand},
  {Hinkley}, {Carpenter}, \& {Brewer}}]{Crepp2012}
{Crepp}, J.~R., {Johnson}, J.~A., {Fischer}, D.~A., {et~al.} 2012,
  {\mhref{http://doi.org/10.1088/0004-637X/751/2/97}{\apj}},
  {\href{https://ui.adsabs.harvard.edu/abs/2012ApJ...751...97C}{751}}{\href{https://ui.adsabs.harvard.edu/abs/2012ApJ...751...97C}{,
  97}}

\bibitem[{J.~R. {Crepp} {et~al.}(2015){Crepp}, {Rice}, {Veicht}, {Aguilar},
  {Pueyo}, {Giorla}, {Nilsson}, {Luszcz-Cook}, {Oppenheimer}, {Hinkley},
  {Brenner}, {Vasisht}, {Cady}, {Beichman}, {Hillenbrand}, {Lockhart},
  {Matthews}, {Roberts}, {Sivaramakrishnan}, {Soummer}, \&
  {Zhai}}]{2015ApJ...798L..43C}
{Crepp}, J.~R., {Rice}, E.~L., {Veicht}, A., {et~al.} 2015,
  {\mhref{http://doi.org/10.1088/2041-8205/798/2/L43}{\apjl}},
  {\href{https://ui.adsabs.harvard.edu/abs/2015ApJ...798L..43C}{798}}{\href{https://ui.adsabs.harvard.edu/abs/2015ApJ...798L..43C}{,
  L43}}

\bibitem[{T.~{Currie} {et~al.}(2020){Currie}, {Brandt}, {Kuzuhara}, {Chilcote},
  {Guyon}, {Marois}, {Groff}, {Lozi}, {Vievard}, {Sahoo}, {Deo}, {Jovanovic},
  {Martinache}, {Wagner}, {Dupuy}, {Wahl}, {Letawsky}, {Li}, {Zeng}, {Brandt},
  {Michalik}, {Grady}, {Janson}, {Knapp}, {Kwon}, {Lawson}, {McElwain},
  {Uyama}, {Wisniewski}, \& {Tamura}}]{2020ApJ...904L..25C}
{Currie}, T., {Brandt}, T.~D., {Kuzuhara}, M., {et~al.} 2020,
  {\mhref{http://doi.org/10.3847/2041-8213/abc631}{\apjl}},
  {\href{https://ui.adsabs.harvard.edu/abs/2020ApJ...904L..25C}{904}}{\href{https://ui.adsabs.harvard.edu/abs/2020ApJ...904L..25C}{,
  L25}}

\bibitem[{T.~{Currie} {et~al.}(2023){Currie}, {Brandt}, {Brandt}, {Lacy},
  {Burrows}, {Guyon}, {Tamura}, {Liu}, {Sagynbayeva}, {Tobin}, {Chilcote},
  {Groff}, {Marois}, {Thompson}, {Murphy}, {Kuzuhara}, {Lawson}, {Lozi}, {Deo},
  {Vievard}, {Skaf}, {Uyama}, {Jovanovic}, {Martinache}, {Kasdin}, {Kudo},
  {McElwain}, {Janson}, {Wisniewski}, {Hodapp}, {Nishikawa}, {He{\l}miniak},
  {Kwon}, \& {Hayashi}}]{2023Sci...380..198C}
{Currie}, T., {Brandt}, G.~M., {Brandt}, T.~D., {et~al.} 2023,
  {\mhref{http://doi.org/10.1126/science.abo6192}{Science}},
  {\href{https://ui.adsabs.harvard.edu/abs/2023Sci...380..198C}{380}}{\href{https://ui.adsabs.harvard.edu/abs/2023Sci...380..198C}{,
  198}}

\bibitem[{T.~{Currie} {et~al.}(2025){Currie}, {Li}, {El Morsy}, {Lacy},
  {Vincent}, {Tobin}, {Kuzuhara}, {Chilcote}, {Guyon}, {Gu}, {Bovie}, {Peng},
  {An}, {Brandt}, {De Rosa}, {Deo}, {Groff}, {Janson}, {Kasdin}, {Lozi},
  {Marois}, {Mennesson}, {Murakami}, {Nielsen}, {Sagynbayeva}, {Skaf},
  {Thompson}, {Tamura}, {Uyama}, {Vievard}, \& {Zurlo}}]{2025arXiv251202159C}
{Currie}, T., {Li}, Y., {El Morsy}, M., {et~al.} 2025,
  {\mhref{http://doi.org/10.48550/arXiv.2512.02159}{arXiv
  e-prints}}{\href{https://ui.adsabs.harvard.edu/abs/2025arXiv251202159C}{,
  arXiv:2512.02159}}

\bibitem[{J.~L. {Curtis} {et~al.}(2019){Curtis}, {Ag{\"u}eros}, {Douglas}, \&
  {Meibom}}]{curtis2019-6811}
{Curtis}, J.~L., {Ag{\"u}eros}, M.~A., {Douglas}, S.~T., \& {Meibom}, S. 2019,
  {\mhref{http://doi.org/10.3847/1538-4357/ab2393}{\apj}},
  {\href{https://ui.adsabs.harvard.edu/abs/2019ApJ...879...49C}{879}}{\href{https://ui.adsabs.harvard.edu/abs/2019ApJ...879...49C}{,
  49}}

\bibitem[{J.~L. {Curtis} {et~al.}(2020){Curtis}, {Ag{\"u}eros}, {Matt},
  {Covey}, {Douglas}, {Angus}, {Saar}, {Cody}, {Vanderburg}, {Law}, {Kraus},
  {Latham}, {Baranec}, {Riddle}, {Ziegler}, {Lund}, {Torres}, {Meibom},
  {Aguirre}, \& {Wright}}]{curtis2020-147}
{Curtis}, J.~L., {Ag{\"u}eros}, M.~A., {Matt}, S.~P., {et~al.} 2020,
  {\mhref{http://doi.org/10.3847/1538-4357/abbf58}{\apj}},
  {\href{https://ui.adsabs.harvard.edu/abs/2020ApJ...904..140C}{904}}{\href{https://ui.adsabs.harvard.edu/abs/2020ApJ...904..140C}{,
  140}}

\bibitem[{R.~M. {Cutri} {et~al.}(2003){Cutri}, {Skrutskie}, {van Dyk},
  {Beichman}, {Carpenter}, {Chester}, {Cambresy}, {Evans}, {Fowler}, {Gizis},
  {Howard}, {Huchra}, {Jarrett}, {Kopan}, {Kirkpatrick}, {Light}, {Marsh},
  {McCallon}, {Schneider}, {Stiening}, {Sykes}, {Weinberg}, {Wheaton},
  {Wheelock}, \& {Zacarias}}]{Cutri2003}
{Cutri}, R.~M., {Skrutskie}, M.~F., {van Dyk}, S., {et~al.} 2003, {2MASS All
  Sky Catalog of point sources.}

\bibitem[{R.~H. {Cyburt} {et~al.}(2010){Cyburt}, {Amthor}, {Ferguson},
  {Meisel}, {Smith}, {Warren}, {Heger}, {Hoffman}, {Rauscher}, {Sakharuk},
  {Schatz}, {Thielemann}, \& {Wiescher}}]{jinareaclib2010}
{Cyburt}, R.~H., {Amthor}, A.~M., {Ferguson}, R., {et~al.} 2010,
  {\mhref{http://doi.org/10.1088/0067-0049/189/1/240}{\apjs}},
  {\href{https://ui.adsabs.harvard.edu/abs/2010ApJS..189..240C}{189}}{\href{https://ui.adsabs.harvard.edu/abs/2010ApJS..189..240C}{,
  240}}

\bibitem[{G.~R. {Davies} {et~al.}(2014){Davies}, {Handberg}, {Miglio},
  {Campante}, {Chaplin}, \& {Elsworth}}]{davies2014-los}
{Davies}, G.~R., {Handberg}, R., {Miglio}, A., {et~al.} 2014,
  {\mhref{http://doi.org/10.1093/mnrasl/slu143}{\mnras}},
  {\href{https://ui.adsabs.harvard.edu/abs/2014MNRAS.445L..94D}{445}}{\href{https://ui.adsabs.harvard.edu/abs/2014MNRAS.445L..94D}{,
  L94}}

\bibitem[{R.~J. {De Rosa} {et~al.}(2023){De Rosa}, {Nielsen}, {Wahhaj},
  {Ruffio}, {Kalas}, {Peck}, {Hirsch}, \& {Roberson}}]{2023A&A...672A..94D}
{De Rosa}, R.~J., {Nielsen}, E.~L., {Wahhaj}, Z., {et~al.} 2023,
  {\mhref{http://doi.org/10.1051/0004-6361/202345877}{\aap}},
  {\href{https://ui.adsabs.harvard.edu/abs/2023A&A...672A..94D}{672}}{\href{https://ui.adsabs.harvard.edu/abs/2023A&A...672A..94D}{,
  A94}}

\bibitem[{N.~R. {Deacon} {et~al.}(2014){Deacon}, {Liu}, {Magnier}, {Aller},
  {Best}, {Dupuy}, {Bowler}, {Mann}, {Redstone}, {Burgett}, {Chambers},
  {Draper}, {Flewelling}, {Hodapp}, {Kaiser}, {Kudritzki}, {Morgan},
  {Metcalfe}, {Price}, {Tonry}, \& {Wainscoat}}]{2014ApJ...792..119D}
{Deacon}, N.~R., {Liu}, M.~C., {Magnier}, E.~A., {et~al.} 2014,
  {\mhref{http://doi.org/10.1088/0004-637X/792/2/119}{\apj}},
  {\href{https://ui.adsabs.harvard.edu/abs/2014ApJ...792..119D}{792}}{\href{https://ui.adsabs.harvard.edu/abs/2014ApJ...792..119D}{,
  119}}

\bibitem[{P.~{Delorme} {et~al.}(2017){Delorme}, {Schmidt}, {Bonnefoy},
  {Desidera}, {Ginski}, {Charnay}, {Lazzoni}, {Christiaens}, {Messina},
  {D'Orazi}, {Milli}, {Schlieder}, {Gratton}, {Rodet}, {Lagrange}, {Absil},
  {Vigan}, {Galicher}, {Hagelberg}, {Bonavita}, {Lavie}, {Zurlo}, {Olofsson},
  {Boccaletti}, {Cantalloube}, {Mouillet}, {Chauvin}, {Hambsch}, {Langlois},
  {Udry}, {Henning}, {Beuzit}, {Mordasini}, {Lucas}, {Marocco}, {Biller},
  {Carson}, {Cheetham}, {Covino}, {De Caprio}, {Delboulbe}, {Feldt}, {Girard},
  {Hubin}, {Maire}, {Pavlov}, {Petit}, {Rouan}, {Roelfsema}, \&
  {Wildi}}]{2017A&A...608A..79D}
{Delorme}, P., {Schmidt}, T., {Bonnefoy}, M., {et~al.} 2017,
  {\mhref{http://doi.org/10.1051/0004-6361/201731145}{\aap}},
  {\href{https://ui.adsabs.harvard.edu/abs/2017A&A...608A..79D}{608}}{\href{https://ui.adsabs.harvard.edu/abs/2017A&A...608A..79D}{,
  A79}}

\bibitem[{P.~{Demarque} {et~al.}(2008){Demarque}, {Guenther}, {Li}, {Mazumdar},
  \& {Straka}}]{yrec2008}
{Demarque}, P., {Guenther}, D.~B., {Li}, L.~H., {Mazumdar}, A., \& {Straka},
  C.~W. 2008, {\mhref{http://doi.org/10.1007/s10509-007-9698-y}{\apss}},
  {\href{https://ui.adsabs.harvard.edu/abs/2008Ap&SS.316...31D}{316}}{\href{https://ui.adsabs.harvard.edu/abs/2008Ap&SS.316...31D}{,
  31}}

\bibitem[{R.~A. {Donahue} {et~al.}(1996){Donahue}, {Saar}, \&
  {Baliunas}}]{Donahue1996}
{Donahue}, R.~A., {Saar}, S.~H., \& {Baliunas}, S.~L. 1996,
  {\mhref{http://doi.org/10.1086/177517}{\apj}},
  {\href{https://ui.adsabs.harvard.edu/abs/1996ApJ...466..384D}{466}}{\href{https://ui.adsabs.harvard.edu/abs/1996ApJ...466..384D}{,
  384}}

\bibitem[{S.~T. {Douglas} {et~al.}(2017){Douglas}, {Ag{\"u}eros}, {Covey}, \&
  {Kraus}}]{douglas2017-praesepe}
{Douglas}, S.~T., {Ag{\"u}eros}, M.~A., {Covey}, K.~R., \& {Kraus}, A. 2017,
  {\mhref{http://doi.org/10.3847/1538-4357/aa6e52}{\apj}},
  {\href{https://ui.adsabs.harvard.edu/abs/2017ApJ...842...83D}{842}}{\href{https://ui.adsabs.harvard.edu/abs/2017ApJ...842...83D}{,
  83}}

\bibitem[{S.~T. {Douglas} {et~al.}(2019){Douglas}, {Curtis}, {Ag{\"u}eros},
  {Cargile}, {Brewer}, {Meibom}, \& {Jansen}}]{douglas2019-praesepe}
{Douglas}, S.~T., {Curtis}, J.~L., {Ag{\"u}eros}, M.~A., {et~al.} 2019,
  {\mhref{http://doi.org/10.3847/1538-4357/ab2468}{\apj}},
  {\href{https://ui.adsabs.harvard.edu/abs/2019ApJ...879..100D}{879}}{\href{https://ui.adsabs.harvard.edu/abs/2019ApJ...879..100D}{,
  100}}

\bibitem[{R.~{Dungee} {et~al.}(2022){Dungee}, {van Saders}, {Gaidos}, {Chun},
  {Garc{\'\i}a}, {Magnier}, {Mathur}, \& {Santos}}]{dungee2022-m67}
{Dungee}, R., {van Saders}, J., {Gaidos}, E., {et~al.} 2022,
  {\mhref{http://doi.org/10.3847/1538-4357/ac90be}{\apj}},
  {\href{https://ui.adsabs.harvard.edu/abs/2022ApJ...938..118D}{938}}{\href{https://ui.adsabs.harvard.edu/abs/2022ApJ...938..118D}{,
  118}}

\bibitem[{T.~J. {Dupuy} {et~al.}(2016){Dupuy}, {Kratter}, {Kraus}, {Isaacson},
  {Mann}, {Ireland}, {Howard}, \& {Huber}}]{Dupuy2016}
{Dupuy}, T.~J., {Kratter}, K.~M., {Kraus}, A.~L., {et~al.} 2016,
  {\mhref{http://doi.org/10.3847/0004-637X/817/1/80}{\apj}},
  {\href{https://ui.adsabs.harvard.edu/abs/2016ApJ...817...80D}{817}}{\href{https://ui.adsabs.harvard.edu/abs/2016ApJ...817...80D}{,
  80}}

\bibitem[{T.~J. {Dupuy} \& A.~L. {Kraus}(2013){Dupuy} \&
  {Kraus}}]{2013Sci...341.1492D}
{Dupuy}, T.~J., \& {Kraus}, A.~L. 2013,
  {\mhref{http://doi.org/10.1126/science.1241917}{Science}},
  {\href{http://adsabs.harvard.edu/abs/2013Sci...341.1492D}{341}}{\href{http://adsabs.harvard.edu/abs/2013Sci...341.1492D}{,
  1492}}

\bibitem[{T.~J. Dupuy \& M.~C. {Liu}(2012)Dupuy \& {Liu}}]{2012ApJS..201...19D}
Dupuy, T.~J., \& {Liu}, M.~C. 2012,
  {\mhref{http://doi.org/10.1088/0067-0049/201/2/19}{\apjs}},
  {\href{http://adsabs.harvard.edu/abs/2012ApJS..201...19D}{201}}{\href{http://adsabs.harvard.edu/abs/2012ApJS..201...19D}{,
  19}}

\bibitem[{T.~J. {Dupuy} \& M.~C. {Liu}(2017{\natexlab{a}}){Dupuy} \&
  {Liu}}]{Dupuy2017}
{Dupuy}, T.~J., \& {Liu}, M.~C. 2017{\natexlab{a}},
  {\mhref{http://doi.org/10.3847/1538-4365/aa5e4c}{\apjs}},
  {\href{https://ui.adsabs.harvard.edu/abs/2017ApJS..231...15D}{231}}{\href{https://ui.adsabs.harvard.edu/abs/2017ApJS..231...15D}{,
  15}}

\bibitem[{T.~J. {Dupuy} \& M.~C. {Liu}(2017{\natexlab{b}}){Dupuy} \&
  {Liu}}]{2017ApJS..231...15D}
---. 2017{\natexlab{b}},
  {\mhref{http://doi.org/10.3847/1538-4365/aa5e4c}{\apjs}},
  {\href{http://adsabs.harvard.edu/abs/2017ApJS..231...15D}{231}}{\href{http://adsabs.harvard.edu/abs/2017ApJS..231...15D}{,
  15}}

\bibitem[{T.~J. {Dupuy} {et~al.}(2023){Dupuy}, {Liu}, {Evans}, {Best},
  {Pearce}, {Sanghi}, {Phillips}, \& {Bardalez Gagliuffi}}]{Dupuy2023}
{Dupuy}, T.~J., {Liu}, M.~C., {Evans}, E.~L., {et~al.} 2023,
  {\mhref{http://doi.org/10.1093/mnras/stac3557}{\mnras}},
  {\href{https://ui.adsabs.harvard.edu/abs/2023MNRAS.519.1688D}{519}}{\href{https://ui.adsabs.harvard.edu/abs/2023MNRAS.519.1688D}{,
  1688}}

\bibitem[{T.~J. {Dupuy} {et~al.}(2009){Dupuy}, {Liu}, \&
  {Ireland}}]{2009ApJ...692..729D}
{Dupuy}, T.~J., {Liu}, M.~C., \& {Ireland}, M.~J. 2009,
  {\mhref{http://doi.org/10.1088/0004-637X/692/1/729}{\apj}},
  {\href{http://adsabs.harvard.edu/abs/2009ApJ...692..729D}{692}}{\href{http://adsabs.harvard.edu/abs/2009ApJ...692..729D}{,
  729}}

\bibitem[{T.~J. Dupuy {et~al.}(2014)Dupuy, {Liu}, \&
  {Ireland}}]{2014ApJ...790..133D}
Dupuy, T.~J., {Liu}, M.~C., \& {Ireland}, M.~J. 2014,
  {\mhref{http://doi.org/10.1088/0004-637X/790/2/133}{\apj}},
  {\href{http://adsabs.harvard.edu/abs/2014ApJ...790..133D}{790}}{\href{http://adsabs.harvard.edu/abs/2014ApJ...790..133D}{,
  133}}

\bibitem[{T.~J. Dupuy {et~al.}(2015)Dupuy, {Liu}, {Leggett}, {Ireland}, {Chiu},
  \& {Golimowski}}]{2015ApJ...805...56D}
Dupuy, T.~J., {Liu}, M.~C., {Leggett}, S.~K., {et~al.} 2015,
  {\mhref{http://doi.org/10.1088/0004-637X/805/1/56}{\apj}},
  {\href{http://adsabs.harvard.edu/abs/2015ApJ...805...56D}{805}}{\href{http://adsabs.harvard.edu/abs/2015ApJ...805...56D}{,
  56}}

\bibitem[{A.~S. {Eddington}(1926){Eddington}}]{eddington1926}
{Eddington}, A.~S. 1926, {The Internal Constitution of the Stars}

\bibitem[{R.~{Egeland} {et~al.}(2017){Egeland}, {Soon}, {Baliunas}, {Hall},
  {Pevtsov}, \& {Bertello}}]{Egeland2017}
{Egeland}, R., {Soon}, W., {Baliunas}, S., {et~al.} 2017,
  {\mhref{http://doi.org/10.3847/1538-4357/835/1/25}{\apj}},
  {\href{https://ui.adsabs.harvard.edu/abs/2017ApJ...835...25E}{835}}{\href{https://ui.adsabs.harvard.edu/abs/2017ApJ...835...25E}{,
  25}}

\bibitem[{M.~{El Morsy} {et~al.}(2025){El Morsy}, {Currie}, {Bovie},
  {Kuzuhara}, {Lacy}, {Li}, {Tobin}, {Brandt}, {Chilcote}, {Guyon}, {Groff},
  {Lozi}, {Vievard}, {Deo}, {Skaf}, {Bouchy}, {Boisse}, {Dykes}, {Kasdin}, \&
  {Tamura}}]{2025ApJ...981...20E}
{El Morsy}, M., {Currie}, T., {Bovie}, D., {et~al.} 2025,
  {\mhref{http://doi.org/10.3847/1538-4357/ada3be}{\apj}},
  {\href{https://ui.adsabs.harvard.edu/abs/2025ApJ...981...20E}{981}}{\href{https://ui.adsabs.harvard.edu/abs/2025ApJ...981...20E}{,
  20}}

\bibitem[{C.~R. {Epstein} \& M.~H. {Pinsonneault}(2014){Epstein} \&
  {Pinsonneault}}]{Epstein2014}
{Epstein}, C.~R., \& {Pinsonneault}, M.~H. 2014,
  {\mhref{http://doi.org/10.1088/0004-637X/780/2/159}{\apj}},
  {\href{https://ui.adsabs.harvard.edu/abs/2014ApJ...780..159E}{780}}{\href{https://ui.adsabs.harvard.edu/abs/2014ApJ...780..159E}{,
  159}}

\bibitem[{{ESA}(1997)}]{ESA1997}
{ESA}, ed. 1997, ESA Special Publication, Vol. 1200, {The HIPPARCOS and TYCHO
  catalogues. Astrometric and photometric star catalogues derived from the ESA
  HIPPARCOS Space Astrometry Mission}

\bibitem[{F.~{Feng} {et~al.}(2021){Feng}, {Butler}, {Jones}, {Phillips},
  {Vogt}, {Oppenheimer}, {Holden}, {Burt}, \& {Boss}}]{Feng2021}
{Feng}, F., {Butler}, R.~P., {Jones}, H. R.~A., {et~al.} 2021,
  {\mhref{http://doi.org/10.1093/mnras/stab2225}{\mnras}},
  {\href{https://ui.adsabs.harvard.edu/abs/2021MNRAS.507.2856F}{507}}{\href{https://ui.adsabs.harvard.edu/abs/2021MNRAS.507.2856F}{,
  2856}}

\bibitem[{D.~A. {Fischer} {et~al.}(2014){Fischer}, {Marcy}, \&
  {Spronck}}]{Fischer2014}
{Fischer}, D.~A., {Marcy}, G.~W., \& {Spronck}, J. F.~P. 2014,
  {\mhref{http://doi.org/10.1088/0067-0049/210/1/5}{\apjs}},
  {\href{https://ui.adsabs.harvard.edu/abs/2014ApJS..210....5F}{210}}{\href{https://ui.adsabs.harvard.edu/abs/2014ApJS..210....5F}{,
  5}}

\bibitem[{D.~A. {Fischer} {et~al.}(2016){Fischer}, {Anglada-Escude},
  {Arriagada}, {Baluev}, {Bean}, {Bouchy}, {Buchhave}, {Carroll},
  {Chakraborty}, {Crepp}, {Dawson}, {Diddams}, {Dumusque}, {Eastman}, {Endl},
  {Figueira}, {Ford}, {Foreman-Mackey}, {Fournier}, {F{\H{u}}r{\'e}sz},
  {Gaudi}, {Gregory}, {Grundahl}, {Hatzes}, {H{\'e}brard}, {Herrero}, {Hogg},
  {Howard}, {Johnson}, {Jorden}, {Jurgenson}, {Latham}, {Laughlin}, {Loredo},
  {Lovis}, {Mahadevan}, {McCracken}, {Pepe}, {Perez}, {Phillips}, {Plavchan},
  {Prato}, {Quirrenbach}, {Reiners}, {Robertson}, {Santos}, {Sawyer},
  {Segransan}, {Sozzetti}, {Steinmetz}, {Szentgyorgyi}, {Udry}, {Valenti},
  {Wang}, {Wittenmyer}, \& {Wright}}]{Fischer2016}
{Fischer}, D.~A., {Anglada-Escude}, G., {Arriagada}, P., {et~al.} 2016,
  {\mhref{http://doi.org/10.1088/1538-3873/128/964/066001}{\pasp}},
  {\href{https://ui.adsabs.harvard.edu/abs/2016PASP..128f6001F}{128}}{\href{https://ui.adsabs.harvard.edu/abs/2016PASP..128f6001F}{,
  066001}}

\bibitem[{D.~{Foreman-Mackey}(2018){Foreman-Mackey}}]{celerite2}
{Foreman-Mackey}, D. 2018,
  {\mhref{http://doi.org/10.3847/2515-5172/aaaf6c}{Research Notes of the
  American Astronomical Society}},
  {\href{http://adsabs.harvard.edu/abs/2018RNAAS...2a..31F}{2}}{\href{http://adsabs.harvard.edu/abs/2018RNAAS...2a..31F}{,
  31}}

\bibitem[{A.~{Formicola} {et~al.}(2004){Formicola}, {Imbriani}, {Costantini},
  {Angulo}, {Bemmerer}, {Bonetti}, {Broggini}, {Corvisiero}, {Cruz},
  {Descouvemont}, {F{\"u}l{\"o}p}, {Gervino}, {Guglielmetti}, {Gustavino},
  {Gy{\"u}rky}, {Jesus}, {Junker}, {Lemut}, {Menegazzo}, {Prati}, {Roca},
  {Rolfs}, {Romano}, {Rossi Alvarez}, {Sch{\"u}mann}, {Somorjai}, {Straniero},
  {Strieder}, {Terrasi}, {Trautvetter}, {Vomiero}, \&
  {Zavatarelli}}]{formicola2004}
{Formicola}, A., {Imbriani}, G., {Costantini}, H., {et~al.} 2004,
  {\mhref{http://doi.org/10.1016/j.physletb.2004.03.092}{Physics Letters B}},
  {\href{https://ui.adsabs.harvard.edu/abs/2004PhLB..591...61F}{591}}{\href{https://ui.adsabs.harvard.edu/abs/2004PhLB..591...61F}{,
  61}}

\bibitem[{S.~{Frandsen} {et~al.}(1995){Frandsen}, {Jones}, {Kjeldsen},
  {Viskum}, {Hjorth}, {Andersen}, \& {Thomsen}}]{Frandsen1995}
{Frandsen}, S., {Jones}, A., {Kjeldsen}, H., {et~al.} 1995, \aap,
  {\href{https://ui.adsabs.harvard.edu/abs/1995A&A...301..123F}{301}}{\href{https://ui.adsabs.harvard.edu/abs/1995A&A...301..123F}{,
  123}}

\bibitem[{K.~{Franson} \& B.~P. {Bowler}(2023){Franson} \&
  {Bowler}}]{2023AJ....165..246F}
{Franson}, K., \& {Bowler}, B.~P. 2023,
  {\mhref{http://doi.org/10.3847/1538-3881/acca18}{\aj}},
  {\href{https://ui.adsabs.harvard.edu/abs/2023AJ....165..246F}{165}}{\href{https://ui.adsabs.harvard.edu/abs/2023AJ....165..246F}{,
  246}}

\bibitem[{K.~{Franson} {et~al.}(2022){Franson}, {Bowler}, {Brandt}, {Dupuy},
  {Tran}, {Brandt}, {Li}, \& {Kraus}}]{2022AJ....163...50F}
{Franson}, K., {Bowler}, B.~P., {Brandt}, T.~D., {et~al.} 2022,
  {\mhref{http://doi.org/10.3847/1538-3881/ac35e8}{\aj}},
  {\href{https://ui.adsabs.harvard.edu/abs/2022AJ....163...50F}{163}}{\href{https://ui.adsabs.harvard.edu/abs/2022AJ....163...50F}{,
  50}}

\bibitem[{K.~{Franson} {et~al.}(2023{\natexlab{a}}){Franson}, {Bowler}, {Zhou},
  {Pearce}, {Bardalez Gagliuffi}, {Biddle}, {Brandt}, {Crepp}, {Dupuy},
  {Faherty}, {Jensen-Clem}, {Morgan}, {Sanghi}, {Theissen}, {Tran}, \&
  {Wolf}}]{2023ApJ...950L..19F}
{Franson}, K., {Bowler}, B.~P., {Zhou}, Y., {et~al.} 2023{\natexlab{a}},
  {\mhref{http://doi.org/10.3847/2041-8213/acd6f6}{\apjl}},
  {\href{https://ui.adsabs.harvard.edu/abs/2023ApJ...950L..19F}{950}}{\href{https://ui.adsabs.harvard.edu/abs/2023ApJ...950L..19F}{,
  L19}}

\bibitem[{K.~{Franson} {et~al.}(2023{\natexlab{b}}){Franson}, {Bowler},
  {Bonavita}, {Brandt}, {Chen}, {Samland}, {Zhang}, {Lueber}, {Heng},
  {Kitzmann}, {Wolf}, {Jones}, {Tran}, {Bardalez Gagliuffi}, {Biller},
  {Chilcote}, {Crepp}, {Dupuy}, {Faherty}, {Fontanive}, {Groff}, {Gratton},
  {Guyon}, {Jensen-Clem}, {Jovanovic}, {Kasdin}, {Lozi}, {Magnier},
  {Mu{v{z}}i{'c}}, {Sanghi}, \& {Theissen}}]{2023AJ....165...39F}
{Franson}, K., {Bowler}, B.~P., {Bonavita}, M., {et~al.} 2023{\natexlab{b}},
  {\mhref{http://doi.org/10.3847/1538-3881/aca408}{\aj}},
  {\href{https://ui.adsabs.harvard.edu/abs/2023AJ....165...39F}{165}}{\href{https://ui.adsabs.harvard.edu/abs/2023AJ....165...39F}{,
  39}}

\bibitem[{{Gaia Collaboration}(2022)}]{GaiaCollaboration2022}
{Gaia Collaboration}. 2022, {VizieR Online Data Catalog: Gaia DR3 Part 1. Main
  source (Gaia Collaboration, 2022)}, VizieR On-line Data Catalog: I/355.
  Originally published in: doi:10.1051/0004-63,
  \dodoi{10.26093/cds/vizier.1355}

\bibitem[{{Gaia Collaboration} {et~al.}(2023){Gaia Collaboration}, {Vallenari},
  {Brown}, {Prusti}, {de Bruijne}, {Arenou}, {Babusiaux}, {Biermann},
  {Creevey}, {Ducourant}, \& et~al.}]{GaiaCollaboration2023}
{Gaia Collaboration}, {Vallenari}, A., {Brown}, A.~G.~A., {et~al.} 2023,
  {\mhref{http://doi.org/10.1051/0004-6361/202243940}{\aap}},
  {\href{https://ui.adsabs.harvard.edu/abs/2023A&A...674A...1G}{674}}{\href{https://ui.adsabs.harvard.edu/abs/2023A&A...674A...1G}{,
  A1}}

\bibitem[{S.~R. {Gibson}(2016){Gibson}}]{gibson2016-kpf}
{Gibson}, S.~R. 2016, in Society of Photo-Optical Instrumentation Engineers
  (SPIE) Conference Series, Vol. 9911, Modeling, Systems Engineering, and
  Project Management for Astronomy VI, ed. G.~Z. {Angeli} \& P.~{Dierickx},
  {\href{https://ui.adsabs.harvard.edu/abs/2016SPIE.9911E..2CG}{99112C}}

\bibitem[{S.~R. {Gibson} {et~al.}(2018){Gibson}, {Howard}, {Roy}, {Smith},
  {Halverson}, {Edelstein}, {Kassis}, {Wishnow}, {Raffanti}, {Allen}, {Chin},
  {Coutts}, {Cowley}, {Curtis}, {Deich}, {Feger}, {Finstad}, {Gurevich},
  {Ishikawa}, {James}, {Jhoti}, {Lanclos}, {Lilley}, {Miller}, {Milner},
  {Payne}, {Rider}, {Rockosi}, {Sandford}, {Schwab}, {Seifahrt}, {Sirk},
  {Smith}, {Stuermer}, {Weisfeiler}, {Wilcox}, {Vandenberg}, \&
  {Wizinowich}}]{gibson2018-kpf}
{Gibson}, S.~R., {Howard}, A.~W., {Roy}, A., {et~al.} 2018, in Society of
  Photo-Optical Instrumentation Engineers (SPIE) Conference Series, Vol. 10702,
  Ground-based and Airborne Instrumentation for Astronomy VII, ed. C.~J.
  {Evans}, L.~{Simard}, \& H.~{Takami},
  {\href{https://ui.adsabs.harvard.edu/abs/2018SPIE10702E..5XG}{107025X}}

\bibitem[{S.~R. {Gibson} {et~al.}(2020){Gibson}, {Howard}, {Rider}, {Roy},
  {Edelstein}, {Kassis}, {Grillo}, {Halverson}, {Sirk}, {Smith}, {Allen},
  {Baker}, {Beichman}, {Berriman}, {Brown}, {Casey}, {Chin}, {Coutts},
  {Cowley}, {Deich}, {Feger}, {Fulton}, {Gers}, {Gurevich}, {Ishikawa},
  {James}, {Jelinsky}, {Kaye}, {Lanclos}, {Li}, {Lilley}, {McCarney}, {Miller},
  {Milner}, {O'Hanlon}, {Pember}, {Raffanti}, {Rockosi}, {Rubenzahl}, {Rumph},
  {Sandford}, {Savage}, {Schwab}, {Seifahrt}, {Shaum}, {Smith}, {Stuermer},
  {Thorne}, {Vandenberg}, {Von Boeckmann}, {Wang}, {Wang}, {Weisfeiler},
  {Wilcox}, {Wishnow}, {Wizinowich}, {Wold}, \&
  {Wolfenberger}}]{gibson2020-kpf}
{Gibson}, S.~R., {Howard}, A.~W., {Rider}, K., {et~al.} 2020, in Society of
  Photo-Optical Instrumentation Engineers (SPIE) Conference Series, Vol. 11447,
  Ground-based and Airborne Instrumentation for Astronomy VIII, ed. C.~J.
  {Evans}, J.~J. {Bryant}, \& K.~{Motohara},
  {\href{https://ui.adsabs.harvard.edu/abs/2020SPIE11447E..42G}{1144742}}

\bibitem[{S.~R. {Gibson} {et~al.}(2024){Gibson}, {Howard}, {Rider},
  {Halverson}, {Roy}, {Baker}, {Edelstein}, {Smith}, {Fulton}, {Walawender},
  {Brodheim}, {Brown}, {Chan}, {Dai}, {Deich}, {Gottschalk}, {Grillo}, {Hale},
  {Hill}, {Holden}, {Householder}, {Isaacson}, {Ishikawa}, {Jelinsky},
  {Kassis}, {Kaye}, {Laher}, {Lanclos}, {Lee}, {Lilley}, {McCarney}, {Miller},
  {Payne}, {Petigura}, {Poppett}, {Raffanti}, {Rubenzahl}, {Sandford},
  {Schwab}, {Shaum}, {Sirk}, {Smith}, {Thorne}, {Valliant}, {Vandenberg},
  {Wang}, {Wishnow}, {Wold}, {Yeh}, {Baca}, {Beichman}, {Berriman}, {Brown},
  {Casey}, {Chin}, {Chong}, {Cowley}, {Devenot}, {Elwir}, {Finstad}, {Fraysse},
  {James}, {Jhoti}, {Killian}, {Levine}, {Li}, {Marin}, {Milner}, {Nance},
  {O'Hanlon}, {Orr}, {Ortiz-Soto}, {Payne}, {Pember}, {Raskin}, {Savage},
  {Seifahrt}, {Smith}, {Storesund}, {St{\"u}rmer}, {Suominen}, {Tehero}, {Von
  Boeckmann}, {Wages}, {Weisfeiler}, {Wilcox}, {Wizinowich}, \&
  {Wolfenberger}}]{gibson2024-kpf}
{Gibson}, S.~R., {Howard}, A.~W., {Rider}, K., {et~al.} 2024, in Society of
  Photo-Optical Instrumentation Engineers (SPIE) Conference Series, Vol. 13096,
  Ground-based and Airborne Instrumentation for Astronomy X, ed. J.~J.
  {Bryant}, K.~{Motohara}, \& J.~R.~D. {Vernet},
  {\href{https://ui.adsabs.harvard.edu/abs/2024SPIE13096E..09G}{1309609}}

\bibitem[{R.~L. {Gilliland} {et~al.}(2011){Gilliland}, {Chaplin}, {Dunham},
  {Argabright}, {Borucki}, {Basri}, {Bryson}, {Buzasi}, {Caldwell}, {Elsworth},
  {Jenkins}, {Koch}, {Kolodziejczak}, {Miglio}, {van Cleve}, {Walkowicz}, \&
  {Welsh}}]{Gilliland2011}
{Gilliland}, R.~L., {Chaplin}, W.~J., {Dunham}, E.~W., {et~al.} 2011,
  {\mhref{http://doi.org/10.1088/0067-0049/197/1/6}{\apjs}},
  {\href{https://ui.adsabs.harvard.edu/abs/2011ApJS..197....6G}{197}}{\href{https://ui.adsabs.harvard.edu/abs/2011ApJS..197....6G}{,
  6}}

\bibitem[{L.~{Gizon} \& S.~K. {Solanki}(2003){Gizon} \& {Solanki}}]{Gizon2003}
{Gizon}, L., \& {Solanki}, S.~K. 2003,
  {\mhref{http://doi.org/10.1086/374715}{\apj}},
  {\href{https://ui.adsabs.harvard.edu/abs/2003ApJ...589.1009G}{589}}{\href{https://ui.adsabs.harvard.edu/abs/2003ApJ...589.1009G}{,
  1009}}

\bibitem[{R.~Gold(1964)Gold}]{gold1964}
Gold, R. 1964, {\mhref{http://doi.org/10.2172/4634295}{}}

\bibitem[{M.~Goto {et~al.}(2002)Goto {et~al.}}]{2002ApJ...567L..59G}
Goto, M., {et~al.} 2002, {\mhref{http://doi.org/10.1086/339800}{\apjl}},
  {\href{http://adsabs.harvard.edu/abs/2002ApJ...567L..59G}{567}}{\href{http://adsabs.harvard.edu/abs/2002ApJ...567L..59G}{,
  L59}}

\bibitem[{N.~{Grevesse} \& A.~J. {Sauval}(1998){Grevesse} \& {Sauval}}]{gs98}
{Grevesse}, N., \& {Sauval}, A.~J. 1998,
  {\mhref{http://doi.org/10.1023/A:1005161325181}{\ssr}},
  {\href{https://ui.adsabs.harvard.edu/abs/1998SSRv...85..161G}{85}}{\href{https://ui.adsabs.harvard.edu/abs/1998SSRv...85..161G}{,
  161}}

\bibitem[{A.~F. {Gupta} {et~al.}(2022){Gupta}, {Luhn}, {Wright}, {Mahadevan},
  {Ford}, {Stef{\'a}nsson}, {Bender}, {Blake}, {Halverson}, {Hearty},
  {Kanodia}, {Logsdon}, {McElwain}, {Ninan}, {Robertson}, {Roy}, {Schwab}, \&
  {Terrien}}]{Gupta2022}
{Gupta}, A.~F., {Luhn}, J., {Wright}, J.~T., {et~al.} 2022,
  {\mhref{http://doi.org/10.3847/1538-3881/ac96f3}{\aj}},
  {\href{https://ui.adsabs.harvard.edu/abs/2022AJ....164..254G}{164}}{\href{https://ui.adsabs.harvard.edu/abs/2022AJ....164..254G}{,
  254}}

\bibitem[{O.~J. {Hall} {et~al.}(2021){Hall}, {Davies}, {van Saders}, {Nielsen},
  {Lund}, {Chaplin}, {Garc{\'\i}a}, {Amard}, {Breimann}, {Khan}, {See}, \&
  {Tayar}}]{hall2021}
{Hall}, O.~J., {Davies}, G.~R., {van Saders}, J., {et~al.} 2021,
  {\mhref{http://doi.org/10.1038/s41550-021-01335-x}{Nature Astronomy}},
  {\href{https://ui.adsabs.harvard.edu/abs/2021NatAs...5..707H}{5}}{\href{https://ui.adsabs.harvard.edu/abs/2021NatAs...5..707H}{,
  707}}

\bibitem[{J.~W. {Hammer} {et~al.}(2005){Hammer}, {Fey}, {Kunz}, {Kiener},
  {Tatischeff}, {Haas}, {Weil}, {Assun{\c{c}}{\~a}o}, {Beck},
  {Boukari-Pelissie}, \& {et al.}}]{hammer2005}
{Hammer}, J.~W., {Fey}, M., {Kunz}, R., {et~al.} 2005,
  {\mhref{http://doi.org/10.1016/j.nuclphysa.2005.05.066}{\nphysa}},
  {\href{https://ui.adsabs.harvard.edu/abs/2005NuPhA.758..363H}{758}}{\href{https://ui.adsabs.harvard.edu/abs/2005NuPhA.758..363H}{,
  363}}

\bibitem[{P.~H. {Hauschildt} {et~al.}(1999{\natexlab{a}}){Hauschildt},
  {Allard}, \& {Baron}}]{hauschildt1999a}
{Hauschildt}, P.~H., {Allard}, F., \& {Baron}, E. 1999{\natexlab{a}},
  {\mhref{http://doi.org/10.1086/306745}{\apj}},
  {\href{https://ui.adsabs.harvard.edu/abs/1999ApJ...512..377H}{512}}{\href{https://ui.adsabs.harvard.edu/abs/1999ApJ...512..377H}{,
  377}}

\bibitem[{P.~H. {Hauschildt} {et~al.}(1999{\natexlab{b}}){Hauschildt},
  {Allard}, {Ferguson}, {Baron}, \& {Alexander}}]{hauschildt1999b}
{Hauschildt}, P.~H., {Allard}, F., {Ferguson}, J., {Baron}, E., \& {Alexander},
  D.~R. 1999{\natexlab{b}}, {\mhref{http://doi.org/10.1086/307954}{\apj}},
  {\href{https://ui.adsabs.harvard.edu/abs/1999ApJ...525..871H}{525}}{\href{https://ui.adsabs.harvard.edu/abs/1999ApJ...525..871H}{,
  871}}

\bibitem[{A.~{Helmi} {et~al.}(2018){Helmi}, {Babusiaux}, {Koppelman},
  {Massari}, {Veljanoski}, \& {Brown}}]{2018Natur.563...85H}
{Helmi}, A., {Babusiaux}, C., {Koppelman}, H.~H., {et~al.} 2018,
  {\mhref{http://doi.org/10.1038/s41586-018-0625-x}{\nat}},
  {\href{https://ui.adsabs.harvard.edu/abs/2018Natur.563...85H}{563}}{\href{https://ui.adsabs.harvard.edu/abs/2018Natur.563...85H}{,
  85}}

\bibitem[{T.~J. Henry \& J.~D. {Kirkpatrick}(1990)Henry \&
  {Kirkpatrick}}]{1990ApJ...354L..29H}
Henry, T.~J., \& {Kirkpatrick}, J.~D. 1990,
  {\mhref{http://doi.org/10.1086/185715}{\apjl}},
  {\href{http://cdsads.u-strasbg.fr/abs/1990ApJ...354L..29H}{354}}{\href{http://cdsads.u-strasbg.fr/abs/1990ApJ...354L..29H}{,
  L29}}

\bibitem[{L.~{Henyey} {et~al.}(1965){Henyey}, {Vardya}, \&
  {Bodenheimer}}]{henyey1965}
{Henyey}, L., {Vardya}, M.~S., \& {Bodenheimer}, P. 1965,
  {\mhref{http://doi.org/10.1086/148357}{\apj}},
  {\href{https://ui.adsabs.harvard.edu/abs/1965ApJ...142..841H}{142}}{\href{https://ui.adsabs.harvard.edu/abs/1965ApJ...142..841H}{,
  841}}

\bibitem[{F.~{Herwig}(2000){Herwig}}]{herwig2000}
{Herwig}, F. 2000,
  {\mhref{http://doi.org/10.48550/arXiv.astro-ph/0007139}{\aap}},
  {\href{https://ui.adsabs.harvard.edu/abs/2000A&A...360..952H}{360}}{\href{https://ui.adsabs.harvard.edu/abs/2000A&A...360..952H}{,
  952}}

\bibitem[{S.~{Hinkley} {et~al.}(2023){Hinkley}, {Lacour}, {Marleau},
  {Lagrange}, {Wang}, {Kammerer}, {Cumming}, {Nowak}, {Rodet}, {Stolker},
  {Balmer}, {Ray}, {Bonnefoy}, {Molli{`e}re}, {Lazzoni}, {Kennedy},
  {Mordasini}, {Abuter}, {Aigrain}, {Amorim}, {Asensio-Torres}, {Babusiaux},
  {Benisty}, {Berger}, {Beust}, {Blunt}, {Boccaletti}, {Bohn}, {Bonnet},
  {Bourdarot}, {Brandner}, {Cantalloube}, {Caselli}, {Charnay}, {Chauvin},
  {Chomez}, {Choquet}, {Christiaens}, {Cl{'e}net}, {Coud{'e} du Foresto},
  {Cridland}, {Delorme}, {Dembet}, {Drescher}, {Duvert}, {Eckart},
  {Eisenhauer}, {Feuchtgruber}, {Galland}, {Garcia}, {Garcia Lopez}, {Gardner},
  {Gendron}, {Genzel}, {Gillessen}, {Girard}, {Grandjean}, {Haubois},
  {Hei{ss}el}, {Henning}, {Hippler}, {Horrobin}, {Houll{'e}}, {Hubert},
  {Jocou}, {Keppler}, {Kervella}, {Kreidberg}, {Lapeyr{`e}re}, {Le Bouquin},
  {L{'e}na}, {Lutz}, {Maire}, {Mang}, {M{'e}rand}, {Meunier}, {Monnier},
  {Mouillet}, {Nasedkin}, {Ott}, {Otten}, {Paladini}, {Paumard}, {Perraut},
  {Perrin}, {Philipot}, {Pfuhl}, {Pourr{'e}}, {Pueyo}, {Rameau}, {Rickman},
  {Rubini}, {Rustamkulov}, {Samland}, {Shangguan}, {Shimizu}, {Sing},
  {Straubmeier}, {Sturm}, {Tacconi}, {van Dishoeck}, {Vigan}, {Vincent},
  {Ward-Duong}, {Widmann}, {Wieprecht}, {Wiezorrek}, {Woillez}, {Yazici},
  {Young}, \& {Zicher}}]{2023A&A...671L...5H}
{Hinkley}, S., {Lacour}, S., {Marleau}, G.-D., {et~al.} 2023,
  {\mhref{http://doi.org/10.1051/0004-6361/202244727}{\aap}},
  {\href{https://ui.adsabs.harvard.edu/abs/2023A&A...671L...5H}{671}}{\href{https://ui.adsabs.harvard.edu/abs/2023A&A...671L...5H}{,
  L5}}

\bibitem[{M.~{Hon} {et~al.}(2024{\natexlab{a}}){Hon}, {Li}, \&
  {Ong}}]{Hon2024-flow}
{Hon}, M., {Li}, Y., \& {Ong}, J. 2024{\natexlab{a}},
  {\mhref{http://doi.org/10.3847/1538-4357/ad6320}{\apj}},
  {\href{https://ui.adsabs.harvard.edu/abs/2024ApJ...973..154H}{973}}{\href{https://ui.adsabs.harvard.edu/abs/2024ApJ...973..154H}{,
  154}}

\bibitem[{M.~{Hon} {et~al.}(2024{\natexlab{b}}){Hon}, {Huber}, {Li},
  {Metcalfe}, {Bedding}, {Ong}, {Chontos}, {Rubenzahl}, {Halverson},
  {Garc{\'\i}a}, {Kjeldsen}, {Stello}, {Hey}, {Campante}, {Howard}, {Gibson},
  {Rider}, {Roy}, {Baker}, {Edelstein}, {Smith}, {Fulton}, {Walawender},
  {Brodheim}, {Brown}, {Chan}, {Dai}, {Deich}, {Gottschalk}, {Grillo}, {Hale},
  {Hill}, {Holden}, {Householder}, {Isaacson}, {Ishikawa}, {Jelinsky},
  {Kassis}, {Kaye}, {Laher}, {Lanclos}, {Lee}, {Lilley}, {McCarney}, {Miller},
  {Payne}, {Petigura}, {Poppett}, {Raffanti}, {Rockosi}, {Sanford}, {Schwab},
  {Shaum}, {Sirk}, {Smith}, {Thorne}, {Valliant}, {Vandenberg}, {Ywan Wang},
  {Wishnow}, {Wold}, {Yeh}, {Baker}, {Basu}, {Bedell}, {Cegla}, {Crossfield},
  {Dressing}, {Dumusque}, {Knutson}, {Mawet}, {O'Meara}, {Stef{\'a}nsson},
  {Teske}, {Vasisht}, {Xuesong Wang}, {Weiss}, {Winn}, \&
  {Wright}}]{Hon2024-sig-dra}
{Hon}, M., {Huber}, D., {Li}, Y., {et~al.} 2024{\natexlab{b}},
  {\mhref{http://doi.org/10.48550/arXiv.2407.21234}{arXiv
  e-prints}}{\href{https://ui.adsabs.harvard.edu/abs/2024arXiv240721234H}{,
  arXiv:2407.21234}}

\bibitem[{R.~{Howe} {et~al.}(2020){Howe}, {Chaplin}, {Basu}, {Ball}, {Davies},
  {Elsworth}, {Hale}, {Miglio}, {Nielsen}, \& {Viani}}]{Howe2020}
{Howe}, R., {Chaplin}, W.~J., {Basu}, S., {et~al.} 2020,
  {\mhref{http://doi.org/10.1093/mnrasl/slaa006}{\mnras}},
  {\href{https://ui.adsabs.harvard.edu/abs/2020MNRAS.493L..49H}{493}}{\href{https://ui.adsabs.harvard.edu/abs/2020MNRAS.493L..49H}{,
  L49}}

\bibitem[{D.~{Huber} {et~al.}(2011{\natexlab{a}}){Huber}, {Bedding}, {Stello},
  {Hekker}, {Mathur}, {Mosser}, {Verner}, {Bonanno}, {Buzasi}, {Campante},
  {Elsworth}, {Hale}, {Kallinger}, {Silva Aguirre}, {Chaplin}, {De Ridder},
  {Garc{\'\i}a}, {Appourchaux}, {Frandsen}, {Houdek}, {Molenda-{\.Z}akowicz},
  {Monteiro}, {Christensen-Dalsgaard}, {Gilliland}, {Kawaler}, {Kjeldsen},
  {Broomhall}, {Corsaro}, {Salabert}, {Sanderfer}, {Seader}, \&
  {Smith}}]{Huber2011}
{Huber}, D., {Bedding}, T.~R., {Stello}, D., {et~al.} 2011{\natexlab{a}},
  {\mhref{http://doi.org/10.1088/0004-637X/743/2/143}{\apj}},
  {\href{https://ui.adsabs.harvard.edu/abs/2011ApJ...743..143H}{743}}{\href{https://ui.adsabs.harvard.edu/abs/2011ApJ...743..143H}{,
  143}}

\bibitem[{D.~{Huber} {et~al.}(2011{\natexlab{b}}){Huber}, {Bedding},
  {Arentoft}, {Gruberbauer}, {Guenther}, {Houdek}, {Kallinger}, {Kjeldsen},
  {Matthews}, {Stello}, \& {Weiss}}]{Huber2011-procyon}
{Huber}, D., {Bedding}, T.~R., {Arentoft}, T., {et~al.} 2011{\natexlab{b}},
  {\mhref{http://doi.org/10.1088/0004-637X/731/2/94}{\apj}},
  {\href{https://ui.adsabs.harvard.edu/abs/2011ApJ...731...94H}{731}}{\href{https://ui.adsabs.harvard.edu/abs/2011ApJ...731...94H}{,
  94}}

\bibitem[{D.~{Huber} {et~al.}(2017){Huber}, {Zinn}, {Bojsen-Hansen},
  {Pinsonneault}, {Sahlholdt}, {Serenelli}, {Silva Aguirre}, {Stassun},
  {Stello}, {Tayar}, {Bastien}, {Bedding}, {Buchhave}, {Chaplin}, {Davies},
  {Garc{\'\i}a}, {Latham}, {Mathur}, {Mosser}, \& {Sharma}}]{Huber2017}
{Huber}, D., {Zinn}, J., {Bojsen-Hansen}, M., {et~al.} 2017,
  {\mhref{http://doi.org/10.3847/1538-4357/aa75ca}{\apj}},
  {\href{https://ui.adsabs.harvard.edu/abs/2017ApJ...844..102H}{844}}{\href{https://ui.adsabs.harvard.edu/abs/2017ApJ...844..102H}{,
  102}}

\bibitem[{M.~J. {Ireland} {et~al.}(2008){Ireland}, {Kraus}, {Martinache},
  {Lloyd}, \& {Tuthill}}]{2008ApJ...678..463I}
{Ireland}, M.~J., {Kraus}, A., {Martinache}, F., {Lloyd}, J.~P., \& {Tuthill},
  P.~G. 2008, {\mhref{http://doi.org/10.1086/529578}{\apj}},
  {\href{https://ui.adsabs.harvard.edu/abs/2008ApJ...678..463I}{678}}{\href{https://ui.adsabs.harvard.edu/abs/2008ApJ...678..463I}{,
  463}}

\bibitem[{A.~S. {Jermyn} {et~al.}(2023){Jermyn}, {Bauer}, {Schwab}, {Farmer},
  {Ball}, {Bellinger}, {Dotter}, {Joyce}, {Marchant}, {Mombarg}, {Wolf}, {Sunny
  Wong}, {Cinquegrana}, {Farrell}, {Smolec}, {Thoul}, {Cantiello}, {Herwig},
  {Toloza}, {Bildsten}, {Townsend}, \& {Timmes}}]{mesa2023}
{Jermyn}, A.~S., {Bauer}, E.~B., {Schwab}, J., {et~al.} 2023,
  {\mhref{http://doi.org/10.3847/1538-4365/acae8d}{\apjs}},
  {\href{https://ui.adsabs.harvard.edu/abs/2023ApJS..265...15J}{265}}{\href{https://ui.adsabs.harvard.edu/abs/2023ApJS..265...15J}{,
  15}}

\bibitem[{C.~{Jiang} {et~al.}(2023){Jiang}, {Wu}, {Feinstein}, {Stassun},
  {Bedding}, {Veras}, {Corsaro}, {Buzasi}, {Stello}, {Li}, {Mathur},
  {Garc{\'\i}a}, {Breton}, {Lundkvist}, {Miko{\l}ajczyk}, {Gehan}, {Campante},
  {Bossini}, {Kane}, {Joel Ong}, {Y{\i}ld{\i}z}, {Kayhan}, {{\c{C}}elik Orhan},
  {{\"O}rtel}, {Zhang}, {Cunha}, {de Moura}, {Yu}, {Huber}, {Ou}, {Wittenmyer},
  {Gizon}, \& {Chaplin}}]{Jiangc2023}
{Jiang}, C., {Wu}, T., {Feinstein}, A.~D., {et~al.} 2023,
  {\mhref{http://doi.org/10.3847/1538-4357/acb8ac}{\apj}},
  {\href{https://ui.adsabs.harvard.edu/abs/2023ApJ...945...20J}{945}}{\href{https://ui.adsabs.harvard.edu/abs/2023ApJ...945...20J}{,
  20}}

\bibitem[{J.~{Kammerer} {et~al.}(2024){Kammerer}, {Lawson}, {Perrin},
  {Rebollido}, {Stark}, {Stolker}, {Girard}, {Pueyo}, {Balmer}, {Worthen},
  {Chen}, {van der Marel}, {Lewis}, {Ward-Duong}, {Valenti}, {Clampin}, \&
  {Mountain}}]{2024AJ....168...51K}
{Kammerer}, J., {Lawson}, K., {Perrin}, M.~D., {et~al.} 2024,
  {\mhref{http://doi.org/10.3847/1538-3881/ad4ffe}{\aj}},
  {\href{https://ui.adsabs.harvard.edu/abs/2024AJ....168...51K}{168}}{\href{https://ui.adsabs.harvard.edu/abs/2024AJ....168...51K}{,
  51}}

\bibitem[{J.~{Kammerer} {et~al.}(2025){Kammerer}, {Winterhalder}, {Lacour},
  {Stolker}, {Marleau}, {Balmer}, {Moore}, {Piscarreta}, {Toci}, {M{'e}rand},
  {Nowak}, {Rickman}, {Pueyo}, {Pourr{'e}}, {Nasedkin}, {Wang}, {Bourdarot},
  {Eisenhauer}, {Henning}, {Garcia Lopez}, {van Dishoeck}, {Forveille},
  {Monnier}, {Abuter}, {Amorim}, {Benisty}, {Berger}, {Beust}, {Blunt},
  {Boccaletti}, {Bonnefoy}, {Bonnet}, {Bordoni}, {Brandner}, {Cantalloube},
  {Caselli}, {Ceva}, {Charnay}, {Chauvin}, {Chavez}, {Chomez}, {Choquet},
  {Christiaens}, {Cl{'e}net}, {Coud{'e} du Foresto}, {Cridland}, {Davies},
  {Dembet}, {Dexter}, {Drescher}, {Duvert}, {Eckart}, {Fontanive}, {F{"o}rster
  Schreiber}, {Garcia}, {Gendron}, {Genzel}, {Gillessen}, {Girard}, {Grant},
  {Hagelberg}, {Haubois}, {Hei{ss}el}, {Hinkley}, {Hippler}, {Houll{'e}},
  {Hubert}, {Jocou}, {Keppler}, {Kervella}, {Kreidberg}, {Kurtovic},
  {Lagrange}, {Lapeyr{`e}re}, {Le Bouquin}, {Lutz}, {Maire}, {Mang},
  {Matthews}, {Molli{`e}re}, {Mordasini}, {Mouillet}, {Ott}, {Otten},
  {Paladini}, {Paumard}, {Perraut}, {Perrin}, {Pfuhl}, {Ribeiro},
  {Rustamkulov}, {S{'e}gransan}, {Shangguan}, {Shimizu}, {Samland}, {Sing},
  {Stadler}, {Straub}, {Straubmeier}, {Sturm}, {Tacconi}, {Udry}, {Vigan},
  {Vincent}, {von Fellenberg}, {Widmann}, {Woillez}, {Yazici}, \& {the GRAVITY
  Collaboration}}]{2025arXiv251008691K}
{Kammerer}, J., {Winterhalder}, T.~O., {Lacour}, S., {et~al.} 2025,
  {\mhref{http://doi.org/10.48550/arXiv.2510.08691}{arXiv
  e-prints}}{\href{https://ui.adsabs.harvard.edu/abs/2025arXiv251008691K}{,
  arXiv:2510.08691}}

\bibitem[{Y.~{Kasagi} {et~al.}(2025){Kasagi}, {Kawashima}, {Kawahara},
  {Kotani}, {Masuda}, {Ahn}, {Guyon}, {Hirano}, {Jovanovic}, {Kuzuhara},
  {Lozi}, {Tamura}, {Uyama}, {Vievard}, \& {Yoneta}}]{Kasagi2025}
{Kasagi}, Y., {Kawashima}, Y., {Kawahara}, H., {et~al.} 2025,
  {\mhref{http://doi.org/10.48550/arXiv.2508.01281}{arXiv
  e-prints}}{\href{https://ui.adsabs.harvard.edu/abs/2025arXiv250801281K}{,
  arXiv:2508.01281}}

\bibitem[{R.~R. King {et~al.}(2010)King, {McCaughrean}, {Homeier}, {Allard},
  {Scholz}, \& {Lodieu}}]{2010A&A...510A..99K}
King, R.~R., {McCaughrean}, M.~J., {Homeier}, D., {et~al.} 2010,
  {\mhref{http://doi.org/10.1051/0004-6361/200912981}{\aap}},
  {\href{http://adsabs.harvard.edu/abs/2010A%26A...510A..99K}{510}}{\href{http://adsabs.harvard.edu/abs/2010A%26A...510A..99K}{,
  A99}}

\bibitem[{J.~D. Kirkpatrick {et~al.}(2000)Kirkpatrick, {Reid}, {Liebert},
  {Gizis}, {Burgasser}, {Monet}, {Dahn}, {Nelson}, \&
  {Williams}}]{2000AJ....120..447K}
Kirkpatrick, J.~D., {Reid}, I.~N., {Liebert}, J., {et~al.} 2000, \aj, 120, 447

\bibitem[{H.~{Kjeldsen} \& T.~R. {Bedding}(1995){Kjeldsen} \& {Bedding}}]{kb95}
{Kjeldsen}, H., \& {Bedding}, T.~R. 1995,
  {\mhref{http://doi.org/10.48550/arXiv.astro-ph/9403015}{\aap}},
  {\href{https://ui.adsabs.harvard.edu/abs/1995A&A...293...87K}{293}}{\href{https://ui.adsabs.harvard.edu/abs/1995A&A...293...87K}{,
  87}}

\bibitem[{H.~{Kjeldsen} \& T.~R. {Bedding}(2012){Kjeldsen} \&
  {Bedding}}]{Kjeldsen2012}
{Kjeldsen}, H., \& {Bedding}, T.~R. 2012, in IAU Symposium, Vol. 285, New
  Horizons in Time Domain Astronomy, ed. E.~{Griffin}, R.~{Hanisch}, \&
  R.~{Seaman},
  {\href{https://ui.adsabs.harvard.edu/abs/2012IAUS..285...17K}{17--22}}

\bibitem[{H.~{Kjeldsen} {et~al.}(2008){Kjeldsen}, {Bedding}, {Arentoft},
  {Butler}, {Dall}, {Karoff}, {Kiss}, {Tinney}, \& {Chaplin}}]{Kjeldsen2008}
{Kjeldsen}, H., {Bedding}, T.~R., {Arentoft}, T., {et~al.} 2008,
  {\mhref{http://doi.org/10.1086/589142}{\apj}},
  {\href{https://ui.adsabs.harvard.edu/abs/2008ApJ...682.1370K}{682}}{\href{https://ui.adsabs.harvard.edu/abs/2008ApJ...682.1370K}{,
  1370}}

\bibitem[{H.~{Kjeldsen} {et~al.}(2025){Kjeldsen}, {Bedding}, {Li}, {Grundahl},
  {Fredslund Andersen}, {Wright}, {Soutter}, {Wittenmyer}, {Reyes}, {Stello},
  {Crawford}, {Zhou}, {Clerte}, {Palle}, {Simon-Diaz}, {Christensen-Dalsgaard},
  {Handberg}, {Hansen}, {Heeren}, {Jessen-Hansen}, {Lund}, {Lundkvist},
  {Brogaard}, {Tronsgaard}, {Rudrasingam}, {Casagrande}, {Horner}, {Huber},
  {Lattanzio}, {Martell}, \& {Murphy}}]{Kjeldsen2025}
{Kjeldsen}, H., {Bedding}, T.~R., {Li}, Y., {et~al.} 2025,
  {\mhref{http://doi.org/10.48550/arXiv.2506.00493}{arXiv
  e-prints}}{\href{https://ui.adsabs.harvard.edu/abs/2025arXiv250600493K}{,
  arXiv:2506.00493}}

\bibitem[{Q.~M. {Konopacky} {et~al.}(2010){Konopacky}, {Ghez}, {Barman},
  {Rice}, {Bailey}, {White}, {McLean}, \& {Duch{\^e}ne}}]{2010ApJ...711.1087K}
{Konopacky}, Q.~M., {Ghez}, A.~M., {Barman}, T.~S., {et~al.} 2010,
  {\mhref{http://doi.org/10.1088/0004-637X/711/2/1087}{apj}},
  {\href{https://ui.adsabs.harvard.edu/abs/2010ApJ...711.1087K}{711}}{\href{https://ui.adsabs.harvard.edu/abs/2010ApJ...711.1087K}{,
  1087}}

\bibitem[{K.~S. {Krishna Swamy}(1966){Krishna Swamy}}]{ks1966}
{Krishna Swamy}, K.~S. 1966, {\mhref{http://doi.org/10.1086/148752}{\apj}},
  {\href{https://ui.adsabs.harvard.edu/abs/1966ApJ...145..174K}{145}}{\href{https://ui.adsabs.harvard.edu/abs/1966ApJ...145..174K}{,
  174}}

\bibitem[{M.~{Kuzuhara} {et~al.}(2022){Kuzuhara}, {Currie}, {Takarada},
  {Brandt}, {Sato}, {Uyama}, {Janson}, {Chilcote}, {Tobin}, {Lawson}, {Hori},
  {Guyon}, {Groff}, {Lozi}, {Vievard}, {Sahoo}, {Deo}, {Jovanovic}, {Ahn},
  {Martinache}, {Skaf}, {Akiyama}, {Norris}, {Bonnefoy}, {He{l}miniak}, {Kudo},
  {McElwain}, {Samland}, {Wagner}, {Wisniewski}, {Knapp}, {Kwon}, {Nishikawa},
  {Serabyn}, {Hayashi}, \& {Tamura}}]{2022ApJ...934L..18K}
{Kuzuhara}, M., {Currie}, T., {Takarada}, T., {et~al.} 2022,
  {\mhref{http://doi.org/10.3847/2041-8213/ac772f}{\apjl}},
  {\href{https://ui.adsabs.harvard.edu/abs/2022ApJ...934L..18K}{934}}{\href{https://ui.adsabs.harvard.edu/abs/2022ApJ...934L..18K}{,
  L18}}

\bibitem[{A.~Lagrange {et~al.}(2010)Lagrange {et~al.}}]{2010Sci...329...57L}
Lagrange, A., {et~al.} 2010,
  {\mhref{http://doi.org/10.1126/science.1187187}{Science}},
  {\href{http://adsabs.harvard.edu/abs/2010Sci...329...57L}{329}}{\href{http://adsabs.harvard.edu/abs/2010Sci...329...57L}{,
  57}}

\bibitem[{A.-M. {Lagrange} {et~al.}(2019){Lagrange}, {Meunier}, {Rubini},
  {Keppler}, {Galland}, {Chapellier}, {Michel}, {Balona}, {Beust}, {Guillot},
  {Grandjean}, {Borgniet}, {M{'e}karnia}, {Wilson}, {Kiefer}, {Bonnefoy},
  {Lillo-Box}, {Pantoja}, {Jones}, {Iglesias}, {Rodet}, {Diaz}, {Zapata},
  {Abe}, \& {Schmider}}]{2019NatAs...3.1135L}
{Lagrange}, A.-M., {Meunier}, N., {Rubini}, P., {et~al.} 2019,
  {\mhref{http://doi.org/10.1038/s41550-019-0857-1}{Nature Astronomy}},
  {\href{https://ui.adsabs.harvard.edu/abs/2019NatAs...3.1135L}{3}}{\href{https://ui.adsabs.harvard.edu/abs/2019NatAs...3.1135L}{,
  1135}}

\bibitem[{B.~F. Lane {et~al.}(2001)Lane, {Zapatero Osorio}, {Britton},
  {Mart{\'{\i}}n}, \& {Kulkarni}}]{2001ApJ...560..390L}
Lane, B.~F., {Zapatero Osorio}, M.~R., {Britton}, M.~C., {Mart{\'{\i}}n},
  E.~L., \& {Kulkarni}, S.~R. 2001, \apj,
  {\href{http://adsabs.harvard.edu/cgi-bin/nph-bib_query?bibcode=2001ApJ...560..390L&amp;db_key=AST}{560}}{\href{http://adsabs.harvard.edu/cgi-bin/nph-bib_query?bibcode=2001ApJ...560..390L&amp;db_key=AST}{,
  390}}

\bibitem[{P.~{Ledoux}(1951){Ledoux}}]{Ledoux1951}
{Ledoux}, P. 1951, {\mhref{http://doi.org/10.1086/145477}{\apj}},
  {\href{https://ui.adsabs.harvard.edu/abs/1951ApJ...114..373L}{114}}{\href{https://ui.adsabs.harvard.edu/abs/1951ApJ...114..373L}{,
  373}}

\bibitem[{Y.~{Li} {et~al.}(2023{\natexlab{a}}){Li}, {Bedding}, {Stello},
  {Huber}, {Hon}, {Joyce}, {Li}, {Perkins}, {White}, {Zinn}, {Howard},
  {Isaacson}, {Hey}, \& {Kjeldsen}}]{liyg2023}
{Li}, Y., {Bedding}, T.~R., {Stello}, D., {et~al.} 2023{\natexlab{a}},
  {\mhref{http://doi.org/10.1093/mnras/stad1445}{\mnras}},
  {\href{https://ui.adsabs.harvard.edu/abs/2023MNRAS.523..916L}{523}}{\href{https://ui.adsabs.harvard.edu/abs/2023MNRAS.523..916L}{,
  916}}

\bibitem[{Y.~{Li} {et~al.}(2023{\natexlab{b}}){Li}, {Brandt}, {Brandt}, {An},
  {Franson}, {Dupuy}, {Chen}, {Bowens-Rubin}, {Lewis}, {Bowler}, {Gibbs},
  {Kiman}, {Faherty}, {Currie}, {Jensen-Clem}, {Zhang}, {Contreras-Martinez},
  {Fitzgerald}, {Mazin}, \& {Millar-Blanchaer}}]{2023MNRAS.522.5622L}
{Li}, Y., {Brandt}, T.~D., {Brandt}, G.~M., {et~al.} 2023{\natexlab{b}},
  {\mhref{http://doi.org/10.1093/mnras/stad1315}{\mnras}},
  {\href{https://ui.adsabs.harvard.edu/abs/2023MNRAS.522.5622L}{522}}{\href{https://ui.adsabs.harvard.edu/abs/2023MNRAS.522.5622L}{,
  5622}}

\bibitem[{Y.~{Li} {et~al.}(2024){Li}, {Brandt}, {Franson}, {An}, {Tobin},
  {Currie}, {Chen}, {Wang}, {Dupuy}, {Bowens-Rubin}, {Salama}, {Lewis},
  {Gibbs}, {Bowler}, {Jensen-Clem}, {Faherty}, {Fitzgerald}, \&
  {Mazin}}]{2024MNRAS.533.3501L}
{Li}, Y., {Brandt}, T.~D., {Franson}, K., {et~al.} 2024,
  {\mhref{http://doi.org/10.1093/mnras/stae1903}{\mnras}},
  {\href{https://ui.adsabs.harvard.edu/abs/2024MNRAS.533.3501L}{533}}{\href{https://ui.adsabs.harvard.edu/abs/2024MNRAS.533.3501L}{,
  3501}}

\bibitem[{Y.~{Li} {et~al.}(2025){Li}, {Huber}, {Ong}, {van Saders}, {Costa},
  {Larsen}, {Basu}, {Bedding}, {Dai}, {Chontos}, {Carmichael}, {Hey},
  {Kjeldsen}, {Hon}, {Campante}, {Monteiro}, {Lundkvist}, {Saunders},
  {Isaacson}, {Howard}, {Gibson}, {Halverson}, {Rider}, {Roy}, {Baker},
  {Edelstein}, {Smith}, {Fulton}, \& {Walawender}}]{Liyg2025}
{Li}, Y., {Huber}, D., {Ong}, J.~M.~J., {et~al.} 2025,
  {\mhref{http://doi.org/10.3847/1538-4357/adc737}{\apj}},
  {\href{https://ui.adsabs.harvard.edu/abs/2025ApJ...984..125L}{984}}{\href{https://ui.adsabs.harvard.edu/abs/2025ApJ...984..125L}{,
  125}}

\bibitem[{M.~C. {Liu} {et~al.}(2016){Liu}, {Dupuy}, \&
  {Allers}}]{2016ApJ...833...96L}
{Liu}, M.~C., {Dupuy}, T.~J., \& {Allers}, K.~N. 2016,
  {\mhref{http://doi.org/10.3847/1538-4357/833/1/96}{\apj}},
  {\href{http://adsabs.harvard.edu/abs/2016ApJ...833...96L}{833}}{\href{http://adsabs.harvard.edu/abs/2016ApJ...833...96L}{,
  96}}

\bibitem[{M.~C. Liu {et~al.}(2008)Liu, {Dupuy}, \&
  {Ireland}}]{liu08-2m1534orbit}
Liu, M.~C., {Dupuy}, T.~J., \& {Ireland}, M.~J. 2008,
  {\mhref{http://doi.org/10.1086/591837}{\apj}},
  {\href{http://adsabs.harvard.edu/abs/2008ApJ...689..436L}{689}}{\href{http://adsabs.harvard.edu/abs/2008ApJ...689..436L}{,
  436}}

\bibitem[{M.~C. {Liu} {et~al.}(2002{\natexlab{a}}){Liu}, {Fischer}, {Graham},
  {Lloyd}, {Marcy}, \& {Butler}}]{Liu2002}
{Liu}, M.~C., {Fischer}, D.~A., {Graham}, J.~R., {et~al.} 2002{\natexlab{a}},
  {\mhref{http://doi.org/10.1086/339845}{\apj}},
  {\href{https://ui.adsabs.harvard.edu/abs/2002ApJ...571..519L}{571}}{\href{https://ui.adsabs.harvard.edu/abs/2002ApJ...571..519L}{,
  519}}

\bibitem[{M.~C. {Liu} {et~al.}(2002{\natexlab{b}}){Liu}, {Fischer}, {Graham},
  {Lloyd}, {Marcy}, \& {Butler}}]{2002ApJ...571..519L}
---. 2002{\natexlab{b}}, {\mhref{http://doi.org/10.1086/339845}{\apj}},
  {\href{https://ui.adsabs.harvard.edu/abs/2002ApJ...571..519L}{571}}{\href{https://ui.adsabs.harvard.edu/abs/2002ApJ...571..519L}{,
  519}}

\bibitem[{N.~R. {Lomb}(1976){Lomb}}]{lomb1976}
{Lomb}, N.~R. 1976, {\mhref{http://doi.org/10.1007/BF00648343}{\apss}},
  {\href{https://ui.adsabs.harvard.edu/abs/1976Ap&SS..39..447L}{39}}{\href{https://ui.adsabs.harvard.edu/abs/1976Ap&SS..39..447L}{,
  447}}

\bibitem[{L.~{Long} {et~al.}(2023){Long}, {Bi}, {Zhang}, {Zhang}, {Zhang},
  {Ge}, {Li}, {Chen}, {Li}, {Ye}, {Sun}, \& {Zhou}}]{long2023-cl}
{Long}, L., {Bi}, S., {Zhang}, J., {et~al.} 2023,
  {\mhref{http://doi.org/10.3847/1538-4365/ace5af}{\apjs}},
  {\href{https://ui.adsabs.harvard.edu/abs/2023ApJS..268...30L}{268}}{\href{https://ui.adsabs.harvard.edu/abs/2023ApJS..268...30L}{,
  30}}

\bibitem[{Y.~{Lu} {et~al.}(2024){Lu}, {Angus}, {Foreman-Mackey}, \&
  {Hattori}}]{Lu2024}
{Lu}, Y., {Angus}, R., {Foreman-Mackey}, D., \& {Hattori}, S. 2024,
  {\mhref{http://doi.org/10.3847/1538-3881/ad28b9}{\aj}},
  {\href{https://ui.adsabs.harvard.edu/abs/2024AJ....167..159L}{167}}{\href{https://ui.adsabs.harvard.edu/abs/2024AJ....167..159L}{,
  159}}

\bibitem[{M.~N. {Lund}(2019){Lund}}]{Lund2019}
{Lund}, M.~N. 2019, {\mhref{http://doi.org/10.1093/mnras/stz2010}{\mnras}},
  {\href{https://ui.adsabs.harvard.edu/abs/2019MNRAS.489.1072L}{489}}{\href{https://ui.adsabs.harvard.edu/abs/2019MNRAS.489.1072L}{,
  1072}}

\bibitem[{M.~S. {Lundkvist} {et~al.}(2024){Lundkvist}, {Kjeldsen}, {Bedding},
  {McCaughrean}, {Butler}, {Slumstrup}, {Campante}, {Aerts}, {Arentoft},
  {Bruntt}, {Cardoso}, {Carrier}, {Close}, {Gomes da Silva}, {Kallinger},
  {King}, {Li}, {Murphy}, {R{\o}rsted}, \& {Stello}}]{Lundkvist2024}
{Lundkvist}, M.~S., {Kjeldsen}, H., {Bedding}, T.~R., {et~al.} 2024,
  {\mhref{http://doi.org/10.3847/1538-4357/ad25f2}{\apj}},
  {\href{https://ui.adsabs.harvard.edu/abs/2024ApJ...964..110L}{964}}{\href{https://ui.adsabs.harvard.edu/abs/2024ApJ...964..110L}{,
  110}}

\bibitem[{A.-L. {Maire} {et~al.}(2020{\natexlab{a}}){Maire}, {Baudino},
  {Desidera}, {Messina}, {Brandner}, {Godoy}, {Cantalloube}, {Galicher},
  {Bonnefoy}, {Hagelberg}, {Olofsson}, {Absil}, {Chauvin}, {Henning}, \&
  {Langlois}}]{2020A&A...633L...2M}
{Maire}, A.-L., {Baudino}, J.-L., {Desidera}, S., {et~al.} 2020{\natexlab{a}},
  {\mhref{http://doi.org/10.1051/0004-6361/201937134}{\aap}},
  {\href{https://ui.adsabs.harvard.edu/abs/2020A&A...633L...2M}{633}}{\href{https://ui.adsabs.harvard.edu/abs/2020A&A...633L...2M}{,
  L2}}

\bibitem[{A.-L. {Maire} {et~al.}(2020{\natexlab{b}}){Maire}, {Molaverdikhani},
  {Desidera}, {Trifonov}, {Molli{`e}re}, {D'Orazi}, {Frankel}, {Baudino},
  {Messina}, {M{"u}ller}, {Charnay}, {Cheetham}, {Delorme}, {Ligi}, {Bonnefoy},
  {Brandner}, {Mesa}, {Cantalloube}, {Galicher}, {Henning}, {Biller},
  {Hagelberg}, {Lagrange}, {Lavie}, {Rickman}, {S{'e}gransan}, {Udry},
  {Chauvin}, {Gratton}, {Langlois}, {Vigan}, {Meyer}, {Beuzit}, {Bhowmik},
  {Boccaletti}, {Lazzoni}, {Perrot}, {Schmidt}, {Zurlo}, {Gluck}, {Pragt},
  {Ramos}, {Roelfsema}, {Roux}, \& {Sauvage}}]{2020A&A...639A..47M}
{Maire}, A.-L., {Molaverdikhani}, K., {Desidera}, S., {et~al.}
  2020{\natexlab{b}},
  {\mhref{http://doi.org/10.1051/0004-6361/202037984}{\aap}},
  {\href{https://ui.adsabs.harvard.edu/abs/2020A&A...639A..47M}{639}}{\href{https://ui.adsabs.harvard.edu/abs/2020A&A...639A..47M}{,
  A47}}

\bibitem[{A.-L. {Maire} {et~al.}(2024){Maire}, {Leclerc}, {Balmer}, {Desidera},
  {Lacour}, {D'Orazi}, {Samland}, {Langlois}, {Matthews}, {Babusiaux},
  {Kervella}, {Le Bouquin}, {S{'e}gransan}, {Gratton}, {Biller}, {Bonavita},
  {Delorme}, {Messina}, {Udry}, {Janson}, {Henning}, {Wahhaj}, {Zurlo},
  {Bonnefoy}, {Brandner}, {Cantalloube}, {Galicher}, {Kammerer}, {Nowak},
  {Shangguan}, {Stolker}, {Wang}, {Chauvin}, {Hagelberg}, {Lagrange}, {Vigan},
  {Meyer}, {Beuzit}, {Boccaletti}, {Lazzoni}, {Mesa}, {Perrot}, {Squicciarini},
  {Hinkley}, {Nasedkin}, {Abuter}, {Amorim}, {Benisty}, {Berger}, {Blunt},
  {Bonnet}, {Bourdarot}, {Caselli}, {Charnay}, {Choquet}, {Christiaens},
  {Cl{'e}net}, {Coud{'e} Du Foresto}, {Cridland}, {Dembet}, {Dexter},
  {Drescher}, {Duvert}, {Eckart}, {Eisenhauer}, {Gao}, {Garcia}, {Garcia
  Lopez}, {Gendron}, {Genzel}, {Gillessen}, {Girard}, {Haubois}, {Hei{ss}el},
  {Hippler}, {Houll{'e}}, {Hubert}, {Jocou}, {Kreidberg}, {Lapeyr{`e}re},
  {L{'e}na}, {Lutz}, {M{'e}nard}, {M{'e}rand}, {Molli{`e}re}, {Monnier},
  {Mouillet}, {Ott}, {Otten}, {Paladini}, {Paumard}, {Perraut}, {Perrin},
  {Pfuhl}, {Pourr{'e}}, {Pueyo}, {Rickman}, {Rousset}, {Rustamkulov},
  {Shimizu}, {Sing}, {Stadler}, {Straub}, {Straubmeier}, {Sturm}, {Tacconi},
  {van Dishoeck}, {Vincent}, {von Fellenberg}, {Widmann}, {Wieprecht},
  {Woillez}, {Yazici}, \& {The Gravity Collaboration}}]{2024A&A...691A.263M}
{Maire}, A.-L., {Leclerc}, A., {Balmer}, W.~O., {et~al.} 2024,
  {\mhref{http://doi.org/10.1051/0004-6361/202451184}{\aap}},
  {\href{https://ui.adsabs.harvard.edu/abs/2024A&A...691A.263M}{691}}{\href{https://ui.adsabs.harvard.edu/abs/2024A&A...691A.263M}{,
  A263}}

\bibitem[{E.~E. Mamajek \& L.~A. {Hillenbrand}(2008)Mamajek \&
  {Hillenbrand}}]{mam08-ages}
Mamajek, E.~E., \& {Hillenbrand}, L.~A. 2008,
  {\mhref{http://doi.org/10.1086/591785}{\apj}},
  {\href{http://adsabs.harvard.edu/abs/2008ApJ...687.1264M}{687}}{\href{http://adsabs.harvard.edu/abs/2008ApJ...687.1264M}{,
  1264}}

\bibitem[{M.~S. {Marley} {et~al.}(2007){Marley}, {Fortney}, {Hubickyj},
  {Bodenheimer}, \& {Lissauer}}]{2007ApJ...655..541M}
{Marley}, M.~S., {Fortney}, J.~J., {Hubickyj}, O., {Bodenheimer}, P., \&
  {Lissauer}, J.~J. 2007, {\mhref{http://doi.org/10.1086/509759}{\apj}},
  {\href{http://adsabs.harvard.edu/abs/2007ApJ...655..541M}{655}}{\href{http://adsabs.harvard.edu/abs/2007ApJ...655..541M}{,
  541}}

\bibitem[{M.~S. {Marley} {et~al.}(2021){Marley}, {Saumon}, {Visscher}, {Lupu},
  {Freedman}, {Morley}, {Fortney}, {Seay}, {Smith}, {Teal}, \&
  {Wang}}]{2021ApJ...920...85M}
{Marley}, M.~S., {Saumon}, D., {Visscher}, C., {et~al.} 2021,
  {\mhref{http://doi.org/10.3847/1538-4357/ac141d}{\apj}},
  {\href{https://ui.adsabs.harvard.edu/abs/2021ApJ...920...85M}{920}}{\href{https://ui.adsabs.harvard.edu/abs/2021ApJ...920...85M}{,
  85}}

\bibitem[{C.~Marois {et~al.}(2010)Marois, {Zuckerman}, {Konopacky},
  {Macintosh}, \& {Barman}}]{2010Natur.468.1080M}
Marois, C., {Zuckerman}, B., {Konopacky}, Q.~M., {Macintosh}, B., \& {Barman},
  T. 2010, {\mhref{http://doi.org/10.1038/nature09684}{\nat}},
  {\href{http://adsabs.harvard.edu/abs/2010Natur.468.1080M}{468}}{\href{http://adsabs.harvard.edu/abs/2010Natur.468.1080M}{,
  1080}}

\bibitem[{B.~D. Mason {et~al.}(2001)Mason, {Wycoff}, {Hartkopf}, {Douglass}, \&
  {Worley}}]{2001AJ....122.3466M}
Mason, B.~D., {Wycoff}, G.~L., {Hartkopf}, W.~I., {Douglass}, G.~G., \&
  {Worley}, C.~E. 2001, \aj,
  {\href{http://adsabs.harvard.edu/cgi-bin/nph-bib_query?bibcode=2001AJ....122.3466M&amp;db_key=AST}{122}}{\href{http://adsabs.harvard.edu/cgi-bin/nph-bib_query?bibcode=2001AJ....122.3466M&amp;db_key=AST}{,
  3466}}

\bibitem[{S.~{Mathur} {et~al.}(2019){Mathur}, {Garc{\'\i}a}, {Bugnet},
  {Santos}, {Santiago}, \& {Beck}}]{Mathur2019}
{Mathur}, S., {Garc{\'\i}a}, R.~A., {Bugnet}, L., {et~al.} 2019,
  {\mhref{http://doi.org/10.3389/fspas.2019.00046}{Frontiers in Astronomy and
  Space Sciences}},
  {\href{https://ui.adsabs.harvard.edu/abs/2019FrASS...6...46M}{6}}{\href{https://ui.adsabs.harvard.edu/abs/2019FrASS...6...46M}{,
  46}}

\bibitem[{E.~C. {Matthews} {et~al.}(2024){Matthews}, {Carter}, {Pathak},
  {Morley}, {Phillips}, {P.~M.}, {Feng}, {Bonse}, {Boogaard}, {Burt},
  {Crossfield}, {Douglas}, {Henning}, {Hom}, {Ko}, {Kasper}, {Lagrange}, {Petit
  dit de la Roche}, \& {Philipot}}]{2024Natur.633..789M}
{Matthews}, E.~C., {Carter}, A.~L., {Pathak}, P., {et~al.} 2024,
  {\mhref{http://doi.org/10.1038/s41586-024-07837-8}{\nat}},
  {\href{https://ui.adsabs.harvard.edu/abs/2024Natur.633..789M}{633}}{\href{https://ui.adsabs.harvard.edu/abs/2024Natur.633..789M}{,
  789}}

\bibitem[{S.~{Meibom} {et~al.}(2015){Meibom}, {Barnes}, {Platais}, {Gilliland},
  {Latham}, \& {Mathieu}}]{meibom2015-6819}
{Meibom}, S., {Barnes}, S.~A., {Platais}, I., {et~al.} 2015,
  {\mhref{http://doi.org/10.1038/nature14118}{\nat}},
  {\href{https://ui.adsabs.harvard.edu/abs/2015Natur.517..589M}{517}}{\href{https://ui.adsabs.harvard.edu/abs/2015Natur.517..589M}{,
  589}}

\bibitem[{D.~{Mesa} {et~al.}(2020){Mesa}, {D'Orazi}, {Vigan}, {Kitzmann},
  {Heng}, {Gratton}, {Desidera}, {Bonnefoy}, {Lavie}, {Maire}, {Peretti}, \&
  {Boccaletti}}]{2020MNRAS.495.4279M}
{Mesa}, D., {D'Orazi}, V., {Vigan}, A., {et~al.} 2020,
  {\mhref{http://doi.org/10.1093/mnras/staa1444}{mnras}},
  {\href{https://ui.adsabs.harvard.edu/abs/2020MNRAS.495.4279M}{495}}{\href{https://ui.adsabs.harvard.edu/abs/2020MNRAS.495.4279M}{,
  4279}}

\bibitem[{D.~{Mesa} {et~al.}(2023){Mesa}, {Gratton}, {Kervella}, {Bonavita},
  {Desidera}, {D'Orazi}, {Marino}, {Zurlo}, \&
  {Rigliaco}}]{2023A&A...672A..93M}
{Mesa}, D., {Gratton}, R., {Kervella}, P., {et~al.} 2023,
  {\mhref{http://doi.org/10.1051/0004-6361/202345865}{\aap}},
  {\href{https://ui.adsabs.harvard.edu/abs/2023A&A...672A..93M}{672}}{\href{https://ui.adsabs.harvard.edu/abs/2023A&A...672A..93M}{,
  A93}}

\bibitem[{T.~{Meshkat} {et~al.}(2015){Meshkat}, {Bonnefoy}, {Mamajek}, {Quanz},
  {Chauvin}, {Kenworthy}, {Rameau}, {Meyer}, {Lagrange}, {Lannier}, \&
  {Delorme}}]{2015MNRAS.453.2378M}
{Meshkat}, T., {Bonnefoy}, M., {Mamajek}, E.~E., {et~al.} 2015,
  {\mhref{http://doi.org/10.1093/mnras/stv1732}{\mnras}},
  {\href{http://adsabs.harvard.edu/abs/2015MNRAS.453.2378M}{453}}{\href{http://adsabs.harvard.edu/abs/2015MNRAS.453.2378M}{,
  2378}}

\bibitem[{S.~{Messina} {et~al.}(2001){Messina}, {Rodon{\`o}}, \&
  {Guinan}}]{Messina2001}
{Messina}, S., {Rodon{\`o}}, M., \& {Guinan}, E.~F. 2001,
  {\mhref{http://doi.org/10.1051/0004-6361:20000201}{\aap}},
  {\href{https://ui.adsabs.harvard.edu/abs/2001A&A...366..215M}{366}}{\href{https://ui.adsabs.harvard.edu/abs/2001A&A...366..215M}{,
  215}}

\bibitem[{G.~{Michaud} \& C.~R. {Proffitt}(1993){Michaud} \&
  {Proffitt}}]{Michaud1993}
{Michaud}, G., \& {Proffitt}, C.~R. 1993, in Astronomical Society of the
  Pacific Conference Series, Vol.~40, IAU Colloq. 137: Inside the Stars, ed.
  W.~W. {Weiss} \& A.~{Baglin},
  {\href{https://ui.adsabs.harvard.edu/abs/1993ASPC...40..246M}{246--259}}

\bibitem[{E.~{Michel} {et~al.}(2009){Michel}, {Samadi}, {Baudin}, {Barban},
  {Appourchaux}, \& {Auvergne}}]{Michel2009}
{Michel}, E., {Samadi}, R., {Baudin}, F., {et~al.} 2009,
  {\mhref{http://doi.org/10.1051/0004-6361:200810353}{\aap}},
  {\href{https://ui.adsabs.harvard.edu/abs/2009A&A...495..979M}{495}}{\href{https://ui.adsabs.harvard.edu/abs/2009A&A...495..979M}{,
  979}}

\bibitem[{J.~{Milli} {et~al.}(2017){Milli}, {Hibon}, {Christiaens}, {Choquet},
  {Bonnefoy}, {Kennedy}, {Wyatt}, {Absil}, {G{'o}mez Gonz{'a}lez}, {del Burgo},
  {Matr{`a}}, {Augereau}, {Boccaletti}, {Delacroix}, {Ertel}, {Dent},
  {Forsberg}, {Fusco}, {Girard}, {Habraken}, {Huby}, {Karlsson}, {Lagrange},
  {Mawet}, {Mouillet}, {Perrin}, {Pinte}, {Pueyo}, {Reyes}, {Soummer},
  {Surdej}, {Tarricq}, \& {Wahhaj}}]{2017A&A...597L...2M}
{Milli}, J., {Hibon}, P., {Christiaens}, V., {et~al.} 2017,
  {\mhref{http://doi.org/10.1051/0004-6361/201629908}{\aap}},
  {\href{https://ui.adsabs.harvard.edu/abs/2017A&A...597L...2M}{597}}{\href{https://ui.adsabs.harvard.edu/abs/2017A&A...597L...2M}{,
  L2}}

\bibitem[{N.~{Moedas} {et~al.}(2024){Moedas}, {Bossini}, {Deal}, \&
  {Cunha}}]{Moedas2024}
{Moedas}, N., {Bossini}, D., {Deal}, M., \& {Cunha}, M.~S. 2024,
  {\mhref{http://doi.org/10.1051/0004-6361/202348506}{\aap}},
  {\href{https://ui.adsabs.harvard.edu/abs/2024A&A...684A.113M}{684}}{\href{https://ui.adsabs.harvard.edu/abs/2024A&A...684A.113M}{,
  A113}}

\bibitem[{M.~H. {Montgomery} \& D.~{O'Donoghue}(1999){Montgomery} \&
  {O'Donoghue}}]{Montgomery1999}
{Montgomery}, M.~H., \& {O'Donoghue}, D. 1999, Delta Scuti Star Newsletter,
  {\href{https://ui.adsabs.harvard.edu/abs/1999DSSN...13...28M}{13}}{\href{https://ui.adsabs.harvard.edu/abs/1999DSSN...13...28M}{,
  28}}

\bibitem[{M.~Morháč {et~al.}(2003)Morháč, Matoušek, \&
  Kliman}]{morhac2003-gold}
Morháč, M., Matoušek, V., \& Kliman, J. 2003,
  {\mhref{http://doi.org/https://doi.org/10.1016/S1051-2004(02)00011-8}{Digital
  Signal Processing}}, 13, 144

\bibitem[{C.~V. Morley {et~al.}(2012)Morley, {Fortney}, {Marley}, {Visscher},
  {Saumon}, \& {Leggett}}]{2012ApJ...756..172M}
Morley, C.~V., {Fortney}, J.~J., {Marley}, M.~S., {et~al.} 2012,
  {\mhref{http://doi.org/10.1088/0004-637X/756/2/172}{\apj}},
  {\href{http://adsabs.harvard.edu/abs/2012ApJ...756..172M}{756}}{\href{http://adsabs.harvard.edu/abs/2012ApJ...756..172M}{,
  172}}

\bibitem[{C.~V. {Morley} {et~al.}(2024){Morley}, {Mukherjee}, {Marley},
  {Fortney}, {Visscher}, {Lupu}, {Gharib-Nezhad}, {Thorngren}, {Freedman}, \&
  {Batalha}}]{Morley2024}
{Morley}, C.~V., {Mukherjee}, S., {Marley}, M.~S., {et~al.} 2024,
  {\mhref{http://doi.org/10.3847/1538-4357/ad71d5}{\apj}},
  {\href{https://ui.adsabs.harvard.edu/abs/2024ApJ...975...59M}{975}}{\href{https://ui.adsabs.harvard.edu/abs/2024ApJ...975...59M}{,
  59}}

\bibitem[{T.~Nakajima {et~al.}(1995)Nakajima, {Oppenheimer}, {Kulkarni},
  {Golimowski}, {Matthews}, \& {Durrance}}]{1995Natur.378..463N}
Nakajima, T., {Oppenheimer}, B.~R., {Kulkarni}, S.~R., {et~al.} 1995, \nat,
  378, 463

\bibitem[{E.~{Nasedkin} {et~al.}(2024){Nasedkin}, {Molli{`e}re}, {Lacour},
  {Nowak}, {Kreidberg}, {Stolker}, {Wang}, {Balmer}, {Kammerer}, {Shangguan},
  {Abuter}, {Amorim}, {Asensio-Torres}, {Benisty}, {Berger}, {Beust}, {Blunt},
  {Boccaletti}, {Bonnefoy}, {Bonnet}, {Bordoni}, {Bourdarot}, {Brandner},
  {Cantalloube}, {Caselli}, {Charnay}, {Chauvin}, {Chavez}, {Choquet},
  {Christiaens}, {Cl{'e}net}, {Coud{'e} Du Foresto}, {Cridland}, {Davies},
  {Dembet}, {Dexter}, {Drescher}, {Duvert}, {Eckart}, {Eisenhauer}, {F{"o}rster
  Schreiber}, {Garcia}, {Garcia Lopez}, {Gendron}, {Genzel}, {Gillessen},
  {Girard}, {Grant}, {Haubois}, {Hei{ss}el}, {Henning}, {Hinkley}, {Hippler},
  {Houll{'e}}, {Hubert}, {Jocou}, {Keppler}, {Kervella}, {Kurtovic},
  {Lagrange}, {Lapeyr{`e}re}, {Le Bouquin}, {Lutz}, {Maire}, {Mang}, {Marleau},
  {M{'e}rand}, {Monnier}, {Mordasini}, {Ott}, {Otten}, {Paladini}, {Paumard},
  {Perraut}, {Perrin}, {Pfuhl}, {Pourr{'e}}, {Pueyo}, {Ribeiro}, {Rickman},
  {Ruffio}, {Rustamkulov}, {Shimizu}, {Sing}, {Stadler}, {Straub},
  {Straubmeier}, {Sturm}, {Tacconi}, {van Dishoeck}, {Vigan}, {Vincent}, {von
  Fellenberg}, {Widmann}, {Winterhalder}, {Woillez}, {Yazici}, \& {Gravity
  Collaboration}}]{2024A&A...687A.298N}
{Nasedkin}, E., {Molli{`e}re}, P., {Lacour}, S., {et~al.} 2024,
  {\mhref{http://doi.org/10.1051/0004-6361/202449328}{\aap}},
  {\href{https://ui.adsabs.harvard.edu/abs/2024A&A...687A.298N}{687}}{\href{https://ui.adsabs.harvard.edu/abs/2024A&A...687A.298N}{,
  A298}}

\bibitem[{M.~{Nowak} {et~al.}(2020){Nowak}, {Lacour}, {Lagrange}, {Rubini},
  {Wang}, {Stolker}, {Abuter}, {Amorim}, {Asensio-Torres}, {Baub{"o}ck},
  {Benisty}, {Berger}, {Beust}, {Blunt}, {Boccaletti}, {Bonnefoy}, {Bonnet},
  {Brandner}, {Cantalloube}, {Charnay}, {Choquet}, {Christiaens}, {Cl{'e}net},
  {Coud{'e} Du Foresto}, {Cridland}, {de Zeeuw}, {Dembet}, {Dexter},
  {Drescher}, {Duvert}, {Eckart}, {Eisenhauer}, {Gao}, {Garcia}, {Garcia
  Lopez}, {Gardner}, {Gendron}, {Genzel}, {Gillessen}, {Girard}, {Grandjean},
  {Haubois}, {Hei{ss}el}, {Henning}, {Hinkley}, {Hippler}, {Horrobin},
  {Houll{'e}}, {Hubert}, {Jim{'e}nez-Rosales}, {Jocou}, {Kammerer}, {Kervella},
  {Keppler}, {Kreidberg}, {Kulikauskas}, {Lapeyr{`e}re}, {Le Bouquin},
  {L{'e}na}, {M{'e}rand}, {Maire}, {Molli{`e}re}, {Monnier}, {Mouillet},
  {M{"u}ller}, {Nasedkin}, {Ott}, {Otten}, {Paumard}, {Paladini}, {Perraut},
  {Perrin}, {Pueyo}, {Pfuhl}, {Rameau}, {Rodet}, {Rodr{'i}guez-Coira},
  {Rousset}, {Scheithauer}, {Shangguan}, {Stadler}, {Straub}, {Straubmeier},
  {Sturm}, {Tacconi}, {van Dishoeck}, {Vigan}, {Vincent}, {von Fellenberg},
  {Ward-Duong}, {Widmann}, {Wieprecht}, {Wiezorrek}, {Woillez}, \& {GRAVITY
  Collaboration}}]{2020A&A...642L...2N}
{Nowak}, M., {Lacour}, S., {Lagrange}, A.-M., {et~al.} 2020,
  {\mhref{http://doi.org/10.1051/0004-6361/202039039}{\aap}},
  {\href{https://ui.adsabs.harvard.edu/abs/2020A&A...642L...2N}{642}}{\href{https://ui.adsabs.harvard.edu/abs/2020A&A...642L...2N}{,
  L2}}

\bibitem[{J.~M.~J. {Ong} {et~al.}(2021){Ong}, {Basu}, \&
  {McKeever}}]{Ong2021-surferr}
{Ong}, J.~M.~J., {Basu}, S., \& {McKeever}, J.~M. 2021,
  {\mhref{http://doi.org/10.3847/1538-4357/abc7c1}{\apj}},
  {\href{https://ui.adsabs.harvard.edu/abs/2021ApJ...906...54O}{906}}{\href{https://ui.adsabs.harvard.edu/abs/2021ApJ...906...54O}{,
  54}}

\bibitem[{B.~{Paxton} {et~al.}(2011){Paxton}, {Bildsten}, {Dotter}, {Herwig},
  {Lesaffre}, \& {Timmes}}]{mesa2011}
{Paxton}, B., {Bildsten}, L., {Dotter}, A., {et~al.} 2011,
  {\mhref{http://doi.org/10.1088/0067-0049/192/1/3}{\apjs}},
  {\href{https://ui.adsabs.harvard.edu/abs/2011ApJS..192....3P}{192}}{\href{https://ui.adsabs.harvard.edu/abs/2011ApJS..192....3P}{,
  3}}

\bibitem[{B.~{Paxton} {et~al.}(2013){Paxton}, {Cantiello}, {Arras}, {Bildsten},
  {Brown}, {Dotter}, {Mankovich}, {Montgomery}, {Stello}, {Timmes}, \&
  {Townsend}}]{mesa2013}
{Paxton}, B., {Cantiello}, M., {Arras}, P., {et~al.} 2013,
  {\mhref{http://doi.org/10.1088/0067-0049/208/1/4}{\apjs}},
  {\href{https://ui.adsabs.harvard.edu/abs/2013ApJS..208....4P}{208}}{\href{https://ui.adsabs.harvard.edu/abs/2013ApJS..208....4P}{,
  4}}

\bibitem[{B.~{Paxton} {et~al.}(2015){Paxton}, {Marchant}, {Schwab}, {Bauer},
  {Bildsten}, {Cantiello}, {Dessart}, {Farmer}, {Hu}, {Langer}, {Townsend},
  {Townsley}, \& {Timmes}}]{mesa2015}
{Paxton}, B., {Marchant}, P., {Schwab}, J., {et~al.} 2015,
  {\mhref{http://doi.org/10.1088/0067-0049/220/1/15}{\apjs}},
  {\href{https://ui.adsabs.harvard.edu/abs/2015ApJS..220...15P}{220}}{\href{https://ui.adsabs.harvard.edu/abs/2015ApJS..220...15P}{,
  15}}

\bibitem[{B.~{Paxton} {et~al.}(2018){Paxton}, {Schwab}, {Bauer}, {Bildsten},
  {Blinnikov}, {Duffell}, {Farmer}, {Goldberg}, {Marchant}, {Sorokina},
  {Thoul}, {Townsend}, \& {Timmes}}]{mesa2018}
{Paxton}, B., {Schwab}, J., {Bauer}, E.~B., {et~al.} 2018,
  {\mhref{http://doi.org/10.3847/1538-4365/aaa5a8}{\apjs}},
  {\href{https://ui.adsabs.harvard.edu/abs/2018ApJS..234...34P}{234}}{\href{https://ui.adsabs.harvard.edu/abs/2018ApJS..234...34P}{,
  34}}

\bibitem[{B.~{Paxton} {et~al.}(2019){Paxton}, {Smolec}, {Schwab}, {Gautschy},
  {Bildsten}, {Cantiello}, {Dotter}, {Farmer}, {Goldberg}, {Jermyn}, {Kanbur},
  {Marchant}, {Thoul}, {Townsend}, {Wolf}, {Zhang}, \& {Timmes}}]{mesa2019}
{Paxton}, B., {Smolec}, R., {Schwab}, J., {et~al.} 2019,
  {\mhref{http://doi.org/10.3847/1538-4365/ab2241}{\apjs}},
  {\href{https://ui.adsabs.harvard.edu/abs/2019ApJS..243...10P}{243}}{\href{https://ui.adsabs.harvard.edu/abs/2019ApJS..243...10P}{,
  10}}

\bibitem[{M.~J. {Pecaut} \& E.~E. {Mamajek}(2016){Pecaut} \&
  {Mamajek}}]{2016MNRAS.461..794P}
{Pecaut}, M.~J., \& {Mamajek}, E.~E. 2016,
  {\mhref{http://doi.org/10.1093/mnras/stw1300}{\mnras}},
  {\href{http://adsabs.harvard.edu/abs/2016MNRAS.461..794P}{461}}{\href{http://adsabs.harvard.edu/abs/2016MNRAS.461..794P}{,
  794}}

\bibitem[{S.~{Peretti} {et~al.}(2019){Peretti}, {S{\'e}gransan}, {Lavie},
  {Desidera}, {Maire}, {D'Orazi}, {Vigan}, {Baudino}, {Cheetham}, {Janson},
  {Chauvin}, {Hagelberg}, {Menard}, {Heng}, {Udry}, {Boccaletti}, {Daemgen},
  {Le Coroller}, {Mesa}, {Rouan}, {Samland}, {Schmidt}, {Zurlo}, {Bonnefoy},
  {Feldt}, {Gratton}, {Lagrange}, {Langlois}, {Meyer}, {Carbillet}, {Carle},
  {De Caprio}, {Gluck}, {Hugot}, {Magnard}, {Moulin}, {Pavlov}, {Pragt},
  {Rabou}, {Ramos}, {Rousset}, {Sevin}, {Soenke}, {Stadler}, {Weber}, \&
  {Wildi}}]{2019A&A...631A.107P}
{Peretti}, S., {S{\'e}gransan}, D., {Lavie}, B., {et~al.} 2019,
  {\mhref{http://doi.org/10.1051/0004-6361/201732454}{\aap}},
  {\href{https://ui.adsabs.harvard.edu/abs/2019A&A...631A.107P}{631}}{\href{https://ui.adsabs.harvard.edu/abs/2019A&A...631A.107P}{,
  A107}}

\bibitem[{M.~W. {Phillips} {et~al.}(2020{\natexlab{a}}){Phillips}, {Tremblin},
  {Baraffe}, {Chabrier}, {Allard}, {Spiegelman}, {Goyal}, {Drummond}, \&
  {H{'e}brard}}]{2020A&A...637A..38P}
{Phillips}, M.~W., {Tremblin}, P., {Baraffe}, I., {et~al.} 2020{\natexlab{a}},
  {\mhref{http://doi.org/10.1051/0004-6361/201937381}{aap}},
  {\href{https://ui.adsabs.harvard.edu/abs/2020A&A...637A..38P}{637}}{\href{https://ui.adsabs.harvard.edu/abs/2020A&A...637A..38P}{,
  A38}}

\bibitem[{M.~W. {Phillips} {et~al.}(2020{\natexlab{b}}){Phillips}, {Tremblin},
  {Baraffe}, {Chabrier}, {Allard}, {Spiegelman}, {Goyal}, {Drummond}, \&
  {H{\'e}brard}}]{Phillips2020}
---. 2020{\natexlab{b}},
  {\mhref{http://doi.org/10.1051/0004-6361/201937381}{\aap}},
  {\href{https://ui.adsabs.harvard.edu/abs/2020A&A...637A..38P}{637}}{\href{https://ui.adsabs.harvard.edu/abs/2020A&A...637A..38P}{,
  A38}}

\bibitem[{D.~J. Pinfield {et~al.}(2006)Pinfield, {Jones}, {Lucas}, {Kendall},
  {Folkes}, {Day-Jones}, {Chappelle}, \& {Steele}}]{2006MNRAS.368.1281P}
Pinfield, D.~J., {Jones}, H.~R.~A., {Lucas}, P.~W., {et~al.} 2006,
  {\mhref{http://doi.org/10.1111/j.1365-2966.2006.10213.x}{\mnras}},
  {\href{http://adsabs.harvard.edu/abs/2006MNRAS.368.1281P}{368}}{\href{http://adsabs.harvard.edu/abs/2006MNRAS.368.1281P}{,
  1281}}

\bibitem[{D.~Potter {et~al.}(2002)Potter, {Mart{\' i}n}, {Cushing}, {Baudoz},
  {Brandner}, {Guyon}, \& {Neuh{\" a}user}}]{2002ApJ...567L.133P}
Potter, D., {Mart{\' i}n}, E.~L., {Cushing}, M.~C., {et~al.} 2002, \apjl, 567,
  L133

\bibitem[{S.~H. Pravdo {et~al.}(2005)Pravdo, {Shaklan}, \&
  {Lloyd}}]{2005ApJ...630..528P}
Pravdo, S.~H., {Shaklan}, S.~B., \& {Lloyd}, J. 2005,
  {\mhref{http://doi.org/10.1086/431967}{\apj}},
  {\href{http://adsabs.harvard.edu/abs/2005ApJ...630..528P}{630}}{\href{http://adsabs.harvard.edu/abs/2005ApJ...630..528P}{,
  528}}

\bibitem[{R.~{Radick} \& A.~{Pevtsov}(2018){Radick} \& {Pevtsov}}]{Radick2018}
{Radick}, R., \& {Pevtsov}, A. 2018, {HK\_Project\_v1995\_NSO}, Harvard
  Dataverse dataset, \dodoi{10.7910/DVN/ZRJ6NT}

\bibitem[{L.~M. {Rebull} {et~al.}(2016){Rebull}, {Stauffer}, {Bouvier}, {Cody},
  {Hillenbrand}, {Soderblom}, {Valenti}, {Barrado}, {Bouy}, {Ciardi},
  {Pinsonneault}, {Stassun}, {Micela}, {Aigrain}, {Vrba}, {Somers},
  {Christiansen}, {Gillen}, \& {Collier Cameron}}]{rebull2016-pleiades}
{Rebull}, L.~M., {Stauffer}, J.~R., {Bouvier}, J., {et~al.} 2016,
  {\mhref{http://doi.org/10.3847/0004-6256/152/5/113}{\aj}},
  {\href{https://ui.adsabs.harvard.edu/abs/2016AJ....152..113R}{152}}{\href{https://ui.adsabs.harvard.edu/abs/2016AJ....152..113R}{,
  113}}

\bibitem[{G.~R. {Ricker} {et~al.}(2015){Ricker}, {Winn}, {Vanderspek},
  {Latham}, {Bakos}, {Bean}, {Berta-Thompson}, {Brown}, {Buchhave}, {Butler},
  {Butler}, {Chaplin}, {Charbonneau}, {Christensen-Dalsgaard}, {Clampin},
  {Deming}, {Doty}, {De Lee}, {Dressing}, {Dunham}, {Endl}, {Fressin}, {Ge},
  {Henning}, {Holman}, {Howard}, {Ida}, {Jenkins}, {Jernigan}, {Johnson},
  {Kaltenegger}, {Kawai}, {Kjeldsen}, {Laughlin}, {Levine}, {Lin}, {Lissauer},
  {MacQueen}, {Marcy}, {McCullough}, {Morton}, {Narita}, {Paegert}, {Palle},
  {Pepe}, {Pepper}, {Quirrenbach}, {Rinehart}, {Sasselov}, {Sato}, {Seager},
  {Sozzetti}, {Stassun}, {Sullivan}, {Szentgyorgyi}, {Torres}, {Udry}, \&
  {Villasenor}}]{Ricker2015}
{Ricker}, G.~R., {Winn}, J.~N., {Vanderspek}, R., {et~al.} 2015,
  {\mhref{http://doi.org/10.1117/1.JATIS.1.1.014003}{Journal of Astronomical
  Telescopes, Instruments, and Systems}},
  {\href{https://ui.adsabs.harvard.edu/abs/2015JATIS...1a4003R}{1}}{\href{https://ui.adsabs.harvard.edu/abs/2015JATIS...1a4003R}{,
  014003}}

\bibitem[{E.~L. {Rickman} {et~al.}(2020){Rickman}, {S{'e}gransan}, {Hagelberg},
  {Beuzit}, {Cheetham}, {Delisle}, {Forveille}, \&
  {Udry}}]{2020A&A...635A.203R}
{Rickman}, E.~L., {S{'e}gransan}, D., {Hagelberg}, J., {et~al.} 2020,
  {\mhref{http://doi.org/10.1051/0004-6361/202037524}{\aap}},
  {\href{https://ui.adsabs.harvard.edu/abs/2020A&A...635A.203R}{635}}{\href{https://ui.adsabs.harvard.edu/abs/2020A&A...635A.203R}{,
  A203}}

\bibitem[{E.~L. {Rickman} {et~al.}(2019){Rickman}, {S{'e}gransan}, {Marmier},
  {Udry}, {Bouchy}, {Lovis}, {Mayor}, {Pepe}, {Queloz}, {Santos}, {Allart},
  {Bonvin}, {Bratschi}, {Cersullo}, {Chazelas}, {Choplin}, {Conod}, {Deline},
  {Delisle}, {Dos Santos}, {Figueira}, {Giles}, {Girard}, {Lavie}, {Martin},
  {Motalebi}, {Nielsen}, {Osborn}, {Ottoni}, {Raimbault}, {Rey}, {Roger},
  {Seidel}, {Stalport}, {Su{'a}rez Mascare{~n}o}, {Triaud}, {Turner}, {Weber},
  \& {Wyttenbach}}]{2019A&A...625A..71R}
{Rickman}, E.~L., {S{'e}gransan}, D., {Marmier}, M., {et~al.} 2019,
  {\mhref{http://doi.org/10.1051/0004-6361/201935356}{\aap}},
  {\href{https://ui.adsabs.harvard.edu/abs/2019A&A...625A..71R}{625}}{\href{https://ui.adsabs.harvard.edu/abs/2019A&A...625A..71R}{,
  A71}}

\bibitem[{E.~L. {Rickman} {et~al.}(2024){Rickman}, {Ceva}, {Matthews},
  {S{'e}gransan}, {Bowler}, {Forveille}, {Franson}, {Hagelberg}, {Udry}, \&
  {Vigan}}]{2024A&A...684A..88R}
{Rickman}, E.~L., {Ceva}, W., {Matthews}, E.~C., {et~al.} 2024,
  {\mhref{http://doi.org/10.1051/0004-6361/202347906}{\aap}},
  {\href{https://ui.adsabs.harvard.edu/abs/2024A&A...684A..88R}{684}}{\href{https://ui.adsabs.harvard.edu/abs/2024A&A...684A..88R}{,
  A88}}

\bibitem[{M.~{Riello} {et~al.}(2021){Riello}, {De Angeli}, {Evans},
  {Montegriffo}, {Carrasco}, {Busso}, {Palaversa}, {Burgess}, {Diener},
  {Davidson}, {Rowell}, {Fabricius}, {Jordi}, {Bellazzini}, {Pancino},
  {Harrison}, {Cacciari}, {van Leeuwen}, {Hambly}, {Hodgkin}, {Osborne},
  {Altavilla}, {Barstow}, {Brown}, {Castellani}, {Cowell}, {De Luise},
  {Gilmore}, {Giuffrida}, {Hidalgo}, {Holland}, {Marinoni}, {Pagani},
  {Piersimoni}, {Pulone}, {Ragaini}, {Rainer}, {Richards}, {Sanna}, {Walton},
  {Weiler}, \& {Yoldas}}]{Riello2021}
{Riello}, M., {De Angeli}, F., {Evans}, D.~W., {et~al.} 2021,
  {\mhref{http://doi.org/10.1051/0004-6361/202039587}{\aap}},
  {\href{https://ui.adsabs.harvard.edu/abs/2021A&A...649A...3R}{649}}{\href{https://ui.adsabs.harvard.edu/abs/2021A&A...649A...3R}{,
  A3}}

\bibitem[{L.~J. {Rosenthal} {et~al.}(2021){Rosenthal}, {Fulton}, {Hirsch},
  {Isaacson}, {Howard}, {Dedrick}, {Sherstyuk}, {Blunt}, {Petigura}, {Knutson},
  {Behmard}, {Chontos}, {Crepp}, {Crossfield}, {Dalba}, {Fischer}, {Henry},
  {Kane}, {Kosiarek}, {Marcy}, {Rubenzahl}, {Weiss}, \&
  {Wright}}]{Rosenthal2021}
{Rosenthal}, L.~J., {Fulton}, B.~J., {Hirsch}, L.~A., {et~al.} 2021,
  {\mhref{http://doi.org/10.3847/1538-4365/abe23c}{\apjs}},
  {\href{https://ui.adsabs.harvard.edu/abs/2021ApJS..255....8R}{255}}{\href{https://ui.adsabs.harvard.edu/abs/2021ApJS..255....8R}{,
  8}}

\bibitem[{I.~W. {Roxburgh}(2016){Roxburgh}}]{Roxburgh2016}
{Roxburgh}, I.~W. 2016,
  {\mhref{http://doi.org/10.1051/0004-6361/201526593}{\aap}},
  {\href{https://ui.adsabs.harvard.edu/abs/2016A&A...585A..63R}{585}}{\href{https://ui.adsabs.harvard.edu/abs/2016A&A...585A..63R}{,
  A63}}

\bibitem[{I.~W. {Roxburgh} \& S.~V. {Vorontsov}(1994){Roxburgh} \&
  {Vorontsov}}]{Roxburgh1994}
{Roxburgh}, I.~W., \& {Vorontsov}, S.~V. 1994,
  {\mhref{http://doi.org/10.1093/mnras/268.1.143}{\mnras}},
  {\href{https://ui.adsabs.harvard.edu/abs/1994MNRAS.268..143R}{268}}{\href{https://ui.adsabs.harvard.edu/abs/1994MNRAS.268..143R}{,
  143}}

\bibitem[{I.~W. {Roxburgh} \& S.~V. {Vorontsov}(2003){Roxburgh} \&
  {Vorontsov}}]{Roxburgh2003}
---. 2003, {\mhref{http://doi.org/10.1051/0004-6361:20031318}{\aap}},
  {\href{https://ui.adsabs.harvard.edu/abs/2003A&A...411..215R}{411}}{\href{https://ui.adsabs.harvard.edu/abs/2003A&A...411..215R}{,
  215}}

\bibitem[{A.~{Sanghi} {et~al.}(2023){Sanghi}, {Liu}, {Best}, {Dupuy}, {Siverd},
  {Zhang}, {Hurt}, {Magnier}, {Aller}, \& {Deacon}}]{2023ApJ...959...63S}
{Sanghi}, A., {Liu}, M.~C., {Best}, W. M.~J., {et~al.} 2023,
  {\mhref{http://doi.org/10.3847/1538-4357/acff66}{\apj}},
  {\href{https://ui.adsabs.harvard.edu/abs/2023ApJ...959...63S}{959}}{\href{https://ui.adsabs.harvard.edu/abs/2023ApJ...959...63S}{,
  63}}

\bibitem[{D.~{Saumon} \& M.~S. {Marley}(2008){Saumon} \& {Marley}}]{Saumon2008}
{Saumon}, D., \& {Marley}, M.~S. 2008,
  {\mhref{http://doi.org/10.1086/592734}{\apj}},
  {\href{https://ui.adsabs.harvard.edu/abs/2008ApJ...689.1327S}{689}}{\href{https://ui.adsabs.harvard.edu/abs/2008ApJ...689.1327S}{,
  1327}}

\bibitem[{D.~Saumon {et~al.}(2006)Saumon, {Marley}, {Cushing}, {Leggett},
  {Roellig}, {Lodders}, \& {Freedman}}]{2006ApJ...647..552S}
Saumon, D., {Marley}, M.~S., {Cushing}, M.~C., {et~al.} 2006,
  {\mhref{http://doi.org/10.1086/505419}{\apj}},
  {\href{http://adsabs.harvard.edu/cgi-bin/nph-bib_query?bibcode=2006ApJ...647..552S&db_key=AST}{647}}{\href{http://adsabs.harvard.edu/cgi-bin/nph-bib_query?bibcode=2006ApJ...647..552S&db_key=AST}{,
  552}}

\bibitem[{N.~{Saunders} {et~al.}(2024){Saunders}, {van Saders}, {Lyttle},
  {Metcalfe}, {Li}, {Davies}, {Hall}, {Ball}, {Townsend}, {Creevey}, \&
  {Dodds}}]{saunders2024-ro}
{Saunders}, N., {van Saders}, J.~L., {Lyttle}, A.~J., {et~al.} 2024,
  {\mhref{http://doi.org/10.3847/1538-4357/ad1516}{\apj}},
  {\href{https://ui.adsabs.harvard.edu/abs/2024ApJ...962..138S}{962}}{\href{https://ui.adsabs.harvard.edu/abs/2024ApJ...962..138S}{,
  138}}

\bibitem[{M.~{Sayeed} {et~al.}(2025){Sayeed}, {Huber}, {Chontos}, \&
  {Li}}]{Sayeed2025}
{Sayeed}, M., {Huber}, D., {Chontos}, A., \& {Li}, Y. 2025,
  {\mhref{http://doi.org/10.48550/arXiv.2503.15599}{arXiv
  e-prints}}{\href{https://ui.adsabs.harvard.edu/abs/2025arXiv250315599S}{,
  arXiv:2503.15599}}

\bibitem[{J.~D. {Scargle}(1982){Scargle}}]{scargle1982}
{Scargle}, J.~D. 1982, {\mhref{http://doi.org/10.1086/160554}{\apj}},
  {\href{https://ui.adsabs.harvard.edu/abs/1982ApJ...263..835S}{263}}{\href{https://ui.adsabs.harvard.edu/abs/1982ApJ...263..835S}{,
  835}}

\bibitem[{P.~H. {Scherrer} {et~al.}(1983){Scherrer}, {Wilcox},
  {Christensen-Dalsgaard}, \& {Gough}}]{Scherrer1983}
{Scherrer}, P.~H., {Wilcox}, J.~M., {Christensen-Dalsgaard}, J., \& {Gough},
  D.~O. 1983, {\mhref{http://doi.org/10.1007/BF00145547}{\solphys}},
  {\href{https://ui.adsabs.harvard.edu/abs/1983SoPh...82...75S}{82}}{\href{https://ui.adsabs.harvard.edu/abs/1983SoPh...82...75S}{,
  75}}

\bibitem[{A.~C. {Schneider} {et~al.}(2023){Schneider}, {Munn}, {Vrba},
  {Bruursema}, {Dahm}, {Williams}, {Liu}, \& {Dorland}}]{2023AJ....166..103S}
{Schneider}, A.~C., {Munn}, J.~A., {Vrba}, F.~J., {et~al.} 2023,
  {\mhref{http://doi.org/10.3847/1538-3881/ace9bf}{\aj}},
  {\href{https://ui.adsabs.harvard.edu/abs/2023AJ....166..103S}{166}}{\href{https://ui.adsabs.harvard.edu/abs/2023AJ....166..103S}{,
  103}}

\bibitem[{R.-D. Scholz {et~al.}(2003)Scholz, {McCaughrean}, {Lodieu}, \&
  {Kuhlbrodt}}]{2003A&A...398L..29S}
Scholz, R.-D., {McCaughrean}, M.~J., {Lodieu}, N., \& {Kuhlbrodt}, B. 2003,
  {\mhref{http://doi.org/10.1051/0004-6361:20021847}{\aap}},
  {\href{http://adsabs.harvard.edu/cgi-bin/nph-bib_query?bibcode=2003A%26A...398L..29S&db_key=AST}{398}}{\href{http://adsabs.harvard.edu/cgi-bin/nph-bib_query?bibcode=2003A%26A...398L..29S&db_key=AST}{,
  L29}}

\bibitem[{E.~{Serabyn} {et~al.}(2009){Serabyn}, {Mawet}, {Bloemhof},
  {Haguenauer}, {Mennesson}, {Wallace}, \& {Hickey}}]{Serabyn2009}
{Serabyn}, E., {Mawet}, D., {Bloemhof}, E., {et~al.} 2009,
  {\mhref{http://doi.org/10.1088/0004-637X/696/1/40}{\apj}},
  {\href{https://ui.adsabs.harvard.edu/abs/2009ApJ...696...40S}{696}}{\href{https://ui.adsabs.harvard.edu/abs/2009ApJ...696...40S}{,
  40}}

\bibitem[{A.~{Serenelli} {et~al.}(2017){Serenelli}, {Johnson}, {Huber},
  {Pinsonneault}, {Ball}, {Tayar}, {Silva Aguirre}, {Basu}, {Troup}, {Hekker},
  {Kallinger}, {Stello}, {Davies}, {Lund}, {Mathur}, {Mosser}, {Stassun},
  {Chaplin}, {Elsworth}, {Garc{\'\i}a}, {Handberg}, {Holtzman}, {Hearty},
  {Garc{\'\i}a-Hern{\'a}ndez}, {Gaulme}, \& {Zamora}}]{Serenelli2017}
{Serenelli}, A., {Johnson}, J., {Huber}, D., {et~al.} 2017,
  {\mhref{http://doi.org/10.3847/1538-4365/aa97df}{\apjs}},
  {\href{https://ui.adsabs.harvard.edu/abs/2017ApJS..233...23S}{233}}{\href{https://ui.adsabs.harvard.edu/abs/2017ApJS..233...23S}{,
  23}}

\bibitem[{M.~{Service} {et~al.}(2016){Service}, {Lu}, {Campbell}, {Sitarski},
  {Ghez}, \& {Anderson}}]{Service2016}
{Service}, M., {Lu}, J.~R., {Campbell}, R., {et~al.} 2016,
  {\mhref{http://doi.org/10.1088/1538-3873/128/967/095004}{\pasp}},
  {\href{https://ui.adsabs.harvard.edu/abs/2016PASP..128i5004S}{128}}{\href{https://ui.adsabs.harvard.edu/abs/2016PASP..128i5004S}{,
  095004}}

\bibitem[{V.~{Silva Aguirre} {et~al.}(2015){Silva Aguirre}, {Davies}, {Basu},
  {Christensen-Dalsgaard}, {Creevey}, {Metcalfe}, {Bedding}, {Casagrande},
  {Handberg}, {Lund}, {Nissen}, {Chaplin}, {Huber}, {Serenelli}, {Stello}, {Van
  Eylen}, {Campante}, {Elsworth}, {Gilliland}, {Hekker}, {Karoff}, {Kawaler},
  {Kjeldsen}, \& {Lundkvist}}]{silvaaguirre2015-kages}
{Silva Aguirre}, V., {Davies}, G.~R., {Basu}, S., {et~al.} 2015,
  {\mhref{http://doi.org/10.1093/mnras/stv1388}{\mnras}},
  {\href{https://ui.adsabs.harvard.edu/abs/2015MNRAS.452.2127S}{452}}{\href{https://ui.adsabs.harvard.edu/abs/2015MNRAS.452.2127S}{,
  2127}}

\bibitem[{V.~{Silva Aguirre} {et~al.}(2017){Silva Aguirre}, {Lund}, {Antia},
  {Ball}, {Basu}, {Christensen-Dalsgaard}, {Lebreton}, {Reese}, {Verma},
  {Casagrande}, {Justesen}, {Mosumgaard}, {Chaplin}, {Bedding}, {Davies},
  {Handberg}, {Houdek}, {Huber}, {Kjeldsen}, {Latham}, {White}, {Coelho},
  {Miglio}, \& {Rendle}}]{silvaaguirre2017-legacy}
{Silva Aguirre}, V., {Lund}, M.~N., {Antia}, H.~M., {et~al.} 2017,
  {\mhref{http://doi.org/10.3847/1538-4357/835/2/173}{\apj}},
  {\href{https://ui.adsabs.harvard.edu/abs/2017ApJ...835..173S}{835}}{\href{https://ui.adsabs.harvard.edu/abs/2017ApJ...835..173S}{,
  173}}

\bibitem[{M.~F. {Skrutskie} {et~al.}(1987){Skrutskie}, {Forrest}, \&
  {Shure}}]{1987BAAS...19.1128S}
{Skrutskie}, M.~F., {Forrest}, W.~J., \& {Shure}, M.~A. 1987, in Bulletin of
  the American Astronomical Society, Vol.~19,
  {\href{https://ui.adsabs.harvard.edu/abs/1987BAAS...19.1128S}{1128}}

\bibitem[{G.~{Somers} {et~al.}(2017){Somers}, {Stauffer}, {Rebull}, {Cody}, \&
  {Pinsonneault}}]{somers2017}
{Somers}, G., {Stauffer}, J., {Rebull}, L., {Cody}, A.~M., \& {Pinsonneault},
  M. 2017, {\mhref{http://doi.org/10.3847/1538-4357/aa93ed}{\apj}},
  {\href{https://ui.adsabs.harvard.edu/abs/2017ApJ...850..134S}{850}}{\href{https://ui.adsabs.harvard.edu/abs/2017ApJ...850..134S}{,
  134}}

\bibitem[{T.~{Sonoi} {et~al.}(2015){Sonoi}, {Samadi}, {Belkacem}, {Ludwig},
  {Caffau}, \& {Mosser}}]{sonoi2015}
{Sonoi}, T., {Samadi}, R., {Belkacem}, K., {et~al.} 2015,
  {\mhref{http://doi.org/10.1051/0004-6361/201526838}{\aap}},
  {\href{https://ui.adsabs.harvard.edu/abs/2015A&A...583A.112S}{583}}{\href{https://ui.adsabs.harvard.edu/abs/2015A&A...583A.112S}{,
  A112}}

\bibitem[{F.~{Spada} \& A.~C. {Lanzafame}(2020){Spada} \& {Lanzafame}}]{sl20}
{Spada}, F., \& {Lanzafame}, A.~C. 2020,
  {\mhref{http://doi.org/10.1051/0004-6361/201936384}{\aap}},
  {\href{https://ui.adsabs.harvard.edu/abs/2020A&A...636A..76S}{636}}{\href{https://ui.adsabs.harvard.edu/abs/2020A&A...636A..76S}{,
  A76}}

\bibitem[{K.~R. {Sreenivas} {et~al.}(2024){Sreenivas}, {Bedding}, {Li},
  {Huber}, {Crawford}, {Stello}, \& {Yu}}]{Sreenivas2024}
{Sreenivas}, K.~R., {Bedding}, T.~R., {Li}, Y., {et~al.} 2024,
  {\mhref{http://doi.org/10.1093/mnras/stae991}{\mnras}},
  {\href{https://ui.adsabs.harvard.edu/abs/2024MNRAS.530.3477S}{530}}{\href{https://ui.adsabs.harvard.edu/abs/2024MNRAS.530.3477S}{,
  3477}}

\bibitem[{S.~{Sulis} {et~al.}(2023){Sulis}, {Lendl}, {Cegla}, {Rodr{\'\i}guez
  D{\'\i}az}, {Bigot}, {Van Grootel}, {Bekkelien}, {Cameron}, {Maxted},
  {Simon}, {Lovis}, {Scandariato}, {Bruno}, {Nardiello}, {Bonfanti},
  {Fridlund}, {Persson}, {Salmon}, {Sousa}, {Wilson}, {Krenn}, {Hoyer},
  {Santerne}, {Ehrenreich}, {Alibert}, {Alonso}, {Anglada}, {B{\'a}rczy},
  {Barrado y Navascues}, {Barros}, {Baumjohann}, {Beck}, {Beck}, {Benz},
  {Billot}, {Bonfils}, {Borsato}, {Brandeker}, {Broeg}, {Cabrera}, {Charnoz},
  {Corral van Damme}, {Csizmadia}, {Davies}, {Deleuil}, {Deline}, {Delrez},
  {Demangeon}, {Demory}, {Erikson}, {Fortier}, {Fossati}, {Gandolfi}, {Gillon},
  {G{\"u}del}, {Heng}, {Isaak}, {Kiss}, {Laskar}, {Lecavelier des Etangs},
  {Magrin}, {Munari}, {Nascimbeni}, {Olofsson}, {Ottensamer}, {Pagano},
  {Pall{\'e}}, {Peter}, {Piotto}, {Pollacco}, {Queloz}, {Ragazzoni}, {Rando},
  {Rauer}, {Ribas}, {Rieder}, {Santos}, {S{\'e}gransan}, {Smith},
  {Steinberger}, {Steller}, {Szab{\'o}}, {Thomas}, {Udry}, {Walton}, \&
  {Wolter}}]{Sulis2023}
{Sulis}, S., {Lendl}, M., {Cegla}, H.~M., {et~al.} 2023,
  {\mhref{http://doi.org/10.1051/0004-6361/202244223}{\aap}},
  {\href{https://ui.adsabs.harvard.edu/abs/2023A&A...670A..24S}{670}}{\href{https://ui.adsabs.harvard.edu/abs/2023A&A...670A..24S}{,
  A24}}

\bibitem[{M.~{Tassoul}(1980){Tassoul}}]{Tassoul1980}
{Tassoul}, M. 1980, {\mhref{http://doi.org/10.1086/190678}{\apjs}},
  {\href{https://ui.adsabs.harvard.edu/abs/1980ApJS...43..469T}{43}}{\href{https://ui.adsabs.harvard.edu/abs/1980ApJS...43..469T}{,
  469}}

\bibitem[{M.~{Tassoul}(1990){Tassoul}}]{Tassoul1994}
---. 1990, {\mhref{http://doi.org/10.1086/168988}{\apj}},
  {\href{https://ui.adsabs.harvard.edu/abs/1990ApJ...358..313T}{358}}{\href{https://ui.adsabs.harvard.edu/abs/1990ApJ...358..313T}{,
  313}}

\bibitem[{J.~{Tayar} {et~al.}(2022){Tayar}, {Claytor}, {Huber}, \& {van
  Saders}}]{Tayar2022}
{Tayar}, J., {Claytor}, Z.~R., {Huber}, D., \& {van Saders}, J. 2022,
  {\mhref{http://doi.org/10.3847/1538-4357/ac4bbc}{\apj}},
  {\href{https://ui.adsabs.harvard.edu/abs/2022ApJ...927...31T}{927}}{\href{https://ui.adsabs.harvard.edu/abs/2022ApJ...927...31T}{,
  31}}

\bibitem[{C.~Thalmann {et~al.}(2009)Thalmann {et~al.}}]{2009ApJ...707L.123T}
Thalmann, C., {et~al.} 2009,
  {\mhref{http://doi.org/10.1088/0004-637X/707/2/L123}{\apjl}},
  {\href{http://adsabs.harvard.edu/abs/2009ApJ...707L.123T}{707}}{\href{http://adsabs.harvard.edu/abs/2009ApJ...707L.123T}{,
  L123}}

\bibitem[{A.~A. {Thoul} {et~al.}(1994){Thoul}, {Bahcall}, \&
  {Loeb}}]{thoul1994}
{Thoul}, A.~A., {Bahcall}, J.~N., \& {Loeb}, A. 1994,
  {\mhref{http://doi.org/10.1086/173695}{\apj}},
  {\href{https://ui.adsabs.harvard.edu/abs/1994ApJ...421..828T}{421}}{\href{https://ui.adsabs.harvard.edu/abs/1994ApJ...421..828T}{,
  828}}

\bibitem[{T.~L. {Tobin} {et~al.}(2024){Tobin}, {Currie}, {Li}, {Chilcote},
  {Brandt}, {Lacy}, {Kuzuhara}, {Vincent}, {El Morsy}, {Deo}, {Williams},
  {Guyon}, {Lozi}, {Vievard}, {Skaf}, {Ahn}, {Groff}, {Kasdin}, {Uyama},
  {Tamura}, {Gibbs}, {Lewis}, {Bowens-Rubin}, {Salama}, {An}, \&
  {Chen}}]{2024AJ....167..205T}
{Tobin}, T.~L., {Currie}, T., {Li}, Y., {et~al.} 2024,
  {\mhref{http://doi.org/10.3847/1538-3881/ad3077}{\aj}},
  {\href{https://ui.adsabs.harvard.edu/abs/2024AJ....167..205T}{167}}{\href{https://ui.adsabs.harvard.edu/abs/2024AJ....167..205T}{,
  205}}

\bibitem[{A.~{Tokovinin}(2014){Tokovinin}}]{2014AJ....147...86T}
{Tokovinin}, A. 2014, {\mhref{http://doi.org/10.1088/0004-6256/147/4/86}{\aj}},
  {\href{https://ui.adsabs.harvard.edu/abs/2014AJ....147...86T}{147}}{\href{https://ui.adsabs.harvard.edu/abs/2014AJ....147...86T}{,
  86}}

\bibitem[{A.~T. {Tokunaga} {et~al.}(2002){Tokunaga}, {Simons}, \&
  {Vacca}}]{Tokunaga2002}
{Tokunaga}, A.~T., {Simons}, D.~A., \& {Vacca}, W.~D. 2002,
  {\mhref{http://doi.org/10.1086/338545}{\pasp}},
  {\href{https://ui.adsabs.harvard.edu/abs/2002PASP..114..180T}{114}}{\href{https://ui.adsabs.harvard.edu/abs/2002PASP..114..180T}{,
  180}}

\bibitem[{R.~H.~D. {Townsend} \& S.~A. {Teitler}(2013){Townsend} \&
  {Teitler}}]{gyre2013}
{Townsend}, R.~H.~D., \& {Teitler}, S.~A. 2013,
  {\mhref{http://doi.org/10.1093/mnras/stt1533}{\mnras}},
  {\href{https://ui.adsabs.harvard.edu/abs/2013MNRAS.435.3406T}{435}}{\href{https://ui.adsabs.harvard.edu/abs/2013MNRAS.435.3406T}{,
  3406}}

\bibitem[{R.~{Trampedach} {et~al.}(2014){Trampedach}, {Stein},
  {Christensen-Dalsgaard}, {Nordlund}, \& {Asplund}}]{trampedach2014}
{Trampedach}, R., {Stein}, R.~F., {Christensen-Dalsgaard}, J., {Nordlund},
  {\r{A}}., \& {Asplund}, M. 2014,
  {\mhref{http://doi.org/10.1093/mnras/stu2084}{\mnras}},
  {\href{https://ui.adsabs.harvard.edu/abs/2014MNRAS.445.4366T}{445}}{\href{https://ui.adsabs.harvard.edu/abs/2014MNRAS.445.4366T}{,
  4366}}

\bibitem[{P.~{Tremblin} {et~al.}(2015){Tremblin}, {Amundsen}, {Mourier},
  {Baraffe}, {Chabrier}, {Drummond}, {Homeier}, \&
  {Venot}}]{2015ApJ...804L..17T}
{Tremblin}, P., {Amundsen}, D.~S., {Mourier}, P., {et~al.} 2015,
  {\mhref{http://doi.org/10.1088/2041-8205/804/1/L17}{\apjl}},
  {\href{https://ui.adsabs.harvard.edu/abs/2015ApJ...804L..17T}{804}}{\href{https://ui.adsabs.harvard.edu/abs/2015ApJ...804L..17T}{,
  L17}}

\bibitem[{R.~K. {Ulrich}(1986){Ulrich}}]{Ulrich1986}
{Ulrich}, R.~K. 1986, {\mhref{http://doi.org/10.1086/184700}{\apjl}},
  {\href{https://ui.adsabs.harvard.edu/abs/1986ApJ...306L..37U}{306}}{\href{https://ui.adsabs.harvard.edu/abs/1986ApJ...306L..37U}{,
  L37}}

\bibitem[{J.~L. {van Saders} {et~al.}(2016){van Saders}, {Ceillier},
  {Metcalfe}, {Silva Aguirre}, {Pinsonneault}, {Garc{\'\i}a}, {Mathur}, \&
  {Davies}}]{vanSaders2016-wmb}
{van Saders}, J.~L., {Ceillier}, T., {Metcalfe}, T.~S., {et~al.} 2016,
  {\mhref{http://doi.org/10.1038/nature16168}{\nat}},
  {\href{https://ui.adsabs.harvard.edu/abs/2016Natur.529..181V}{529}}{\href{https://ui.adsabs.harvard.edu/abs/2016Natur.529..181V}{,
  181}}

\bibitem[{J.~L. {van Saders} \& M.~H. {Pinsonneault}(2013){van Saders} \&
  {Pinsonneault}}]{vanSaders2013-sg}
{van Saders}, J.~L., \& {Pinsonneault}, M.~H. 2013,
  {\mhref{http://doi.org/10.1088/0004-637X/776/2/67}{\apj}},
  {\href{https://ui.adsabs.harvard.edu/abs/2013ApJ...776...67V}{776}}{\href{https://ui.adsabs.harvard.edu/abs/2013ApJ...776...67V}{,
  67}}

\bibitem[{A.~{Vigan} {et~al.}(2016){Vigan}, {Bonnefoy}, {Ginski}, {Beust},
  {Galicher}, {Janson}, {Baudino}, {Buenzli}, {Hagelberg}, {D'Orazi},
  {Desidera}, {Maire}, {Gratton}, {Sauvage}, {Chauvin}, {Thalmann}, {Malo},
  {Salter}, {Zurlo}, {Antichi}, {Baruffolo}, {Baudoz}, {Blanchard},
  {Boccaletti}, {Beuzit}, {Carle}, {Claudi}, {Costille}, {Delboulb{\'e}},
  {Dohlen}, {Dominik}, {Feldt}, {Fusco}, {Gluck}, {Girard}, {Giro}, {Gry},
  {Henning}, {Hubin}, {Hugot}, {Jaquet}, {Kasper}, {Lagrange}, {Langlois}, {Le
  Mignant}, {Llored}, {Madec}, {Martinez}, {Mawet}, {Mesa}, {Milli},
  {Mouillet}, {Moulin}, {Moutou}, {Orign{\'e}}, {Pavlov}, {Perret}, {Petit},
  {Pragt}, {Puget}, {Rabou}, {Rochat}, {Roelfsema}, {Salasnich}, {Schmid},
  {Sevin}, {Siebenmorgen}, {Smette}, {Stadler}, {Suarez}, {Turatto}, {Udry},
  {Vakili}, {Wahhaj}, {Weber}, \& {Wildi}}]{2016A&A...587A..55V}
{Vigan}, A., {Bonnefoy}, M., {Ginski}, C., {et~al.} 2016,
  {\mhref{http://doi.org/10.1051/0004-6361/201526465}{\aap}},
  {\href{https://ui.adsabs.harvard.edu/abs/2016A&A...587A..55V}{587}}{\href{https://ui.adsabs.harvard.edu/abs/2016A&A...587A..55V}{,
  A55}}

\bibitem[{K.~{Wagner} {et~al.}(2020){Wagner}, {Apai}, {Kasper}, {McClure},
  {Robberto}, \& {Currie}}]{2020ApJ...902L...6W}
{Wagner}, K., {Apai}, D., {Kasper}, M., {et~al.} 2020,
  {\mhref{http://doi.org/10.3847/2041-8213/abb94e}{\apjl}},
  {\href{https://ui.adsabs.harvard.edu/abs/2020ApJ...902L...6W}{902}}{\href{https://ui.adsabs.harvard.edu/abs/2020ApJ...902L...6W}{,
  L6}}

\bibitem[{J.~{Wang} {et~al.}(2022){Wang}, {Kolecki}, {Ruffio}, {Wang}, {Mawet},
  {Baker}, {Bartos}, {Blake}, {Bond}, {Calvin}, {Cetre}, {Delorme}, {Doppmann},
  {Echeverri}, {Finnerty}, {Fitzgerald}, {Jovanovic}, {Liu}, {Lopez}, {Morris},
  {Pai Asnodkar}, {Pezzato}, {Ragland}, {Roy}, {Ruane}, {Sappey}, {Schofield},
  {Skemer}, {Venenciano}, {Kent Wallace}, {Wallack}, {Wizinowich}, \&
  {Xuan}}]{Wangj2022}
{Wang}, J., {Kolecki}, J.~R., {Ruffio}, J.-B., {et~al.} 2022,
  {\mhref{http://doi.org/10.3847/1538-3881/ac56e2}{\aj}},
  {\href{https://ui.adsabs.harvard.edu/abs/2022AJ....163..189W}{163}}{\href{https://ui.adsabs.harvard.edu/abs/2022AJ....163..189W}{,
  189}}

\bibitem[{K.~{Ward-Duong} {et~al.}(2021){Ward-Duong}, {Patience}, {Follette},
  {De Rosa}, {Rameau}, {Marley}, {Saumon}, {Nielsen}, {Rajan}, {Greenbaum},
  {Lee}, {Wang}, {Czekala}, {Duch{\^e}ne}, {Macintosh}, {Ammons}, {Bailey},
  {Barman}, {Bulger}, {Chen}, {Chilcote}, {Cotten}, {Doyon}, {Esposito},
  {Fitzgerald}, {Gerard}, {Goodsell}, {Graham}, {Hibon}, {Hom}, {Hung},
  {Ingraham}, {Kalas}, {Konopacky}, {Larkin}, {Maire}, {Marchis}, {Marois},
  {Metchev}, {Millar-Blanchaer}, {Oppenheimer}, {Palmer}, {Perrin}, {Poyneer},
  {Pueyo}, {Rantakyr{\"o}}, {Ren}, {Ruffio}, {Savransky}, {Schneider},
  {Sivaramakrishnan}, {Song}, {Soummer}, {Tallis}, {Thomas}, {Wallace},
  {Wiktorowicz}, \& {Wolff}}]{2021AJ....161....5W}
{Ward-Duong}, K., {Patience}, J., {Follette}, K., {et~al.} 2021,
  {\mhref{http://doi.org/10.3847/1538-3881/abc263}{\aj}},
  {\href{https://ui.adsabs.harvard.edu/abs/2021AJ....161....5W}{161}}{\href{https://ui.adsabs.harvard.edu/abs/2021AJ....161....5W}{,
  5}}

\bibitem[{A.~{Weiss} \& H.~{Schlattl}(2008){Weiss} \& {Schlattl}}]{garstec2008}
{Weiss}, A., \& {Schlattl}, H. 2008,
  {\mhref{http://doi.org/10.1007/s10509-007-9606-5}{\apss}},
  {\href{https://ui.adsabs.harvard.edu/abs/2008Ap&SS.316...99W}{316}}{\href{https://ui.adsabs.harvard.edu/abs/2008Ap&SS.316...99W}{,
  99}}

\bibitem[{T.~R. {White} {et~al.}(2011){White}, {Bedding}, {Stello},
  {Christensen-Dalsgaard}, {Huber}, \& {Kjeldsen}}]{White2011}
{White}, T.~R., {Bedding}, T.~R., {Stello}, D., {et~al.} 2011,
  {\mhref{http://doi.org/10.1088/0004-637X/743/2/161}{\apj}},
  {\href{https://ui.adsabs.harvard.edu/abs/2011ApJ...743..161W}{743}}{\href{https://ui.adsabs.harvard.edu/abs/2011ApJ...743..161W}{,
  161}}

\bibitem[{T.~R. {White} {et~al.}(2012){White}, {Bedding}, {Gruberbauer},
  {Benomar}, {Stello}, {Appourchaux}, {Chaplin}, {Christensen-Dalsgaard},
  {Elsworth}, {Garc{\'\i}a}, {Hekker}, {Huber}, {Kjeldsen}, {Mosser},
  {Kinemuchi}, {Mullally}, \& {Still}}]{White2012}
{White}, T.~R., {Bedding}, T.~R., {Gruberbauer}, M., {et~al.} 2012,
  {\mhref{http://doi.org/10.1088/2041-8205/751/2/L36}{\apjl}},
  {\href{https://ui.adsabs.harvard.edu/abs/2012ApJ...751L..36W}{751}}{\href{https://ui.adsabs.harvard.edu/abs/2012ApJ...751L..36W}{,
  L36}}

\bibitem[{T.~O. {Winterhalder} {et~al.}(2024){Winterhalder}, {Lacour},
  {M{'e}rand}, {Kammerer}, {Maire}, {Stolker}, {Pourr{'e}}, {Babusiaux},
  {Glindemann}, {Abuter}, {Amorim}, {Asensio-Torres}, {Balmer}, {Benisty},
  {Berger}, {Beust}, {Blunt}, {Boccaletti}, {Bonnefoy}, {Bonnet}, {Bordoni},
  {Bourdarot}, {Brandner}, {Cantalloube}, {Caselli}, {Charnay}, {Chauvin},
  {Chavez}, {Choquet}, {Christiaens}, {Cl{'e}net}, {Coud{'e} du Foresto},
  {Cridland}, {Davies}, {Dembet}, {Dexter}, {Drescher}, {Duvert}, {Eckart},
  {Eisenhauer}, {F{"o}rster Schreiber}, {Garcia}, {Garcia Lopez}, {Gardner},
  {Gendron}, {Genzel}, {Gillessen}, {Girard}, {Grant}, {Haubois}, {Hei{ss}el},
  {Henning}, {Hinkley}, {Hippler}, {Houll{'e}}, {Hubert}, {Jocou}, {Keppler},
  {Kervella}, {Kreidberg}, {Kurtovic}, {Lagrange}, {Lapeyr{`e}re}, {Le
  Bouquin}, {Lutz}, {Mang}, {Marleau}, {Molli{`e}re}, {Monnier}, {Mordasini},
  {Mouillet}, {Nasedkin}, {Nowak}, {Ott}, {Otten}, {Paladini}, {Paumard},
  {Perraut}, {Perrin}, {Pfuhl}, {Pueyo}, {Ribeiro}, {Rickman}, {Rustamkulov},
  {Shangguan}, {Shimizu}, {Sing}, {Stadler}, {Straub}, {Straubmeier}, {Sturm},
  {Tacconi}, {van Dishoeck}, {Vigan}, {Vincent}, {von Fellenberg}, {Wang},
  {Widmann}, {Woillez}, \& {Yazici}}]{2024A&A...688A..44W}
{Winterhalder}, T.~O., {Lacour}, S., {M{'e}rand}, A., {et~al.} 2024,
  {\mhref{http://doi.org/10.1051/0004-6361/202450018}{\aap}},
  {\href{https://ui.adsabs.harvard.edu/abs/2024A&A...688A..44W}{688}}{\href{https://ui.adsabs.harvard.edu/abs/2024A&A...688A..44W}{,
  A44}}

\bibitem[{T.~O. {Winterhalder} {et~al.}(2025){Winterhalder}, {Kammerer},
  {Lacour}, {M{'e}rand}, {Nowak}, {Stolker}, {Balmer}, {Marleau}, {Abuter},
  {Amorim}, {Asensio-Torres}, {Berger}, {Beust}, {Blunt}, {Bonnefoy}, {Bonnet},
  {Bordoni}, {Bourdarot}, {Brandner}, {Cantalloube}, {Caselli}, {Charnay},
  {Chauvin}, {Chavez}, {Choquet}, {Christiaens}, {Cl{'e}net}, {Coud{'e} du
  Foresto}, {Cridland}, {Davies}, {Dembet}, {Dexter}, {Drescher}, {Duvert},
  {Eckart}, {Eisenhauer}, {F{"o}rster Schreiber}, {Garcia}, {Garcia Lopez},
  {Gardner}, {Gendron}, {Genzel}, {Gillessen}, {Girard}, {Grant}, {Haubois},
  {Hei{ss}el}, {Henning}, {Hinkley}, {Hippler}, {Houll{'e}}, {Hubert}, {Jocou},
  {Keppler}, {Kervella}, {Kreidberg}, {Kurtovic}, {Lagrange}, {Lapeyr{`e}re},
  {Le Bouquin}, {Lutz}, {Maire}, {Mang}, {Molli{`e}re}, {Mordasini},
  {Mouillet}, {Nasedkin}, {Ott}, {Otten}, {Paladini}, {Paumard}, {Perraut},
  {Perrin}, {Pourr{'e}}, {Pueyo}, {C Ribeiro}, {Rickman}, {Rustamkulov},
  {Shangguan}, {Shimizu}, {Sing}, {Stadler}, {Straub}, {Straubmeier}, {Sturm},
  {Tacconi}, {van Dishoeck}, {Vigan}, {Vincent}, {von Fellenberg}, {Wang},
  {Widmann}, {Woillez}, \& {Yazici}}]{2025A&A...700A...4W}
{Winterhalder}, T.~O., {Kammerer}, J., {Lacour}, S., {et~al.} 2025,
  {\mhref{http://doi.org/10.1051/0004-6361/202554766}{\aap}},
  {\href{https://ui.adsabs.harvard.edu/abs/2025A&A...700A...4W}{700}}{\href{https://ui.adsabs.harvard.edu/abs/2025A&A...700A...4W}{,
  A4}}

\bibitem[{J.~T. {Wright}(2018){Wright}}]{Wright2018}
{Wright}, J.~T. 2018, in Handbook of Exoplanets, ed. H.~J. {Deeg} \& J.~A.
  {Belmonte}, {\href{https://ui.adsabs.harvard.edu/abs/2018haex.bookE...4W}{4}}

\bibitem[{J.~W. {Xuan} {et~al.}(2024){Xuan}, {M{'e}rand}, {Thompson}, {Zhang},
  {Lacour}, {Blakely}, {Mawet}, {Oppenheimer}, {Kammerer}, {Batygin}, {Sanghi},
  {Wang}, {Ruffio}, {Liu}, {Knutson}, {Brandner}, {Burgasser}, {Rickman},
  {Bowens-Rubin}, {Salama}, {Balmer}, {Blunt}, {Bourdarot}, {Caselli},
  {Chauvin}, {Davies}, {Drescher}, {Eckart}, {Eisenhauer}, {Fabricius},
  {Feuchtgruber}, {Finger}, {F{"o}rster Schreiber}, {Garcia}, {Genzel},
  {Gillessen}, {Grant}, {Hartl}, {Hau{ss}mann}, {Henning}, {Hinkley},
  {H{"o}nig}, {Horrobin}, {Houll{'e}}, {Janson}, {Kervella}, {Kral},
  {Kreidberg}, {Le Bouquin}, {Lutz}, {Mang}, {Marleau}, {Millour}, {More},
  {Nowak}, {Ott}, {Otten}, {Paumard}, {Rabien}, {Rau}, {Ribeiro}, {Sadun
  Bordoni}, {Sauter}, {Shangguan}, {Shimizu}, {Sykes}, {Soulain}, {Spezzano},
  {Straubmeier}, {Stolker}, {Sturm}, {Subroweit}, {Tacconi}, {van Dishoeck},
  {Vigan}, {Widmann}, {Wieprecht}, {Winterhalder}, \&
  {Woillez}}]{2024Natur.634.1070X}
{Xuan}, J.~W., {M{'e}rand}, A., {Thompson}, W., {et~al.} 2024,
  {\mhref{http://doi.org/10.1038/s41586-024-08064-x}{\nat}},
  {\href{https://ui.adsabs.harvard.edu/abs/2024Natur.634.1070X}{634}}{\href{https://ui.adsabs.harvard.edu/abs/2024Natur.634.1070X}{,
  1070}}

\bibitem[{J.~{Zhang} {et~al.}(2024){Zhang}, {Huber}, {Weiss}, {Xuan}, {Burt},
  {Dai}, {Saunders}, {Petigura}, {Rubenzahl}, {Winn}, {Wang}, {Van Zandt},
  {Brodheim}, {Claytor}, {Crossfield}, {Deich}, {Fulton}, {Gibson},
  {Halverson}, {Hill}, {Holden}, {Householder}, {Howard}, {Isaacson}, {Kaye},
  {Lanclos}, {Laher}, {Lubin}, {Payne}, {Roy}, {Schwab}, {Shaum}, {Walawender},
  {Wishnow}, \& {Yeh}}]{Zhangjw2024}
{Zhang}, J., {Huber}, D., {Weiss}, L.~M., {et~al.} 2024,
  {\mhref{http://doi.org/10.3847/1538-3881/ad86c4}{\aj}},
  {\href{https://ui.adsabs.harvard.edu/abs/2024AJ....168..295Z}{168}}{\href{https://ui.adsabs.harvard.edu/abs/2024AJ....168..295Z}{,
  295}}

\bibitem[{Z.~{Zhang} {et~al.}(2020){Zhang}, {Liu}, {Hermes}, {Magnier},
  {Marley}, {Tremblay}, {Tucker}, {Do}, {Payne}, \& {Shappee}}]{Zhangzj2020}
{Zhang}, Z., {Liu}, M.~C., {Hermes}, J.~J., {et~al.} 2020,
  {\mhref{http://doi.org/10.3847/1538-4357/ab765c}{\apj}},
  {\href{https://ui.adsabs.harvard.edu/abs/2020ApJ...891..171Z}{891}}{\href{https://ui.adsabs.harvard.edu/abs/2020ApJ...891..171Z}{,
  171}}

\bibitem[{Y.~{Zhou} {et~al.}(2021){Zhou}, {Nordlander}, {Casagrande}, {Joyce},
  {Li}, {Amarsi}, {Reggiani}, \& {Asplund}}]{zhouyx2021}
{Zhou}, Y., {Nordlander}, T., {Casagrande}, L., {et~al.} 2021,
  {\mhref{http://doi.org/10.1093/mnras/stab337}{\mnras}},
  {\href{https://ui.adsabs.harvard.edu/abs/2021MNRAS.503...13Z}{503}}{\href{https://ui.adsabs.harvard.edu/abs/2021MNRAS.503...13Z}{,
  13}}

\bibitem[{A.~{Zurlo} {et~al.}(2016){Zurlo}, {Vigan}, {Galicher}, {Maire},
  {Mesa}, {Gratton}, {Chauvin}, {Kasper}, {Moutou}, {Bonnefoy}, {Desidera},
  {Abe}, {Apai}, {Baruffolo}, {Baudoz}, {Baudrand}, {Beuzit}, {Blancard},
  {Boccaletti}, {Cantalloube}, {Carle}, {Cascone}, {Charton}, {Claudi},
  {Costille}, {de Caprio}, {Dohlen}, {Dominik}, {Fantinel}, {Feautrier},
  {Feldt}, {Fusco}, {Gigan}, {Girard}, {Gisler}, {Gluck}, {Gry}, {Henning},
  {Hugot}, {Janson}, {Jaquet}, {Lagrange}, {Langlois}, {Llored}, {Madec},
  {Magnard}, {Martinez}, {Maurel}, {Mawet}, {Meyer}, {Milli},
  {Moeller-Nilsson}, {Mouillet}, {Orign{\'e}}, {Pavlov}, {Petit}, {Puget},
  {Quanz}, {Rabou}, {Ramos}, {Rousset}, {Roux}, {Salasnich}, {Salter},
  {Sauvage}, {Schmid}, {Soenke}, {Stadler}, {Suarez}, {Turatto}, {Udry},
  {Vakili}, {Wahhaj}, {Wildi}, \& {Antichi}}]{2016A&A...587A..57Z}
{Zurlo}, A., {Vigan}, A., {Galicher}, R., {et~al.} 2016,
  {\mhref{http://doi.org/10.1051/0004-6361/201526835}{\aap}},
  {\href{https://ui.adsabs.harvard.edu/abs/2016A&A...587A..57Z}{587}}{\href{https://ui.adsabs.harvard.edu/abs/2016A&A...587A..57Z}{,
  A57}}

\end{thebibliography}
\bibliographystyle{aasjournal-compact}

\clearpage

\appendix

\section{Dynamical Orbit Fit}\label{sec:dyn}

\begin{figure*}
    \centering
    \includegraphics[width=0.49\linewidth]{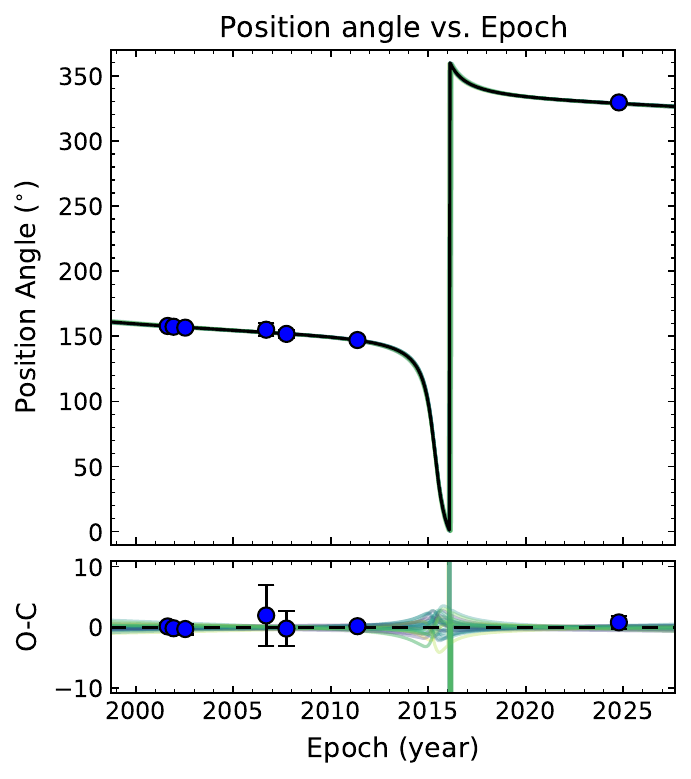}
    \includegraphics[width=0.49\linewidth]{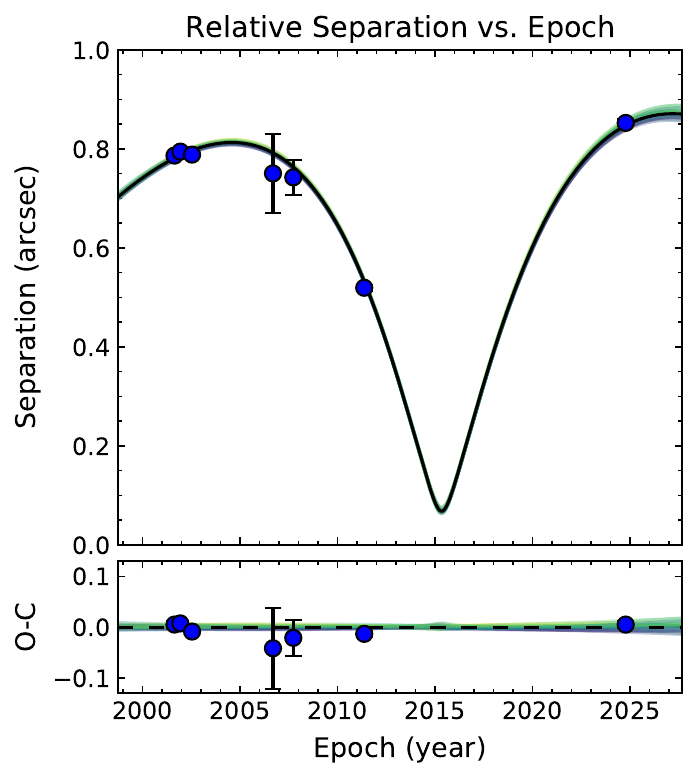}
    \includegraphics[width=0.98\linewidth]{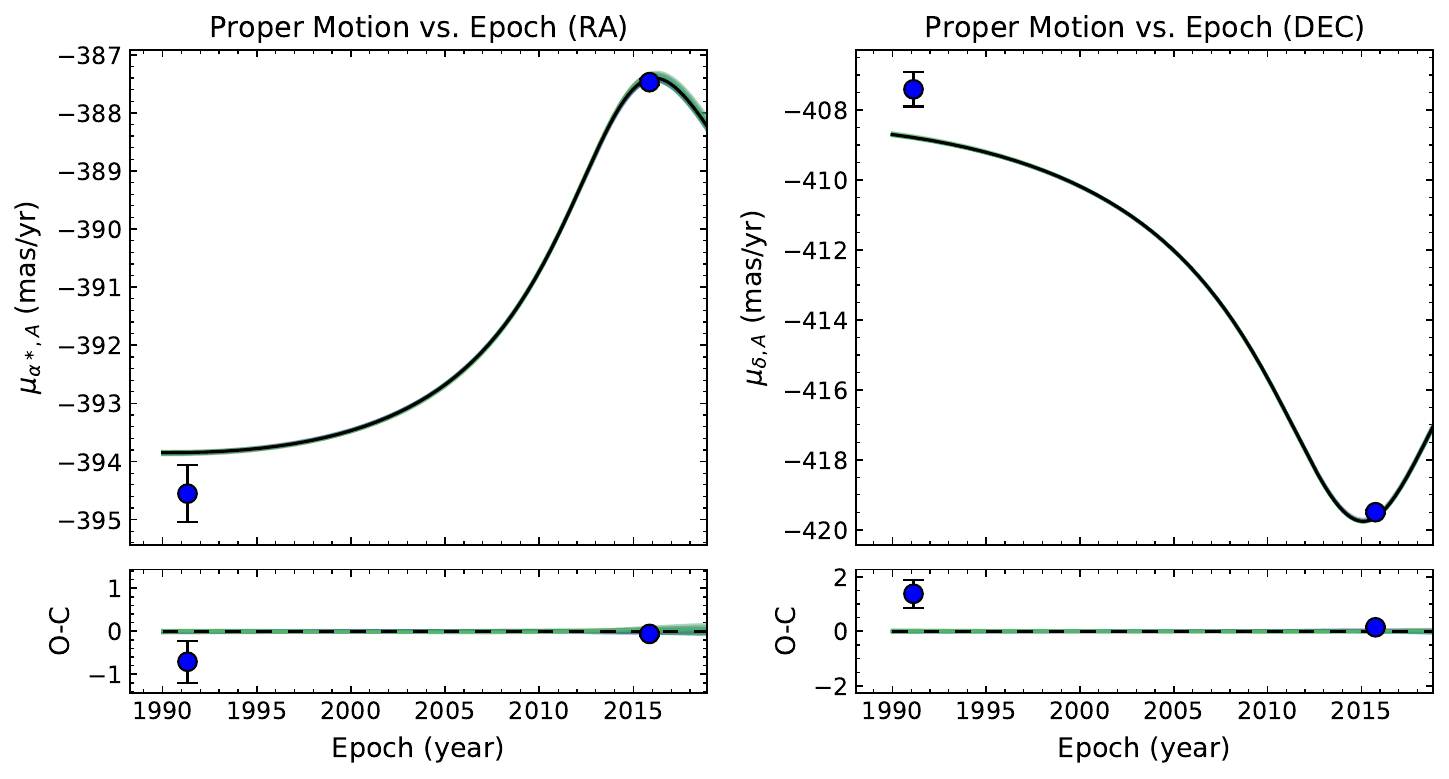}
    \caption{Relative and absolute astrometry measurements of the HR~7672 system, fitted with a dynamical model jointly constrained by the long-baseline radial velocities. Residuals relative to the best-fit model are also shown. }
    \label{fig:astrometry}
\end{figure*}

\begin{figure*}
    \centering
    \includegraphics[width=0.5\linewidth]{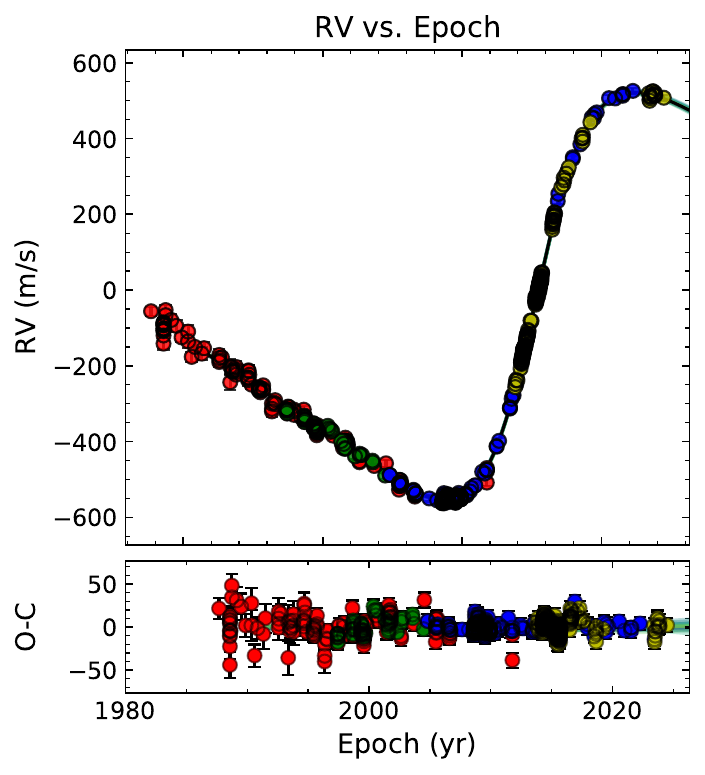}
    \caption{Top: Long-baseline radial velocity measurements of HR~7672 A, fitted with a dynamical model jointly constrained by astrometric data. Colored symbols indicate radial velocities measured by different instruments: Lick-Fischer, HIRES (pre and post upgrade), and APF. Bottom: residuals relative to the best-fit model.}
    \label{fig:rv}
\end{figure*}

Figures~\ref{fig:astrometry} and~\ref{fig:rv} present the best-fitting dynamical orbit model to the astrometric and long-baseline radial velocity data. The corresponding parameter estimates are listed in Table~\ref{table:mcmc}.

\begin{deluxetable*}{lcc}
\Large
\tablecaption{Orbital parameters estimated for HR~7672AB \label{table:mcmc}}
\setlength{\tabcolsep}{0.10in}
\tablewidth{0pt}
\tablehead{
\colhead{Parameter}              &
\colhead{Prior} &
\colhead{Median $\pm$1$\sigma$}}
\startdata
\multicolumn{3}{c}{Stellar parameters} \\[3pt]
\cline{1-3}
\multicolumn{3}{c}{} \\[-5pt]
Host-star mass $\rm{M_{A}}\ (\Msun{}) $                                      &   -      &      ${1.111}_{-0.017}^{+0.017}$                                                                 \\[3pt]
Companion  mass $\rm{M_{B}}\ (\Mjup{})$                                     &  -     &   ${75.39}_{-0.67}^{+0.67}$                    \\[3pt]
\cline{1-3}
\multicolumn{3}{c}{Orbital parameters} \\[3pt]
\cline{1-3}
\multicolumn{3}{c}{} \\[-5pt]
Semi-major axis $\rm a_B$ (au)       &                                -                &   ${16.88}_{-0.10}^{+0.10}$               \\[3pt]
Orbital period $P_B$ (yr)       &                                     -         & ${63.77}_{-0.61}^{+0.63}$                                                                                           \\[3pt]
Inclination $i_B$ (deg)         &                -                          &  ${97.73}_{-0.31}^{+0.31}$                 \\[3pt]
$\sqrt{e_B}\sin{\omega_B}$                           &        -                     & ${-0.6828}_{-0.0021}^{+0.0021}$               \\[3pt]
$\sqrt{e_B}\cos{\omega_B}$                           &         -                    & ${-0.0450}_{-0.0055}^{+0.0054}$         \\[3pt]
Eccentricity $e_B$                                      &          $\mathcal{U}(0,0.99)$             &  ${0.4683}_{-0.0031}^{+0.0032}$                                                                 \\[3pt]
Mean longitude at $t_{\rm ref}=2455197.5$~JD, $\lambda_{\rm ref}$ (deg) &    -        &     ${237.29}_{-0.18}^{+0.18}$   \\[3pt]
Longitide of the ascending node $\Omega_B$ (deg)                                 &   -   &   ${330.88}_{-0.21}^{+0.21}$           \\[3pt]
Parallax (mas)                                                              &  $\mathcal{N}(\varpi_{\rm DR3} ,\sigma[\varpi_{\rm DR3}])^{*}$     &       ${56.2724}_{-0.0094}^{+0.0094}$             \\[3pt]
Argument of periastron $\omega_B$ (deg)                                   &    -   &     ${266.23}_{-0.45}^{+0.45}$                                                              \\[3pt]
Time of periastron $T_0=t_{\rm ref}-P\frac{\lambda-\omega}{360 }$ (JD)&  -   &   ${2457070}_{-11}^{+11}$             \\[3pt]
\cline{1-3}
\multicolumn{3}{c}{} \\[-5pt]
\multicolumn{3}{c}{Other Parameters} \\[1pt]
\cline{1-3}
\multicolumn{3}{c}{} \\[-5pt]
Lick Fischer RV zero point (m\,s$^{-1}$)   &                    -                & $-307.8^{+2.4}_{-2.4}$                             \\[3pt]
HIRES pre RV zero point (m\,s$^{-1}$)    &                      -             & $-417.1^{+2.2}_{-2.2}$                                \\[3pt]
HIRES post RV zero point (m\,s$^{-1}$)         &                 -             & $-414.4^{+2.4}_{-2.3}$                                  \\[3pt]
APF RV zero point (m\,s$^{-1}$)      &                        -         & $ 14.6^{+2.2}_{-2.2}$                             \\[3pt]
\enddata
\tablecomments{* $\varpi_{\rm DR3}$ and $\sigma[\varpi_{\rm DR3}]$ present the parallax and parallax uncertainty of HR7672 from Gaia DR3 observations.    }
\end{deluxetable*}

\clearpage 

\end{CJK}
\end{document}